\documentclass{aa}  
\usepackage{graphicx}
\usepackage{txfonts}
\usepackage{natbib}
\usepackage{longtable}
\usepackage{subcaption}

\newcommand{\msun}{M$_{\odot}$}
\newcommand{\mygi}{MyGIsFOS}
\newcommand{\Teff}{\ensuremath{T_\mathrm{eff}}}

\newcommand{\loggf}{\ensuremath{\log\,gf}}
\newcommand{\logg}{\ensuremath{\log\,g}}
\def\teff{$T\rm_{eff}$}
\newcommand{\kms}{$\rm km s ^{-1}$}

\usepackage{hyperref}
\hypersetup{colorlinks=true,linkcolor=blue,citecolor=blue,filecolor=blue,urlcolor=blue}
\begin{document} 

\title{MINCE II. Neutron capture elements 
\thanks{Based on observations made with HARPSN at TNG, Fies at NOT, Sophie at OHP and ESPaDOnS at CFHT.}
}
\titlerunning{MINCE II}

\author{
P.~Fran\c{c}ois \inst{1,2} \and
G.~Cescutti \inst{3,4,5} \and
P.~Bonifacio \inst{1} \and
E.~Caffau    \inst{1} \and
L. Monaco \inst{6}\and
M. Steffen \inst{7}\and
J. Puschnig \inst{8}\and
F. Calura \inst{9}\and
S.~Cristallo \inst{10,11} \and
P.~Di Marcantonio \inst{4} \and
V. Dobrovolskas\inst{12} \and
M.~Franchini \inst{4} \and
A.~J.~Gallagher\inst{7} \and
C.~J.~Hansen \inst{13} \and
A. Korn \inst{8} \and
A. Ku\v{c}inskas\inst{12} \and
R.~Lallement \inst{4}\and
L.~Lombardo \inst{13}\and
F.~Lucertini \inst{14}\and
L.~Magrini \inst{15} \and
A.M.~Matas Pinto \inst{1} \and
F.~Matteucci \inst{3,4,5} \and
A.~Mucciarelli \inst{16,9} \and
L.~Sbordone \inst{14} \and
M.~Spite \inst{1} \and
E.~Spitoni \inst{4}\and
M.~Valentini \inst{7}
}

\institute{
  GEPI, Observatoire de Paris, Universit\'{e} PSL, CNRS,  5 Place Jules Janssen, 92190 Meudon, France
\and
 UPJV, Universit\'e de Picardie Jules Verne, P\^ole Scientifique, 33 rue St Leu, 80039, Amiens, France
\and
 Dipartimento di Fisica, Sezione di Astronomia, Università di Trieste, Via G. B. Tiepolo 11, 34143 Trieste, Italy
\and
INAF, Osservatorio Astronomico di Trieste, Via Tiepolo 11, I-34143 Trieste, Italy
\and
INFN, Sezione di Trieste, Via A. Valerio 2, I-34127 Trieste, Italy
\and
Instituto de Astrofisica, Departamento de Ciencias Fisicas, Universidad Andres Bello, Autopista Concepcion-Talcahuano, 7100,
Chile
\and
Leibniz-Institut für Astrophysik Potsdam (AIP), An der Sternwarte 16, 14482, Potsdam, Germany
\and
Division of Astronomy and Space Physics, Department of Physics and Astronomy, Uppsala University, Box 516, 75120 Uppsala, Sweden 
\and
INAF - Osservatorio di Astrofisica e Scienza dello Spazio di Bologna, Via Gobetti 93/3, I-40129 Bologna, Italy
\and
INAF, Osservatorio Astronomico d'Abruzzo, Via Mentore Maggini snc, 64100 Teramo, Italy
\and 
INFN, Sezione di Perugia, Via A. Pascoli snc, 06123 Perugia, Italy
\and
Institute of Theoretical Physics and Astronomy, Vilnius University, Saul\.{e}tekio al. 3, Vilnius, LT-10257, Lithuania
\and
Goethe University Frankfurt, Institute for Applied Physics, Max-von-Laue-Str. 12, 60438 Frankfurt am Main, Germany; 
other institutes 
\and
ESO-European Southern Observatory, Alonso de Cordova 3107, Vitacura, Santiago, Chile
\and
INAF, Osservatorio Astrofisico di Arcetri, Largo E. Fermi 5, 50125, Firenze, Italy
\and 
Dipartimento di Fisica e Astronomia, Universit\`a degli Studi di Bologna, Via Gobetti 93/2, I-40129 Bologna, Italy}

\date{  Received \today ;  Accepted   }

\abstract
{  Most of the studies on the determination of the chemical composition of  metal-poor stars  have been focused on the search of the most pristine stars, searching for the imprints of the ejecta of the first supernovae. Apart from the  rare and very interesting r-enriched stars, few elements are measurable in the very metal-poor stars.  On the other hand, a lot of work has been done also on the thin-disc and thick-disc abundance ratios in a metallicity range from [Fe/H]$>-$1.5 dex to solar.  In the available literature, the intermediate metal-poor stars ($-$2.5$<$[Fe/H]$<-$1.5) have been frequently overlooked.  The MINCE  (Measuring at Intermediate metallicity  Neutron-Capture Elements) project aims to gather the abundances of  neutron-capture elements but also of light elements and  iron peak elements in a large sample of  giant stars in this metallicity range. 
{The missing information has consequences for the precise study of the chemical enrichment of our Galaxy in particular for what concerns neutron-capture elements and it  will be only partially covered by future multi object spectroscopic surveys such as WEAVE and 4MOST.}}
{ The aim of this work is to  study the chemical evolution of galactic sub-components recently identified (i.e. Gaia Sausage Enceladus (GSE), Sequoia). }
{ We used  high signal-to-noise ratios, high-resolution spectra and standard 1D LTE spectrum synthesis to determine the detailed abundances. }
{ We  could determine the abundances for  up to 10 neutron-capture elements (Sr, Y, Zr, Ba, La, Ce, Pr, Nd, Sm and Eu) in 33 stars. The  general trends  of  abundance ratios [n-capture element/Fe]  versus [Fe/H] are in agreement with the results found in the literature. When our sample 
is divided in sub-groups depending on their kinematics, we found that  the run of [Sr/Ba] vs [Ba/H] for the stars belonging to the GSE accretion event shows a tight anti-correlation.  The results for the Sequoia stars, although based on a 
very limited sample, shows a [Sr/Ba] systematically higher than the [Sr/Ba] found in the GSE stars at a given [Ba/H] hinting at a different nucleosynthetic history.   
Stochastic chemical evolution models have been computed to understand the evolution of the GSE chemical composition of Sr and Ba. The first conclusions are that the GSE chemical evolution is similar to the  evolution of a dwarf   galaxy with galactic winds and inefficient star formation. }
{ Detailed abundances of neutron-capture  elements have been  measured in high-resolution, high signal-to-noise spectra of intermediate metal-poor stars, the metallicity range  covered  by the MINCE project. 
These abundances have been compared to detailed stochastic models of galactic chemical evolution. 
}

\maketitle

%
\section{Introduction}
   The MINCE (Measuring at Intermediate metallicity
  Neutron-Capture Elements) project aims to gather abundances
  for the neutron-capture elements of several hundred stars
  at intermediate metallicity using different facilities worldwide.
  The idea is to study the nucleosynthetic signatures that can be found in old stars, in particular in the
  specific class of chemical 
elements with Z $>$  30; that is, the neutron-capture elements.
As neutron-capture elements are formed through several nucleosynthetic channels (mainly the s-process and the r-process), they can be used to constrain 
their source of production throughout the history of the Galaxy. In particular, it will be  possible to determine the spread  in the neutron-capture elements as a function of metallicity, revealing the different sites of production of the r-process that has enriched the interstellar medium at different timescales \citep{Cescutti08, Cescutti15, Simonetti19}.

While most of the observational efforts have been put into the search for the most metal-poor stars,  
 several detailed analyses \citep{ishigaki2013,roederer2014}  have considered the full range of metallicities,  including  the stars in the  intermediate range of metallicity between the most metal-poor ones ([Fe/H]$<-$2.5) and thin- or thick-disc stars ([Fe/H]$>-$1.5)   .

  The interest in analysing stars in this metallicity 
  range is that the determination of  their detailed abundance ratios is associated with 
  accurate kinematics derived from the \textit{Gaia} Early Data Release 3 ({\em Gaia}\,EDR3) \citep{Gaia2016A&A...595A...1G,Gaia2021A&A...649A...1G}.
  This information can be used to  constrain the most recent models of galactic chemical evolution but can also help to    
characterise the recently discovered stellar streams  and galactic substructures \citep[see e.g.][and reference therein]{Belokurov18, Haywood18, Helmi18,  myeong19, Kruijssen2020, naidu2020, malhan2021, feuillet21}.
 Although the SAGA database \citep{suda2008} reveals that the study of the chemical composition of stars in the intermediate metallicity range 
has also been well covered, the aim of our project is to analyse a large sample of stars in  the same metallicity range using  the same methods (codes, a line list, the derivation of stellar parameters, and so on) to derive chemical abundances, with the aim of deriving a set of homogeneous abundances.

  This  paper follows the article of \citet[][hereafter MINCE I]{cescutti2022}. In  their paper, the authors  present the first sample of 46 stars. 
  They measured radial velocities and computed Galactic orbits for all of the stars. They found that eight stars belong to the thin disc and 15 to disrupted satellites, and that the remaining ones cannot be associated with the aforementioned structures; they call the latter halo stars. 
  For 33 stars they  provided abundances of a set of elements up to zinc.
  This article  presents the results of the determination of the abundances for  up to ten  neutron-capture elements using  the same set of reduced spectra.  
  
  The paper is structured as follows. In section \ref{sec:dataset} we present the main characteristics of the spectra that have been used in this study.
  In sect. \ref{sec:linelist} we provide details of the lines that were used to compute the abundances. Section \ref{sec:analysis} gives a summary on how the stellar parameters were determined, and discusses the method used to compute the abundances and their associated  errors. In this section we also show how some elemental abundances are affected  by non-local thermodynamic equilibrium (NLTE) effects.
   In sect. \ref{sec:results} we present  the results of the abundance determinations and a comparison with the literature data. 
Section  \ref{sec:streams} shows a comparison between the abundance  ratios found in {\em Gaia}-Sausage-Enceladus \citep[GSE,][]{Belokurov18,Haywood18,Helmi18} and Sequoia \citep[][]{barba19,villanova19,myeong19} in light of
the  recent abundance results from \citet{aguado2021} and \citet{matsuno2022}.
Detailed models of galactic chemical evolution  for the GSE substructure are presented in sect. \ref{sec:GCE}.
Finally,   sect. \ref{sec:Conclusions} summarises the main conclusions that can be drawn from this first set of MINCE data.

\section{Dataset \label{sec:dataset} }
 
 As more detailed information is given in MINCE I, we recall  the main characteristics of the dataset in this section. 
Spectra were taken with several instruments that 
deliver high-resolution spectra. 
From lower  to higher spectral resolution, we have:

\begin{itemize}
 \item {}
ESPaDOnS \citep{Espadons} is  a fibre-fed spectropolarimeter  installed on the  3.6m Canada-France-Hawaii telescope (CFHT) at Mauna Kea Observatory. The observing mode used (star + sky)  gives  a 
resolving power of R=65,000 and the spectral range extends from 370 nm to 1051 nm.

\item {}
FIES \citep{FIES} is a fibre-fed cross-dispersed high-resolution echelle spectrograph with a maximum spectral resolution of R = 67,000 installed at the Nordic Optical Telescope (NOT), which is a 2.56-m telescope located at the  Roque de los Muchachos Observatory in Canarias. The spectral range is  370-830 nm.

\item {}
The Sophie spectrograph \citep{2006tafp.conf..319B}, installed on the Observatoire de Haute Provence (OHP) 1.93m telescope, has been used in  high-resolution mode,  providing a resolving power of R=75,000 and a spectral range
from 387.2 nm to 694.3 nm.

\item {}
 HARPS-N \citep{HARPSN}  is a high-resolution spectrograph with a resolving power R=115,000 and spectral coverage ranging from 383 to 693 nm.  The spectrograph is installed at the  3.58m Telescopio Nazionale Galileo (TNG) 
 at the Roque de Los Muchachos Observatory. 
 
\end{itemize} 

As was mentioned in the first paper of the series (MINCE I), the spectra were obtained thanks to a total of four proposals with three 
different Principle Investigators: Cescutti for HARPS-north at TNG, E. Spitoni for FIES at NOT,
and P. Bonifacio for Sophie at OHP and ESPaDOnS at CFHT.


\section{Line list \label{sec:linelist}}

In the spectra of this work, we could identify the lines of ten neutron-capture elements.  
In order to have a complete description of the lines used in our analysis, we report below the sources of the oscillator strengths and hyperfine structure (HFS) if present for a given element. 

\subsection {Strontium}

We first selected the three ionised strontium lines at 4077.71  $\AA$, 4161.79 $\AA$, and 4215.52 $\AA$, and 
the \ion{Sr}{I} line at 4607.33 $\AA$. Given that the stars in our sample 
are giant stars and that the metallicity range is between $\simeq$ solar and [Fe/H]  $\simeq$ -2.5 dex, the \ion{Sr}{II} lines are  saturated and  the placement of the 
continuum is difficult to evaluate. We find that the Sr I line is  a better choice over this range of stellar parameters. 

\subsection {Yttrium }
Yttrium abundances were determined by fitting the \ion{Y}{II} lines 4854.86 $\AA$, 4883.68 $\AA$, 4900.12 $\AA$, 5087.42 $\AA$,  5119.11 $\AA$, 5123.21 $\AA$, 5200.41 $\AA$, 5402.77 $\AA$, and 5662.93 $\AA$.
We adopted  the oscillator strengths computed by \citet{hannaford1982}. 

\subsection {Zirconium }

The \ion{Zr}{II} lines at 4317.30 \,$\AA$, 4613.95 $\AA$, 5112.27 $\AA$, and 5350.35 $\AA$ 
were used to determine its abundance.  The \loggf s are from \citet{ljung2006}. 

\subsection { Barium}

We used  the \ion{Ba}{II} lines at 5853.68 $\AA$, 6141.71 $\AA$, and    6496.90 $\AA$ 
to derive the Barium abundances. Hyperfine splitting   and isotopic shifts were taken into account, following \citet{gallagher2020}. The isotopic fractions for isotopes 134, 135, 136, 137, and 138 are  the r-process fractions  0\%, 40\%, 0\%, 32\%, and 28\%  respectively, following \citet{mcwilliam98}.

\subsection { Lanthanum }

The \ion{La}{II} lines at 4662.51 $\AA$, 4920.98 $\AA$, 4921.78 $\AA$, 5114.56 $\AA$, 5122.99 $\AA$, and 5290.82 $\AA$ 
were used. The \loggf's are from  \citet{lawler2001}. 
Hyperfine structure was taken from \citet{ivans2006}.

\subsection {Cerium }

The \ion{Ce}{II} lines at  4539.85 $\AA$, 4562.28 $\AA$, 4628.16 $\AA$, and 5187.46 $\AA$ 
were used in this study. The \loggf's are from \citet{palmeri2000}.

\subsection { Praseodymium}

The \ion{Pr}{II} abundances were determined using the \ion{Pr}{II} lines  at 5219.05 $\AA$, 5220.11 $\AA$, 5259.73 $\AA$, 5322.76 $\AA$, and 6165.89 $\AA$. 
The oscillator strengths and HFS are from \citet{li2007}, \citet{ivarsson2001},  and \citet{sneden2009}.

\subsection {Neodymium }

The singly ionised transitions of neodymium (\ion{Nd}{II}) at 4501.81 $\AA$, 4859.03 $\AA$,   4959.12 $\AA$,  5076.58 $\AA$,  5255.50 $\AA$,  5293.16 $\AA$, and      5319.81 $\AA$ 
were used to determine the neodymium abundance.
The atomic data are from \citet{denhartog2003}.

\subsection {Samarium}

The \ion{Sm}{II} lines at  4566.20 $\AA$, 4615.44 $\AA$, 4669.64 $\AA$, 4674.59 $\AA$, 4676.90 $\AA$, 4791.58 $\AA$, and 4913.26 $\AA$ 
were used in this work.
The \loggf s were taken from \citet{lawler2006}. There are no HFSs available for these transitions. 
In any case, they would only affect the odd isotopes, $^{147}$Sm and $^{149}$Sm, which account
for only about 29\% of the Sm abundance in the Solar System.

\subsection {Europium}

The two \ion{Eu}{II}  lines  at 4435.58 $\AA$  and 6645.10 $\AA$ were measured in our spectra. For a couple of stars, we could also measure the transition at  4522.58 $\AA$.
Atomic quantities such as oscillator strengths, HFS, and isotopic shifts were adopted from \citet{lawler2001}. We assumed a 50\%-50\%  mix for the Eu isotopes 151 and 153, following \citet{ivans2006}. We  used  linemake \footnote{https://github.com/vmplacco/linemake} to generate the line list for europium \citep{placco2021}.

\section{Analysis} \label{sec:analysis}
\subsection{Stellar parameters} 

The stellar parameters were taken from MINCE I. 
To summarise, the stellar parameters were derived  using  colours and distances from 
{\em Gaia}\,EDR3 and de-reddened using the maps from  \citet{schlafy};
the process was iterated up to the point when the changes in stellar parameters were less than 50\,K in \Teff and less than 0.05\,dex in \logg.
For the micro-turbulence, we employed the calibration by \citet{mashonkina17} at any iteration, and applied these values as the final choice.
The stellar parameters and derived metallicity are reported in Table\,\ref{T1}.

\begin{table}
\caption{Stellar parameters of the sample.  Signal-to-noise ratios are given per resolution element at 500 nm. } 
\label{T1}
\begin{tabular}{lrrrrr }
\hline
  \multicolumn{1}{c}{Star} &
  \multicolumn{1}{c}{\teff} &
  \multicolumn{1}{c}{\logg} &
  \multicolumn{1}{c}{$\xi$} &
  \multicolumn{1}{c}{[Fe/H]} &
  \multicolumn{1}{c}{SNR}  \\

  \multicolumn{1}{c}{} &
  \multicolumn{1}{c}{[K]} &
  \multicolumn{1}{c}{[gcs]} &
  \multicolumn{1}{c}{\kms} &
  \multicolumn{1}{c}{} &
   \multicolumn{1}{c}{} 
  \\
  
\hline
  HD 115575 & 4393 & 1.08 & 1.94 & -1.99 &  88 \\      
  TYC 4267-2023-1 & 4660 & 0.96 & 2.11 & -1.74 & 68\\ 
  BD+31  2143 & 4565 & 1.15 & 2.03 & -2.37 & 100 \\   
   BD+20  3298 & 4154 & 0.57 & 2.07 & -1.95 & 92\\   
  TYC 1008-1200-1 & 4199 & 0.78 & 2.01 & -2.23 & 50 \\
   HD 238439 & 4154 & 0.53 & 2.10 & -2.09  & 87 \\ 
    HD 142614 & 4316 & 0.87 & 1.96 & -1.46 & 95 \\  
  BD+04    18 & 4053 & 0.74 & 1.9 & -1.48 & 58 \\  
   BD+39  3309 & 4909 & 1.73 & 1.94 & -2.58 &  90 \\   
   TYC 2824-1963-1 & 4036 & 0.64 & 1.95 & -1.60 & 54\\       
  TYC 4001-1161-1 & 4129 & 0.75 & 1.94 & -1.62& 59 \\
  TYC 4221-640-1 & 4295 & 0.66 & 2.12 & -2.27 & 54 \\   
  TYC    4-369-1 & 4234 & 0.89 & 1.94 & -1.84& 50 \\            
    BD-00  4538 & 4482 & 1.29 & 1.88 & -1.9 & 99 \\   
   BD+03  4904 & 4497 & 1.03 & 2.06 & -2.58 &  58\\    
  BD+07  4625 & 4757 & 1.64 & 1.86 & -1.93  & 98 \\     
    BD+11  2896 & 4254 & 1.07 & 1.83 & -1.41 & 85 \\
    BD+21  4759 & 4503 & 1.06 & 2.05 & -2.51 & 60 \\ 
    BD+25  4520 & 4276 & 0.70 & 2.08 & -2.28 & 98 \\ 
     BD+32  2483 & 4516 & 1.17 & 1.99 & -2.25 & 88 \\
    BD+35  4847 & 4237 & 0.76 & 2.01 & -1.92 &  84 \\ 
  BD+48  2167 & 4468 & 1.00 & 2.04 & -2.29 &  99\\ 
    BD-07  3523 & 4193 & 0.71 & 2.02 & -1.95 & 84 \\   
 BD+06  2880 & 4167 & 0.82 & 1.91 & -1.45 & 74\\ 
  HD 139423 & 4287 & 0.70 & 2.05 & -1.71 & 83 \\   
  HD 208316 & 4249 & 0.79 & 1.98 & -1.61 & 90 \\   
   HD 354750 & 4626 & 0.90 & 2.17 & -2.36 & 83 \\  
  TYC 2588-1386-1 & 4130 & 0.66 & 1.99 & -1.74 & 50 \\    
  TYC 3085-119-1 & 4820 & 2.26 & 1.56 & -1.51 & 85 \\   
  TYC   33-446-1 & 4289 & 0.75 & 2.07 & -2.22 & 85 \\     
  TYC 3944-698-1 & 4091 & 0.45 & 2.11 & -2.18 & 50 \\  
  TYC 4331-136-1 & 4133 & 0.50 & 2.13 & -2.53 & 46\\  
  TYC 4584-784-1 & 4232 & 0.80 & 2.00 & -2.04 & 40 \\   

  \hline
  \\

  \end{tabular}

\end{table}

\subsection{Abundances}

We carried out a classical 1D local thermodynamic equilibrium (LTE) analysis using OSMARCS model  atmospheres  \citep{gustafsson1975,   gustafsson2003,   gustafsson2008,  plez1992, edvardsson1993}. The abundances  used  in  the  model  atmospheres  were  solar-scaled with respect to the  \citet{grevesse2000} solar abundances, except for the $\alpha$ elements that are enhanced by 0.4 dex. We corrected the resulting abundances by taking into account the difference  between the solar values of  \citep{grevesse2000} and   \citet{Lodders09}
. The solar abundances we adopted are reported in Table\,\ref{tab:solarabbo}.

The abundance analysis was performed using the LTE spectral  line  analysis code turbospectrum  \citep{alvarez1998,    plez2012}, which treats scattering in detail. The abundances were determined by matching a synthetic spectrum centred on each line of interest to the observed one.  Tables \ref{tab:linelist1} and \ref{tab:linelist2}  list the lines used to measure the abundances  in our sample of stars. Detailed HFS components have also been included in these tables.  For the spectrum synthesis, we took into account  all the known blending lines from the VALD database \citep[][and references therein]{Ryabchikova2015}.

When not specified, we adopted the abundance derived from \ion{Fe}{i} lines as the metallicity .
Since our surface gravities are derived from the parallaxes
and not the Fe ionisation equilibrium, in order to minimise the gravity
dependence in abundance ratios, we used [X/Fe] =  [X/\ion{Fe}{i}], where
X is a neutral species and [X/Fe] = [X/\ion{Fe}{ii}] for ionised species.

\begin{table}
\caption{Solar abundances used throughout this paper  are from \citet{Lodders09}.}
\label{tab:solarabbo}
\center 
\begin{tabular}{ll}
\hline
Element & A(X)  \\
\hline
Sr & 2.90  \\
Y  & 2.20  \\
Zr & 2.57    \\
Ba & 2.18 \\
La & 1.19 \\
Ce & 1.60  \\
Pr & 0.77  \\
Nd & 1.47  \\
Sm & 0.96 \\
Eu & 0.53  \\

\hline
\end{tabular}
\end{table}

\subsection{Non-local thermodynamic equilibrium effects}

Strontium, barium, and europium are known to be sensitive to departures from LTE (or NLTE effects), particularly in metal-poor stars.

In this study, we used the \ion{Sr}{I} strontium line at 4607.33 $\AA$. 
The abundance of strontium in metal-poor stars has been studied in detail 
by \citet{hansen2013}. They  confirmed that the ionisation equilibrium between \ion{Sr}{I}  and \ion{Sr}{II}  
is satisfied in NLTE but not in LTE, where the difference
between neutral and ionised Sr is on average -0.3 dex. 
We applied a correction of -0.3 dex on our \ion{Sr}{I} result to match the literature results that are mostly based on \ion{Sr}{II}  lines.  We note that we used the \ion{Sr}{I} line because the \ion{Sr}{II} lines  visible in our spectra are saturated and the placement of the continuum is difficult to evaluate.

\citet{mashonkina2008} computed the NLTE effect on  barium  and europium abundances in the metal-poor giant HD~122563  ([Fe/H] $\simeq$ -2.6 dex), a metallicity at the lower end of our sample.
 Based on the  \ion{Ba}{II}  lines  at 4554.031 \AA\ and 6496.90  \AA, they derived a correction (NLTE - LTE) of +0.03 dex.
 
\citet{korotin2015} computed NLTE equivalent widths (EW) and NLTE abundance corrections for the four main  \ion{Ba}{II} 
lines:  4554.0, 5853.7, 6141.7,  and 6496.9 \AA. 
By comparing LTE and NLTE abundances, they showed  that the LTE calculations for the weaker 5853.7 \AA\  line tend to yield LTE abundances close to the NLTE ones, and that
the difference between the LTE and NLTE abundance for the three red Ba lines is on average  $\pm$ 0.1 dex. In some cases, the effect can reach 0.2 dex.  They also showed that the 4554.03  \AA\  line is not suitable for abundance determination. 

More recently, NLTE   departure coefficients for the large spectroscopic survey  GALAH have been calculated by \citet{amarsi2020} for 13 elements, including barium. They constructed grids of departure coefficients  that have been implemented into the GALAH
Data Release 3 analysis pipeline in order to complement the existing NLTE grid for iron.  Their grids cover the range of metallicity and gravities encountered in our sample. They studied the BaII lines at 5853.7 \AA\ and 6496.9 \AA\ and derived an abundance correction (NLTE - LTE)  ranging from -0.01 to 0.18 dex.

\begin{table*}[]
    \centering
    \begin{tabular}{lllrrrrrrrrl}

    Star    &    Teff  & log g &[Fe/H] &[Ba/Fe] & Ba corr & [Ba/Fe] & Ba corr & [Ba/Fe] &  Ba corr & Eq. Width   &  Ba corr  \\
           &          &       &       & 5853.7&   5853.7 &  6141.7 & 6141.7 & 6496.9 & 6496.9  &  5853.7  &\\

\hline
Sun           &   5770 &  4.44  & 0.00 & 0.00 &   0.09  &  0.00 &     0.06 & 0.00 &   0.05    & -   & - \\

HD\,115575    &   4393 &  1.50 & -1.99 & -0.32  &   0.17 & -0.22 & 0.11 & -0.21 & 0.10 & 83.0 & 0.2 \\
BD\,+48\,2167 &   4468 &  1.50 & -2.29 & -0.13  &   0.22 &  -0.03 & 0.12 & 0.02 & 0.11 & 83.5 & 0.2\\
BD\,+11\,2896 &   4254 &  1.50 & -1.41 & 0.02 &   0.03 & 0.12 & -0.01 & 0.02 &-0.06 & 120.8 & 0.067\\
BD\,-00\,4538 &   4482 &  1.50 & -1.90 & -0.03  &   0.23 & 0.07 & 0.12 & 0.02 & 0.12 & 98.8 & 0.216\\
BD\,+03\,4904 &   4497 &  1.50 & -2.58 & -0.31  &   0.18 & -0.36 & 0.12 & -0.41 &  0.14 & 49.2 & 0.177\\

\hline
    \end{tabular}
    \caption{Barium 3D NLTE corrections for a subset of stars, computed for the  \ion{Ba}{II} lines at 5853.7\,\AA, 6141.7\,\AA, and 6496.9\,\AA. Given the 1D LTE Ba abundance [Ba/Fe], they are extracted from the 3D NLTE – 1D LTE tables provided by Gallagher et al. (in prep). The last two columns give the measured EW of the barium line at 5853.7\,\AA\ and the corresponding barium abundance correction computed by a slightly different interpolation method from the same tables.} 
    \label{tab:ba_corrections}
\end{table*}

We computed independent 3D NLTE corrections of the barium abundances
based on 3D NLTE – 1D LTE grids recently computed by Gallagher et al. (in prep). The tables representing the Ba abundance correction grids can be
found online at ChETEC-INFRA \footnote{http://chetec-infra.eu/3dnlte}, together with background
information and instructions on how to use the corrections.
The solar 3D NLTE barium abundance used for these correction grids
($A$(Ba)$^\odot_{\rm 3D-NLTE}=2.27$) was taken from \citet{gallagher2020}.  

\begin{figure}
\centering
\includegraphics[width=1.0\linewidth]{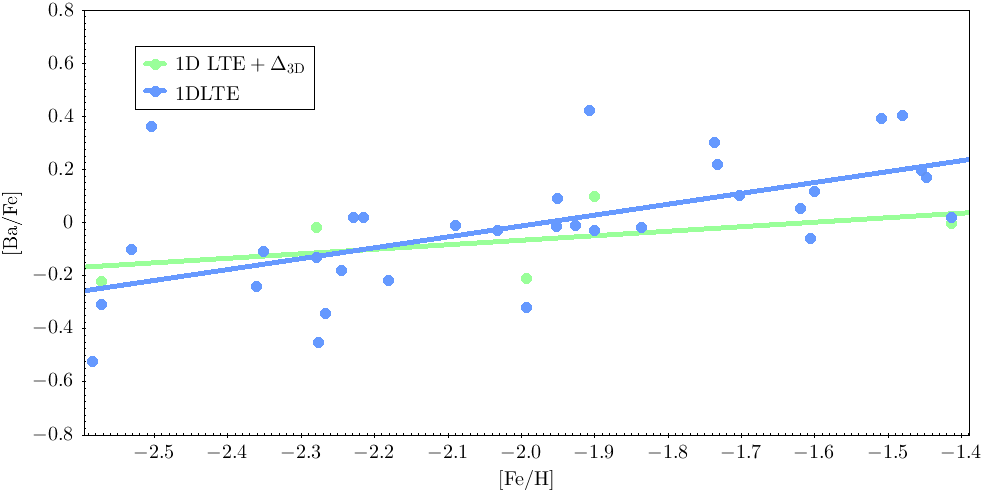}
 \caption{Barium abundance, [Ba/Fe], vs metallicity, [Fe/H]. Blue dots represent the 1D LTE barium abundances derived from the barium line at 5853 \AA. Green dots represent the corrected [Ba/Fe] values obtained for a subset of five stars by adding the 3D NLTE corrections (Ba corr$^{(\ast)}$ - Ba corr$^{(\odot)}$) given in Table \ref{tab:ba_corrections} to the 1D LTE [Ba/Fe] abundance. The blue line (resp. green line) is the linear fit to the 1D LTE (3D NLTE) abundances.}
\label{fig:NLTE_BaH}
\end{figure} 

Unfortunately, most of our MINCE sample of stars have very low $T_{\rm eff}$
and $\log g$ values that are not covered by the Gallagher et al.\ correction
grid. In order to obtain corrections for at least a few targets, we chose to derive corrections using a nearest-neighbour interpolation approach;
that is, we assumed $\log g = 1.5$ for targets with gravities in the range  $1.0 < \log g \le 1.5$. Table~\ref{tab:ba_corrections} shows the 3D NLTE corrections (the Ba corr columns) for the three main \ion{Ba}{II} lines. These corrections are to be added to the 1D LTE barium abundance, $A$(Ba), to obtain the 3D NLTE Ba abundance ($A$(Ba)$\equiv \log(N$(Ba)$-\log(N$(H)$+12$). The corrections for [Ba/H] or [Ba/Fe] are given by the difference of stellar minus solar corrections.

The results are plotted in Fig.\ref{fig:NLTE_BaH}. The blue dots represent the
uncorrected 1D LTE [Ba/Fe] abundance of our sample stars. The green dots indicate
the [Ba/Fe] abundances of the five targets listed in Table \ref{tab:ba_corrections}
after correction for 3D NLTE effects.

The final corrections for [Ba/Fe] are of the order of $\simeq$ 0.1 dex over the
range of metallicity in our sample. Importantly, the 3D NLTE corrections do
not significantly affect the trend of [Ba/Fe] abundances with metallicity.

Europium NLTE corrections have been computed by \citet{mashonkina2008}  for HD~125563,  a cool metal-poor giant ([Fe/H] $\simeq$ -2.6 dex).  
They determined a NLTE - LTE abundance correction of 0.12 dex.

From these studies, we can conclude that NLTE effect corrections for barium and europium are rather small compared to the large range of abundance ratios of [Sr/Fe] and [Ba/Fe]. Moreover, it is unlikely that the dispersion found for these
abundance ratios at a given metallicity can be attributed to the adopted LTE assumption.

\subsection{Error budget}

\begin{table}
\caption{ Sensitivity of abundances on atmospheric parameters. }
\label{tab:errors}
\begin{tabular}{lccc}
\hline
Element & $\Delta T_{eff}$  & $\Delta log~g$   & $\Delta \xi$     \\
               &    +100K &  +0.2 dex  & 0.2  km/s \\
\hline
\ion{Sr}{i}  &  -0.05 &    -0.02 &   +0.05 \\     
\ion{Y}{ii}  &  -0.03 &    -0.12 &   +0.05  \\ 
\ion{Zr}{ii} &  +0.02 &    -0.15 &   +0.04  \\ 
\ion{Ba}{ii} &  -0.05 &    -0.11 &   +0.28  \\ 
\ion{La}{ii} &  -0.05 &    -0.12 &   +0.01  \\ 
\ion{Ce}{ii} &  -0.07 &    -0.11 &   +0.03  \\ 
\ion{Pr}{ii} &  -0.04 &    -0.12 &   +0.00  \\ 
\ion{Nd}{ii} &  -0.04 &    -0.13 &   +0.00  \\ 
\ion{Sm}{ii} &  -0.06 &    -0.12 &   +0.02  \\ 
\ion{Eu}{ii} &  -0.02 &    -0.12 &   +0.01 \\
\hline
\end{tabular}
\end{table}

Table  \ref{tab:errors}   lists an estimate of the errors that are due to typical
uncertainties in the stellar parameters. 
We adopted the uncertainties on the stellar parameters as : $\Delta T_{\rm eff}$ =100 K,  $\Delta$ log~g = 0.20 dex, and $\Delta$  $\xi$  = 0.2  km s$^{-1}$. 
These are typical uncertainties used to estimate the  sensitivity of each parameter on the abundance determination.
These  adopted uncertainties  for $T_{\rm eff}$ and $log~g$ are of the order of the standard deviation found 
between the stellar parameters derived  using   MyGisFos  \citep{Sbordone14} taken in \citet{cescutti2022} and the stellar parameters obtained by Starhorse \citep{Anders19}. More details can be found in \citet{cescutti2022}.
The 0.2 km/s error on the microturbulence velocity corresponds to the acceptable variation of this parameter, giving abundances of  \ion{Fe}{I} independently of the excitation potential of the line. A change in the stellar parameters leads to a change in the [X/Fe] abundance derived for the star.

These errors were estimated by varying $T_{\rm eff}$ , log~g, and $\xi$ in the model atmosphere of
 BD\,+11\,2896.   We chose this star as 
 the  determination  of the  abundances of all the elements analysed in the article was possible.

 As the stars in our sample have stellar parameters close to  BD\,+11\,2896,
the other stars yield similar results.  In particular,  this is also the case for  the lower metallicity range of our sample of stars.

The total error was estimated
by adding the quadratic sum of the uncertainties in the stellar
parameters and the error in the fitting procedure of the synthetic
spectrum and the observed spectrum (the main source of error
comes from the uncertainty in the placement of the continuum). 
The error in the fitting procedure  can be estimated by determining the line-to-line scatter  of the abundance when  several lines of a given element are available. These errors are given in Table \ref{tab:abund}.

\section{Results \label{sec:results}}

In Table \ref{tab:abund}, we present the results of the abundance determination.  For each element, we give the [X/H] ratios and the $\sigma$(X), calculated as the standard deviation of the mean value when abundance determination
from several lines is given. [\ion{Fe}{i}/H] and [\ion{Fe}{ii}/H] are from MINCE I.  Hyphens in the table  means  that the corresponding line was severely blended and  the blend was  dominated by other lines, rendering an abundance determination impossible.

\subsection {General comparison with literature results}

\begin{figure}
 \begin{subfigure}{.5\textwidth}
  \centering
\includegraphics[width=1.0\linewidth]{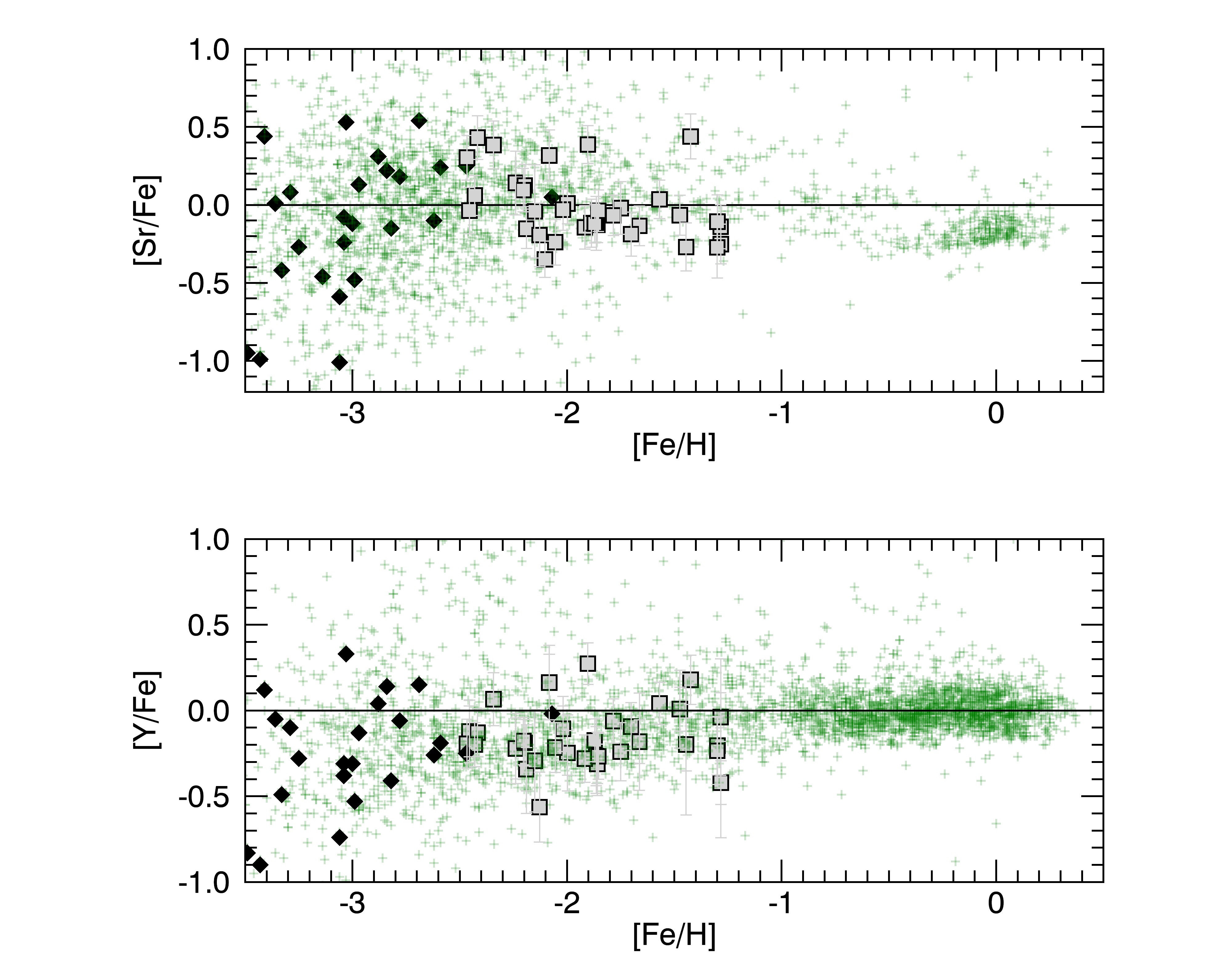}
\end{subfigure} \\%
\begin{subfigure}{.5\textwidth}
 \centering
\includegraphics[width=1.0\linewidth]{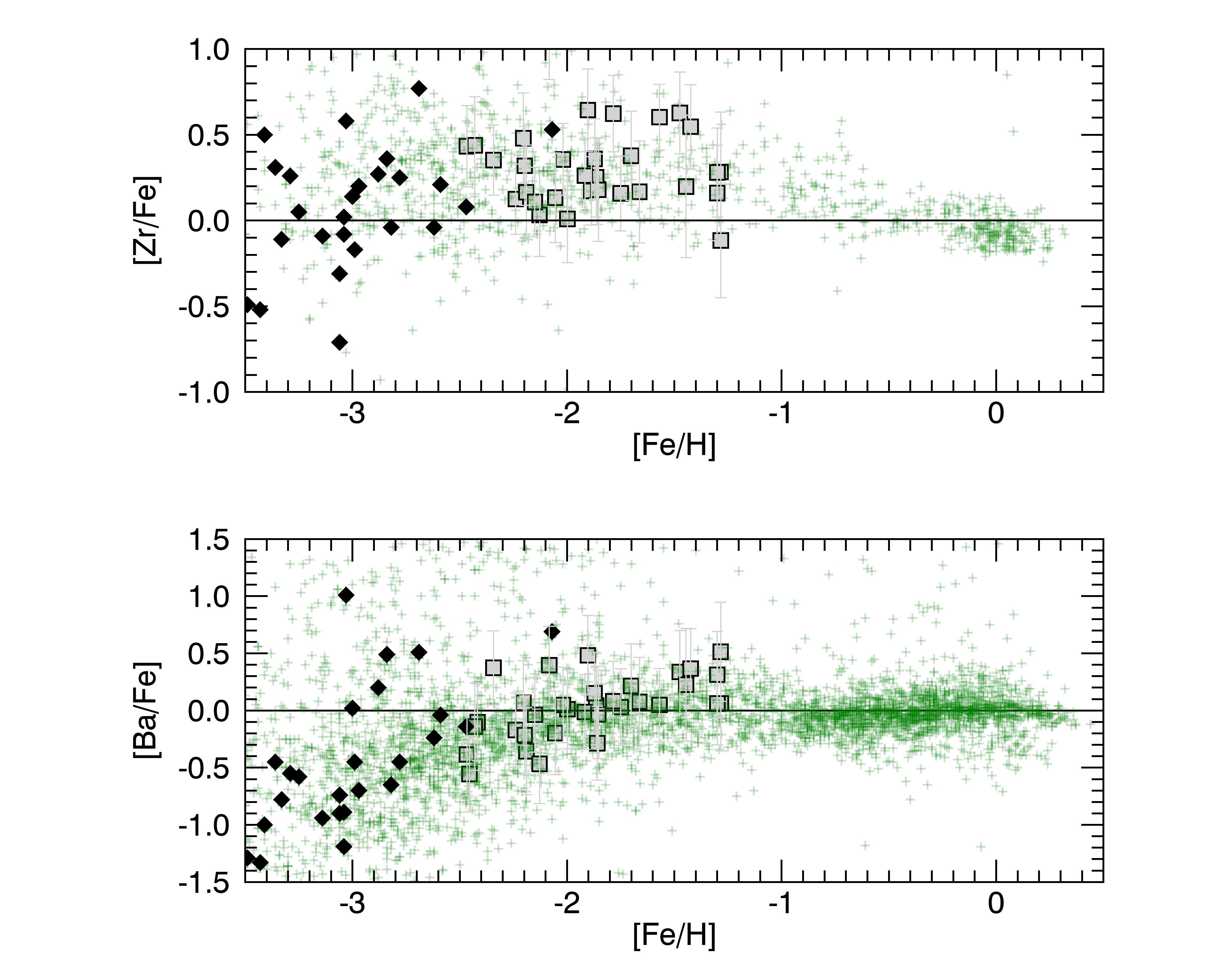}
\end{subfigure} \\
 \caption{Abundance  ratios   of [Sr/Fe],  [Y/Fe],  [Zr/Fe], and  [Ba/Fe]  as a function of [Fe/H] for the stars of our sample,  shown as grey squares.   We added the stars from the  SAGA database \citep{suda2008}, which are represented by green crosses.
 Results for very metal-poor stars from \citet{francois2007} are shown as black diamonds.}
\label{fig:lAbu_plotM1}

 \end{figure}

\begin{figure}
 \begin{subfigure}{.5\textwidth}
 \centering
\includegraphics[width=1.0\linewidth]{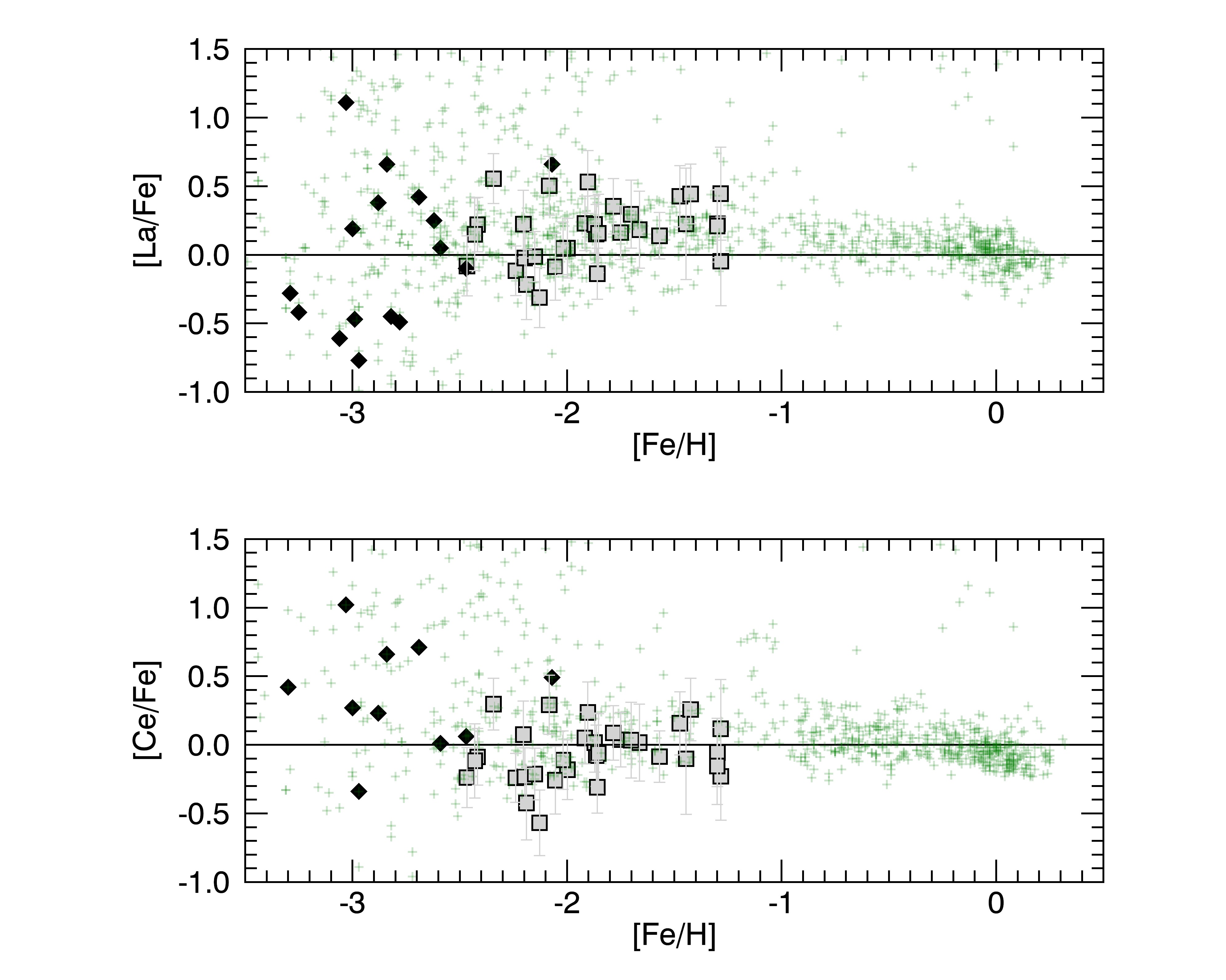}
 \end{subfigure} \\%
\begin{subfigure}{.5\textwidth}
 \centering
\includegraphics[width=1.0\linewidth]{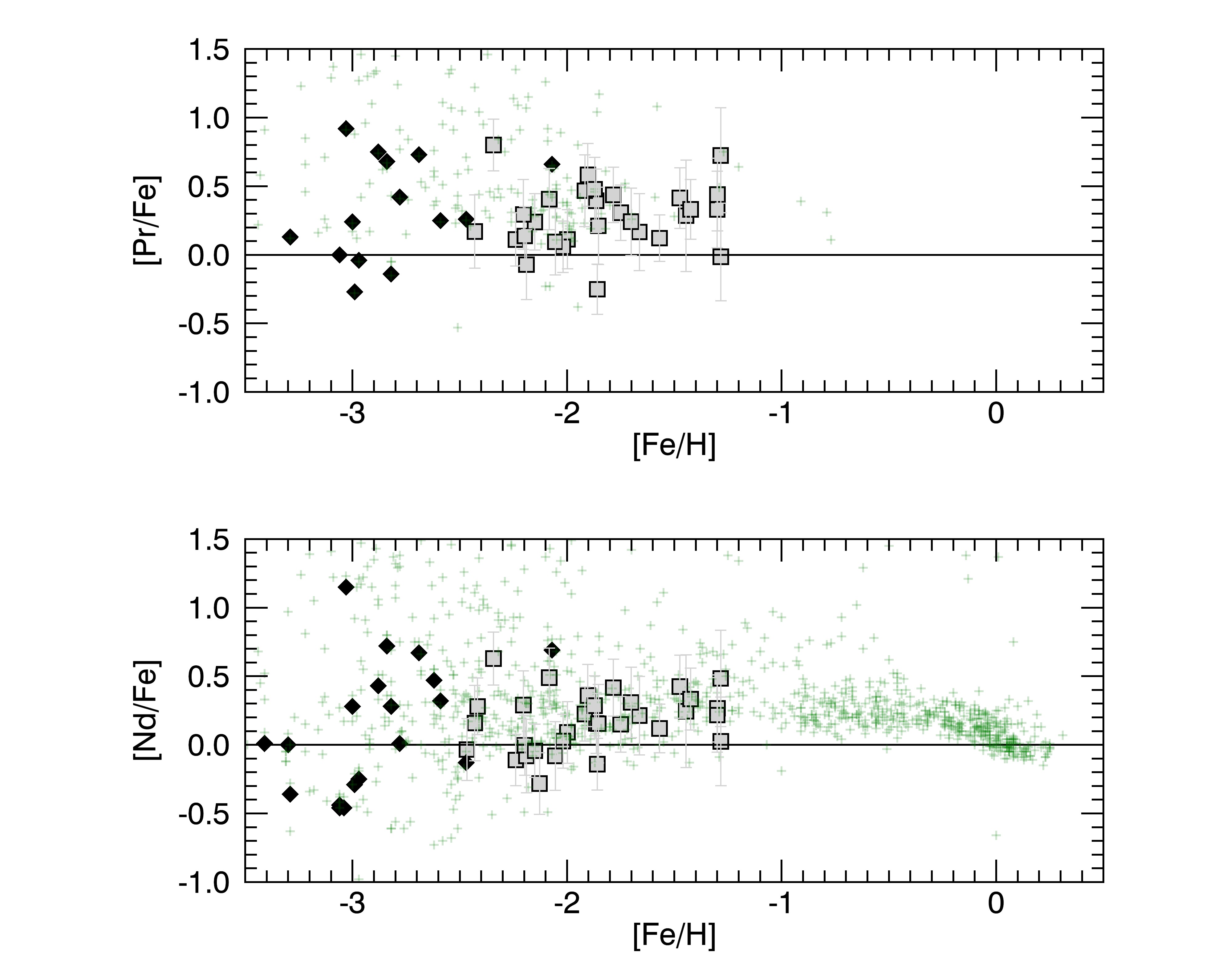}
 \end{subfigure} \\%
 \caption{Abundance ratios   of   [La/Fe], [Ce/Fe], [Pr/Fe], and  [Nd/Fe]  as a function of [Fe/H] for the stars in our sample shown as grey squares.   We added the stars from the  SAGA database \citep{suda2008}, which are represented by green crosses.  Results for very metal-poor stars from \citet{francois2007} are shown as black diamonds.}
\label{fig:lAbu_plotM2}
 \end{figure}

\begin{figure}
 \centering
\includegraphics[width=1.0\linewidth]{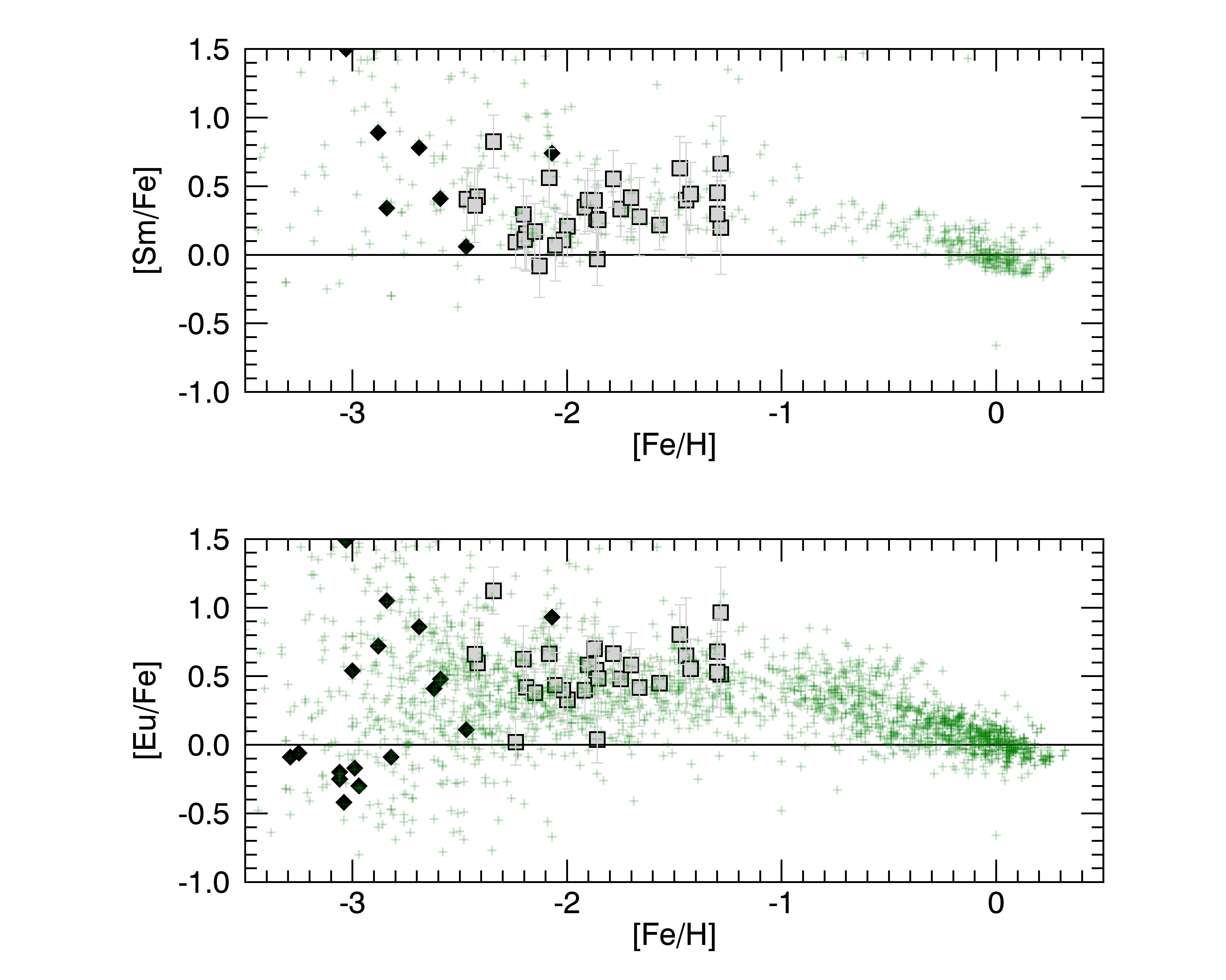}
 \caption{Abundance  ratios   of [Sm/Fe]  and [Eu/Fe]  as a function of [Fe/H] for the stars in our sample shown as grey squares.   We added the stars from the  SAGA database, which are  represented by green crosses. Results for very metal-poor stars from \citet{francois2007} are shown as black diamonds.}
\label{fig:lAbu_plotM3}
 \end{figure}

In Figs. \ref{fig:lAbu_plotM1}  to \ref{fig:lAbu_plotM3}   we show our results, compared with data from a large compilation of results found in the SAGA database \citep{suda2008}.
In this compilation, all  the abundance ratios  [X/Fe]  have  been  recalculated  with  the  
solar  abundances of \citet{asplund2009}.  We did not  apply any  abundance corrections  to take into account the difference between the solar abundances adopted in  our study and the ones used in the SAGA database data. For all the elements but one  (Lanthanum), the difference ranges from 0.00 to 0.05. For La, our adopted solar abundance is 0.09  dex higher than in \citet{asplund2009}.
We have also added (shown with black symbols) the  abundance results from \citet{francois2007}, using the large programme “First Stars”  sample \citep{cayrel2004}. Although the “First Stars” sample is dedicated to extremely metal-poor stars, their abundances have been determined by the same methods (i.e., atomic data, synthesis code). The continuity in the abundance trends, [X/Fe] versus [Fe/H], is a good indicator that no systematic offset is present in the analysis of this sample of moderately metal-poor stars.

The comparison of our abundance ratios with the literature data does not show any offset or a different trend to [Fe/H] or [Ba/H]. The only significant difference is a 
visibly lower dispersion as a function of [Fe/H].


\subsection {Comparison with main-sequence turn-off star abundances}

The main goal of this comparison is to evaluate the possible offsets or trends that could appear from the use of a different set of absorption lines or the same lines with different strengths. 
As our sample is made of mildly metal-poor giant stars, the  lines  we measured are stronger than in dwarf stars. They can also be affected by more severe blends. 

\begin{figure}
 \centering
\includegraphics[width=1.0\linewidth]{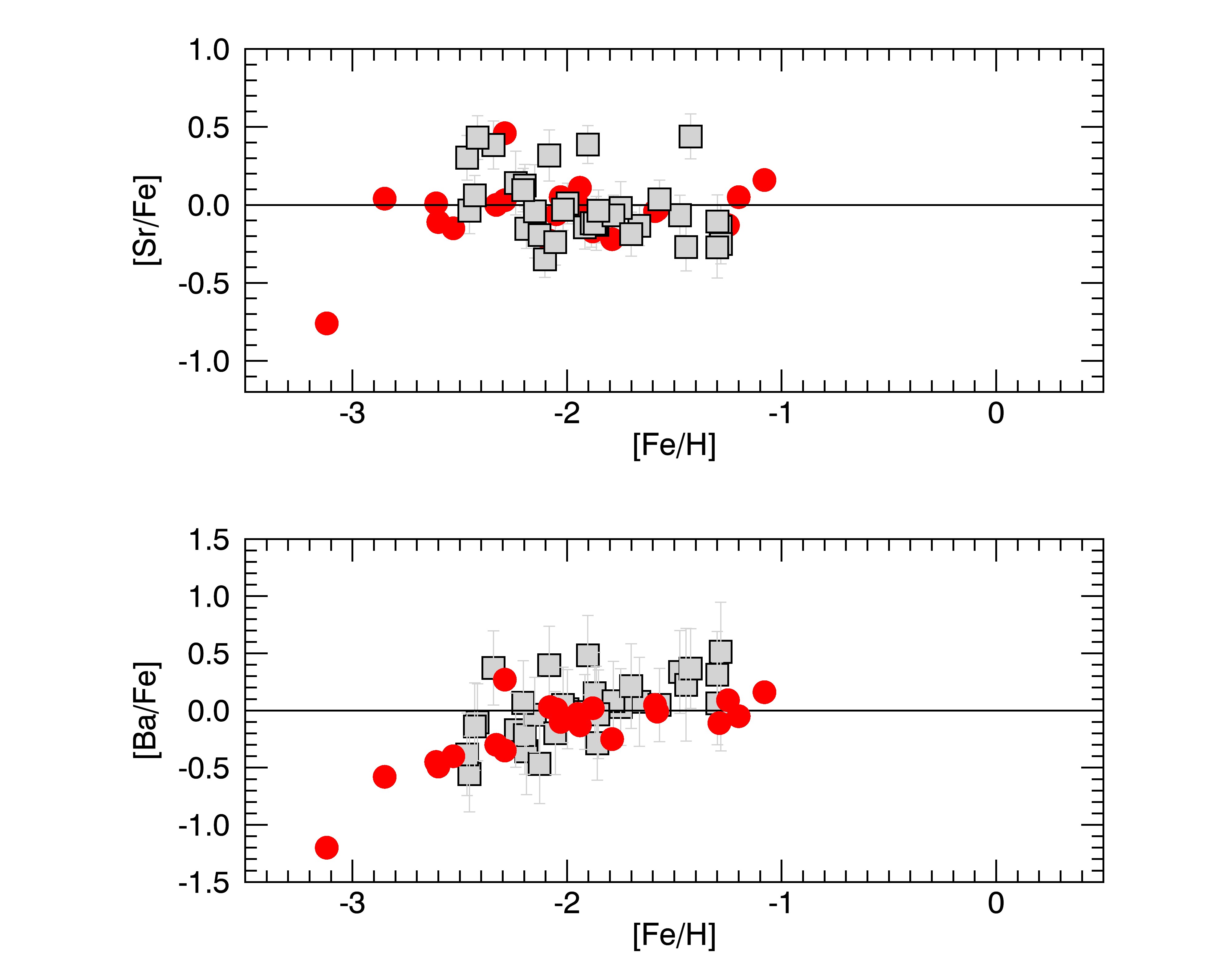}
 \caption{Abundance ratios  of [Sr/Fe]  and [Ba/Fe] as a function of [Fe/H]. Our sample is represented as grey squares. Main-sequence stars from \citet{roederer2014} are shown as red circles.  }
\label{fig:lAbu_SrFe}
 \end{figure}

In Fig. \ref{fig:lAbu_SrFe}, we plot the abundance ratios [Sr/Fe] and [Ba/Fe] as a function of [Fe/H] for our sample of stars.
We have added the abundance results for a sample of main-sequence stars from \citet{roederer2014} to see possible differences between evolved and un-evolved stars.

In our project, we have chosen to observe giant stars for two main reasons.  The first is that  they are intrinsically brighter and easier to observe with 2-meter-class telescopes. Even with telescopes with moderate apertures, 
we can obtain  spectra for a large sample of giant stars  with exposure times not exceeding 1 h.  
 The second and most important reason is  that we can detect more neutron-capture element species  that can only be measured from weak lines not visible in the spectra of main-sequence turn-off stars. 
In  Fig. \ref{fig:lAbu_SrFe}, we highlight two important neutron-capture elements.  These elements are crucial as some of their lines can be detected in the most metal-poor stars.
Our figures show that no difference is found in the location of the abundance trends of [Sr/Fe] and [Ba/Fe] versus [Fe/H].  However, we note that the results for main-sequence stars show a smaller dispersion.

\subsection {Galactic substructures}

 Recent years have witnessed significant strides in our understanding of the accretion history of the Milky Way.
Thanks to Gaia kinematics results, it is now possible to identify stars in the solar vicinity  that may be the result of past  accretion events \citep[see e.g.][and references therein]{Helmi20}. One should however keep in mind that even a single accretion event may result
in multiple dynamical substructures \citep{Jean-Baptiste17} and that a purely dynamical selection is always
subject to contamination \citep[see e.g. the discussion on the metallicity distribution function of GSE in][]{toposVI}; hence, the interest in complementing the dynamical information with chemical information.
Concurrently, advancements in the theoretical framework have paralleled these observations, enhancing our comprehensive understanding of the complex processes governing the Milky Way's accretion history \citep{Calura09,Murphy22,Garcia23}.
In MINCE I, kinematics and action properties  have been used to identify several stars likely belonging to the GSE \citep{Belokurov18,Haywood18,Helmi18} 
and Sequoia \citep[][]{barba19,villanova19,myeong19} accretion events.
In this section, we present our results  and compare them  with literature data, taking into account  the substructure with which the stars can be  associated. For our sample of stars, we adopted the separation into 
GSE, Sequoia, and thin or thick discs, as was suggested by MINCE I. The stars not identified in these three categories are halo stars. In this sample, TYC 3085-119-1 was considered to be a thick-disc star, following the classification of \citet{bensby14}. 
 
Among the large set of results that can be found in the literature, we selected a sample 
where the separation between galactic components (thick disc, inner  halo, and outer halo)  was considered. 
The paper from  \citet{ishigaki2013} gathers a  good sample of stars from the different sub-components of the Galaxy.   

\begin{figure}
 \begin{subfigure}{.5\textwidth}
 \centering
\includegraphics[width=1.0\linewidth]{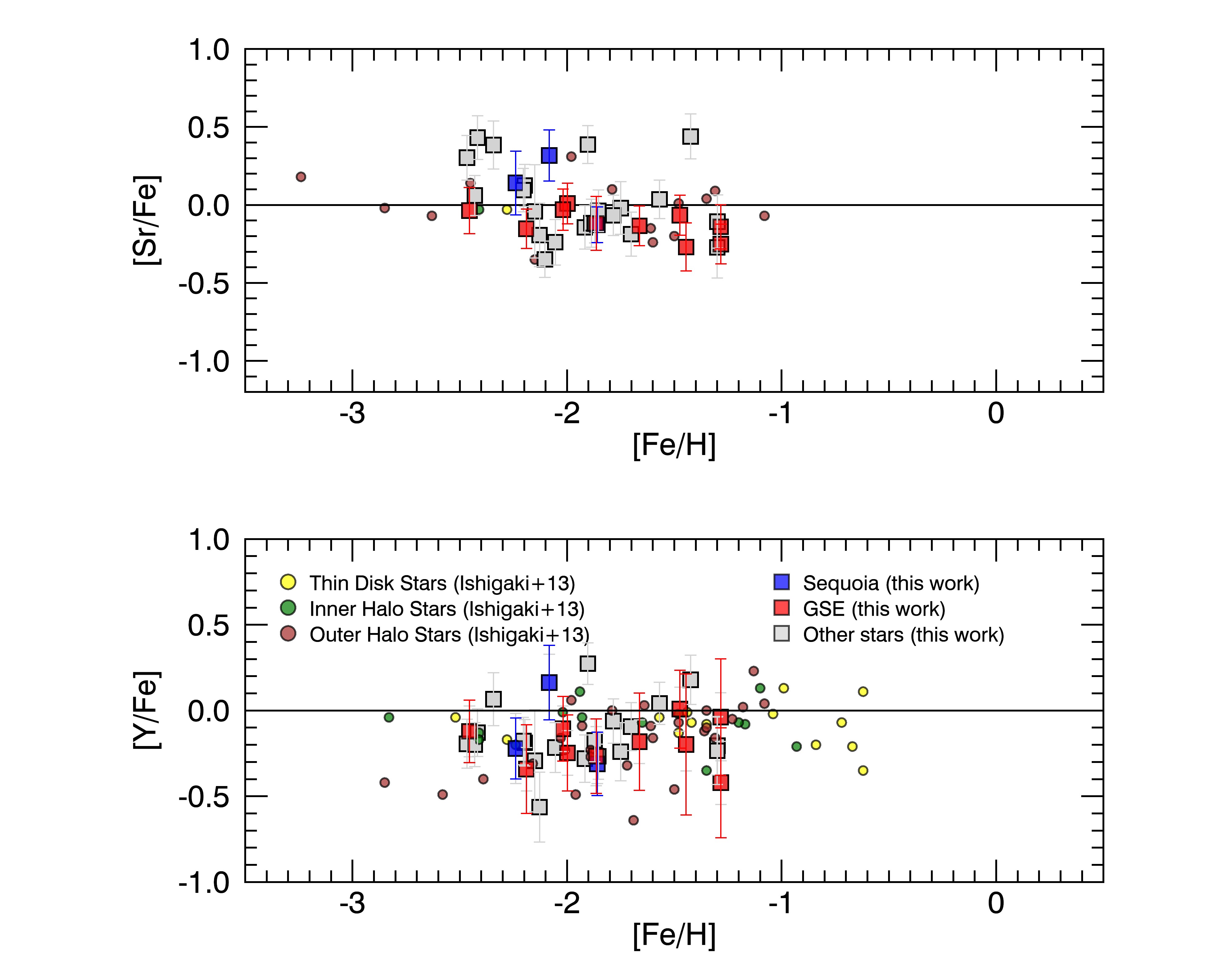}
 \end{subfigure} \\%
\begin{subfigure}{.5\textwidth}
 \centering
\includegraphics[width=1.0\linewidth]{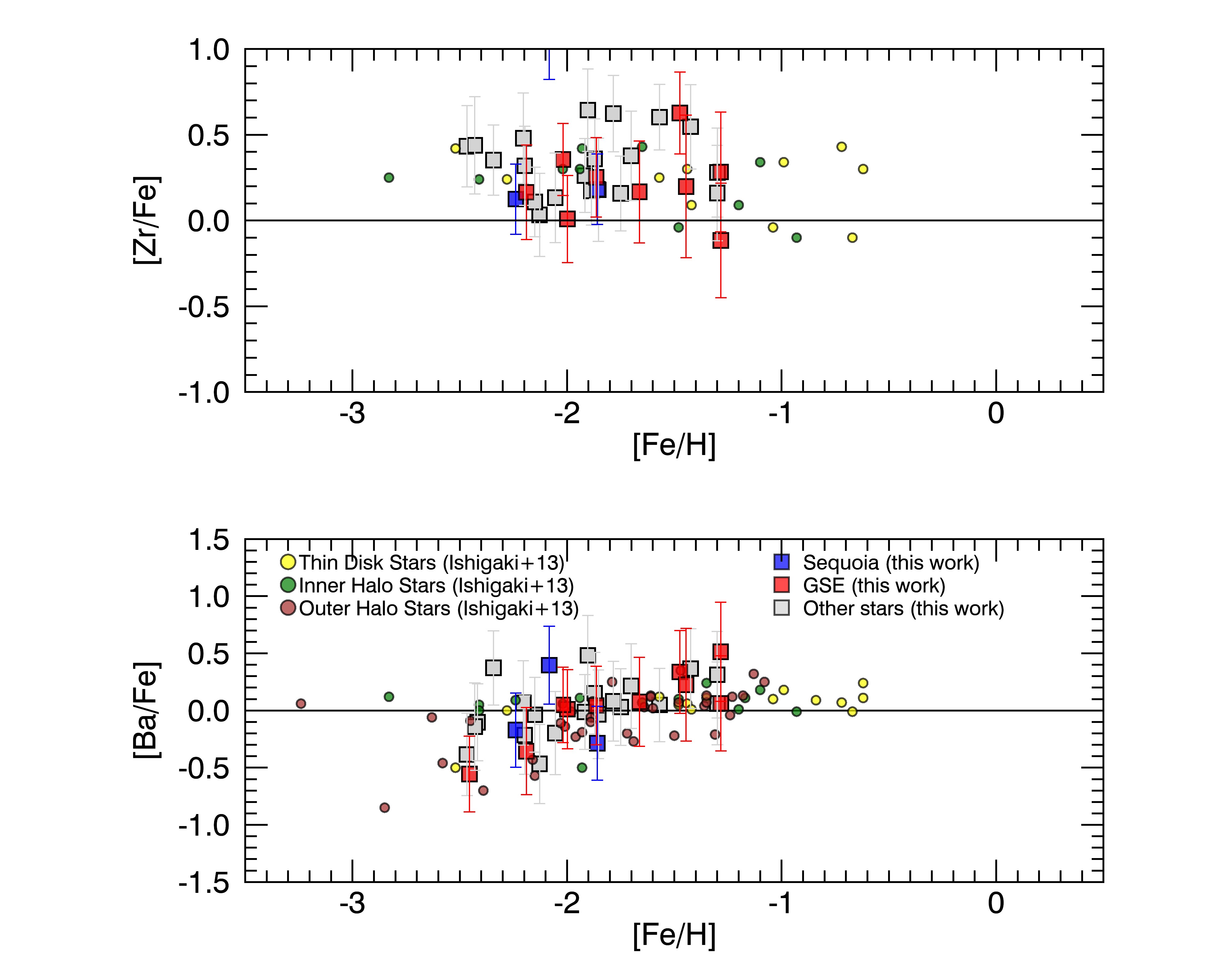}
 \end{subfigure} \\
 \caption{Abundance  ratios   of [Sr/Fe],  [Y/Fe], [Zr/Fe],  
 and [Ba/Fe] as a function of [Fe/H] for the stars in our sample shown as 
 square symbols.  Red squares represent stars from GSE, while blue squares are stars identified as Sequoia stars.  Grey squares are the remaining halo stars.   We added the stars  \citet{ishigaki2013}, which are represented as  circles: respectively, disc stars in yellow, inner halo stars in green, and outer halo stars in brown.}
\label{fig:lAbu_plot1}
 \end{figure}

\begin{figure}
 \begin{subfigure}{.5\textwidth}
 \centering
\includegraphics[width=1.0\linewidth]{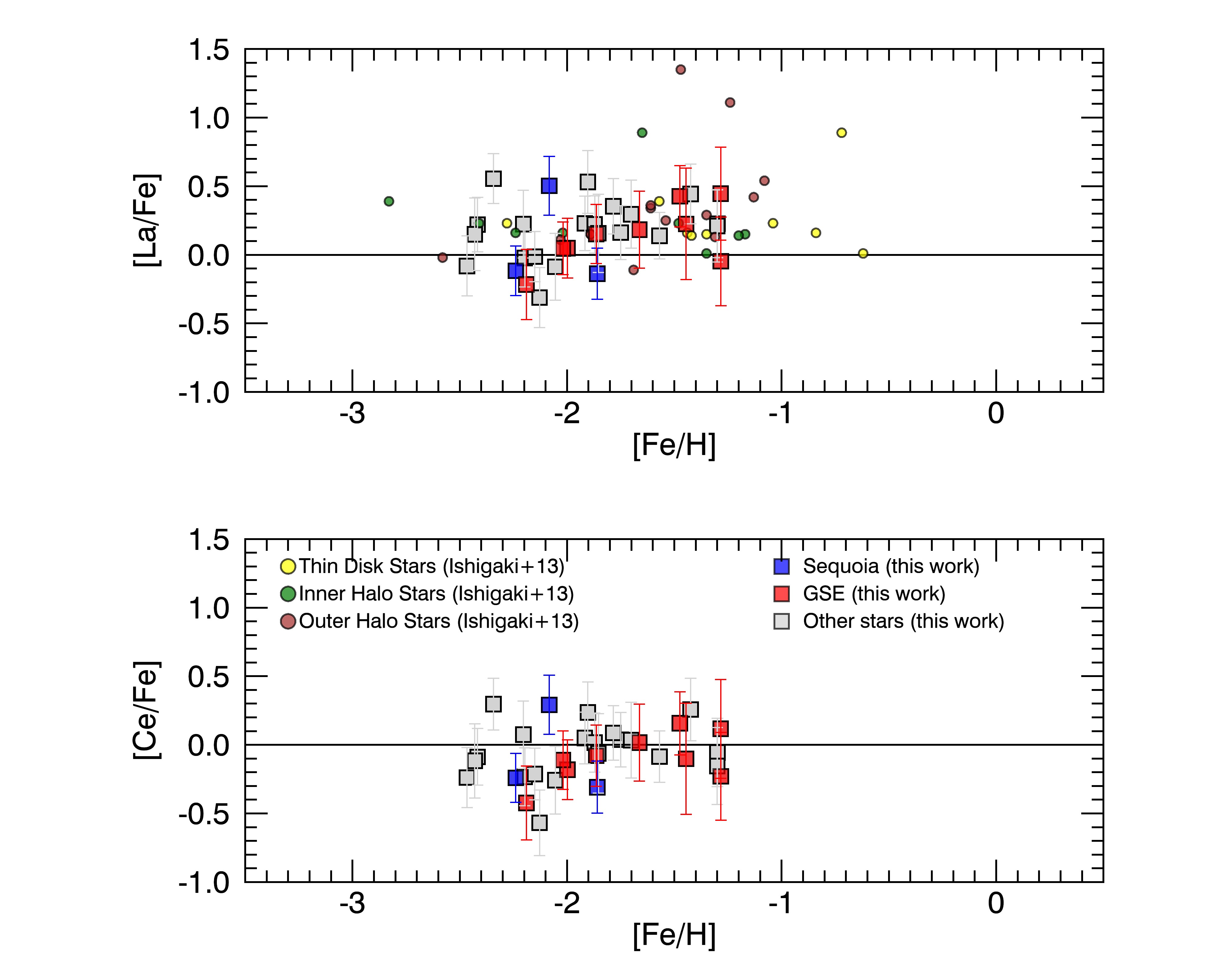}
 \end{subfigure} \\%
 \begin{subfigure}{.5\textwidth}
 \centering
\includegraphics[width=1.0\linewidth]{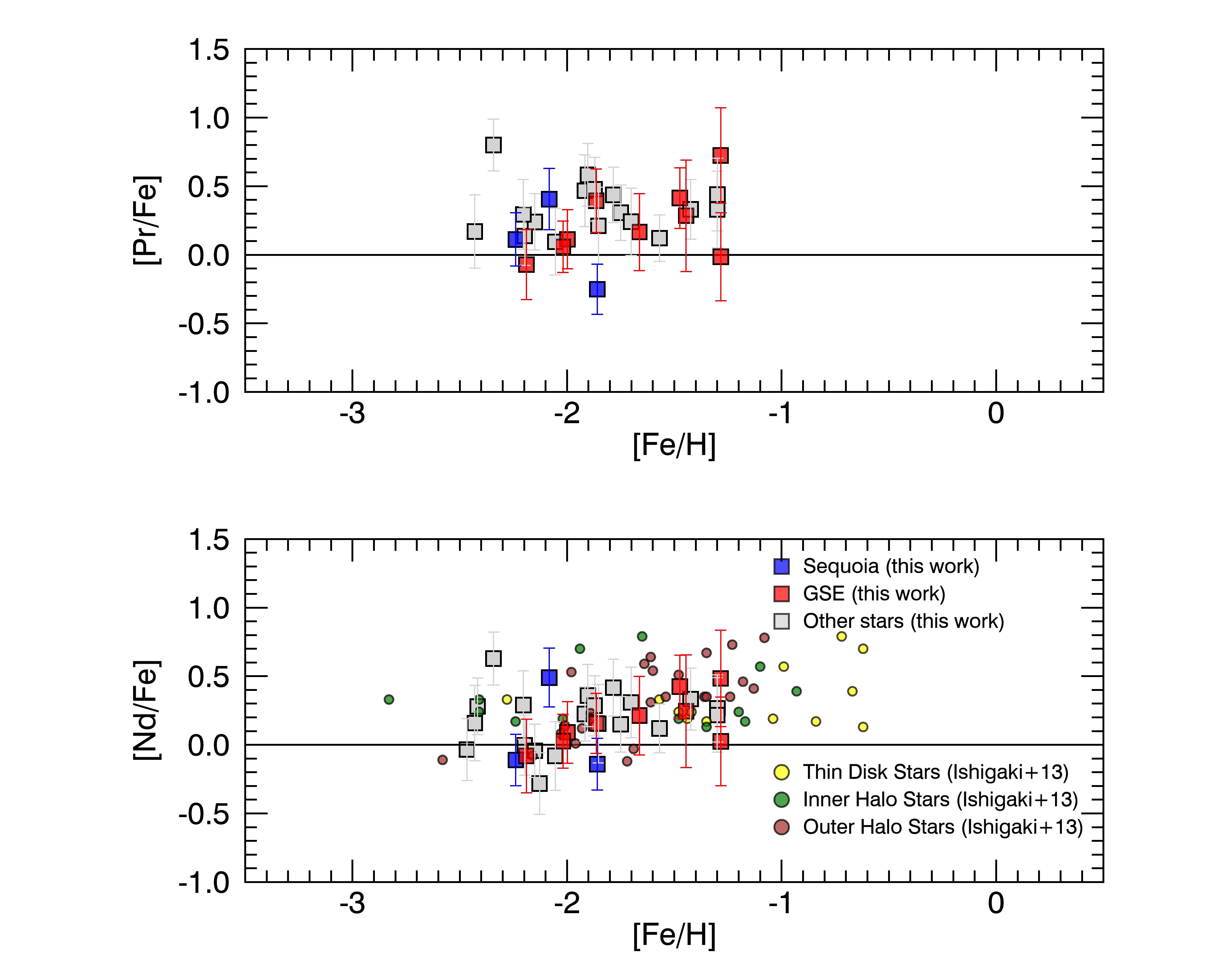}
 \end{subfigure} \\%
 \caption{Abundance ratios   of  [La/Fe], [Ce/Fe], [Pr/Fe], and  [Nd/Fe]  as a function of [Fe/H] for the stars in our sample shown as square symbols.   We added the stars from \citet{ishigaki2013}, which are  represented as circles: respectively, disc stars in  yellow, inner halo stars in green, and outer halo stars in brown.}
\label{fig:lAbu_plot2}
 \end{figure}

\begin{figure}
 \centering
\includegraphics[width=1.0\linewidth]{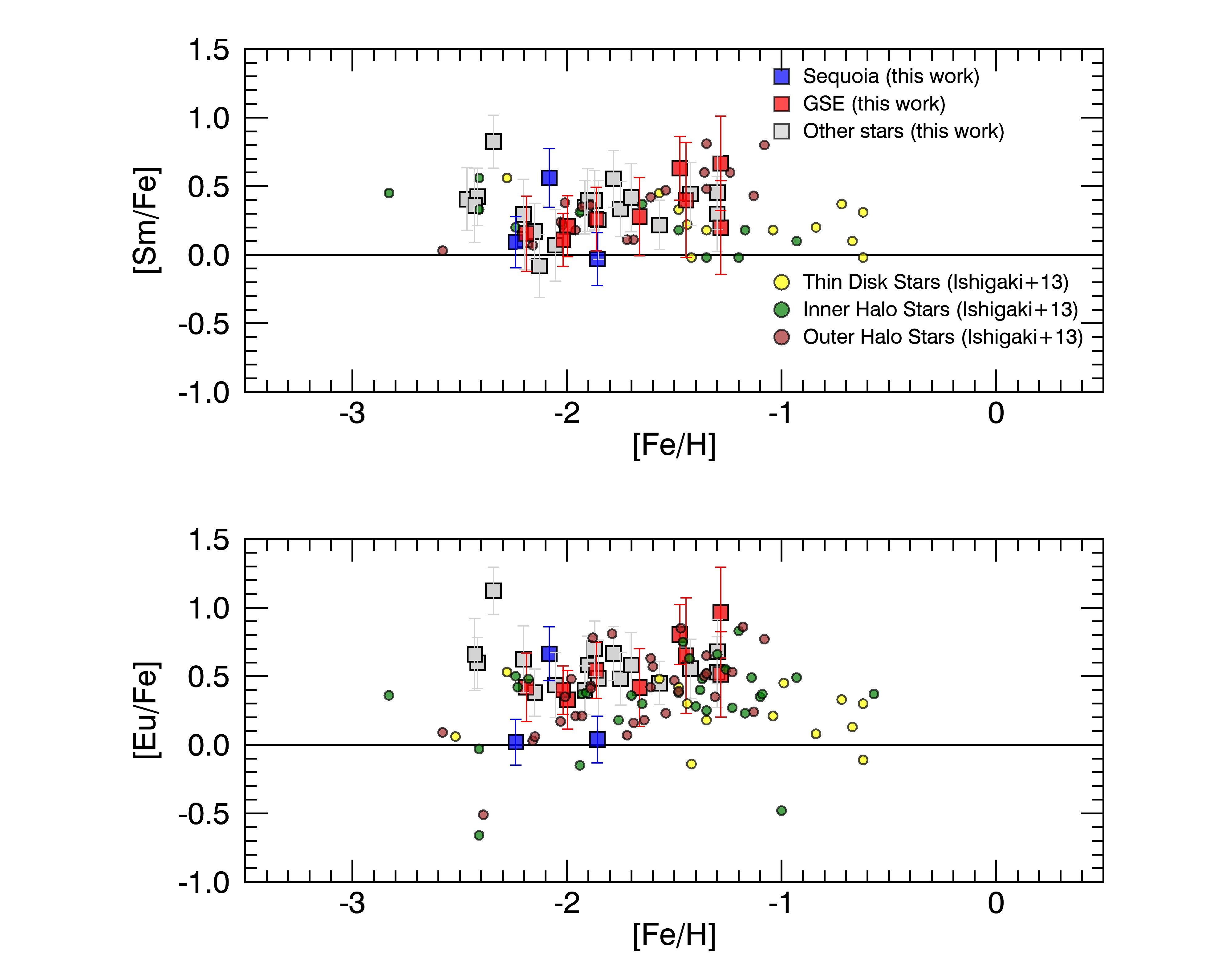}
 \caption{Abundance  ratios   of [Sm/Fe]  and [Eu/Fe]  as a function of [Fe/H] for the stars in our sample shown as  square symbols.   We added the stars from \citet{ishigaki2013}, which are represented as circles: respectively, disc stars in yellow, inner halo stars in green, and outer halo stars in brown.}
\label{fig:lAbu_plot3}
 \end{figure}

The results are presented in Fig. \ref{fig:lAbu_plot1}, \ref{fig:lAbu_plot2}  and \ref{fig:lAbu_plot3}. Our stars are shown as grey squares. Blue squares represent stars from GSE and red squares are stars identified as Sequoia stars. 
 Grey squares are the remaining  stars, mostly halo stars. The  results from \citet{ishigaki2013}  are divided into outer halo stars (brown circles), inner halo stars (green circles), and thick-disc stars (yellow circles).

It is interesting to note that for most of the elements, the [X/Fe] versus [Fe/H] found for the GSE stars seems to follow a rather well-defined trend, with a much smaller dispersion than that of the remaining sample of halo stars.  For the Sequoia stars, it is not possible to 
conclude anything about a peculiar behaviour of the [X/Fe] versus [Fe/H] distribution.  Our results are in agreement with the abundance ratios found by \citet{ishigaki2013} for the outer and inner halo stars.  

\subsection {Neutron-capture element abundance ratios }

\begin{figure}
 \begin{subfigure}{.5\textwidth}
 \centering
\includegraphics[width=1.0\linewidth]{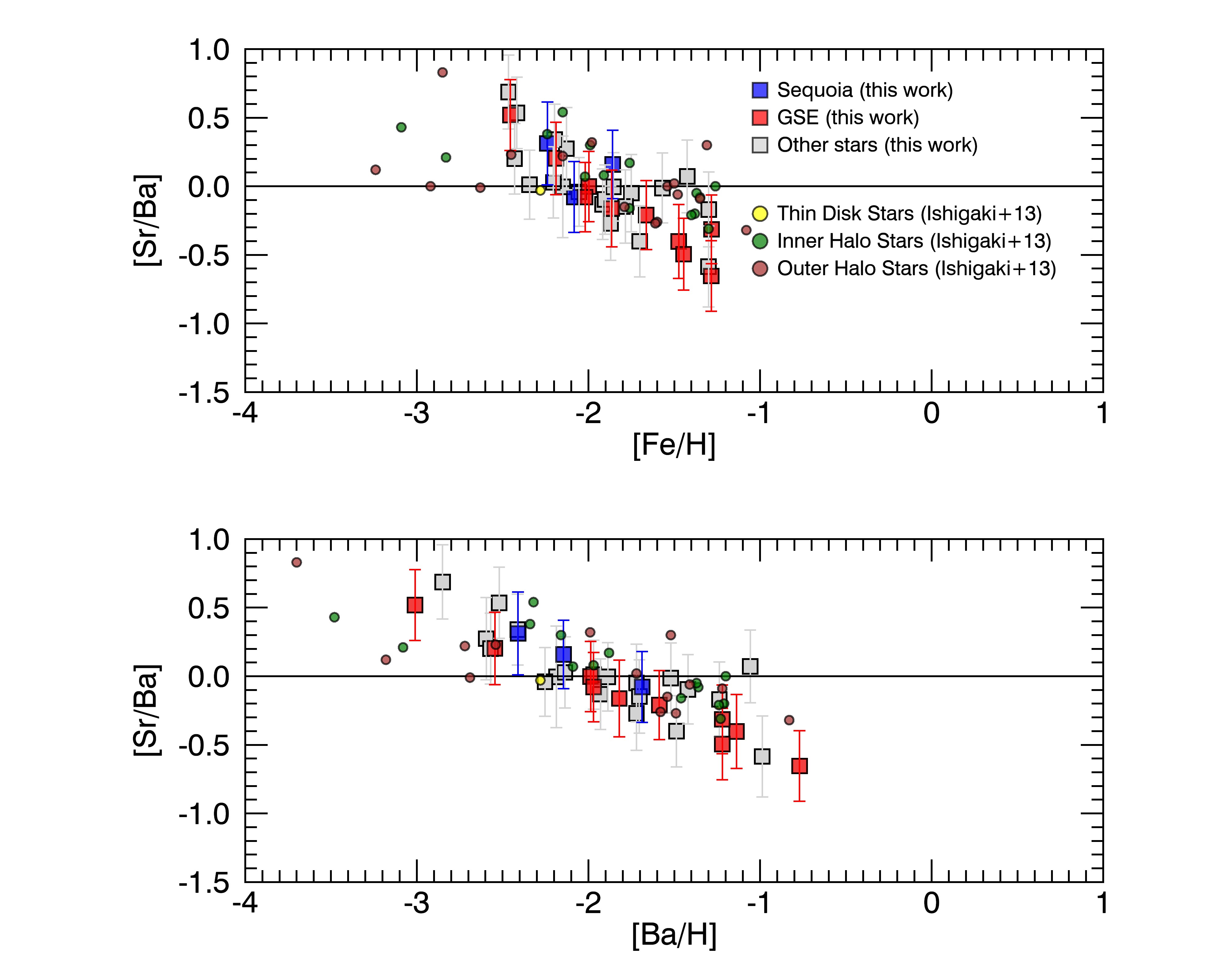}
 \end{subfigure} \\%
\begin{subfigure}{.5\textwidth}
 \centering
\includegraphics[width=1.0\linewidth]{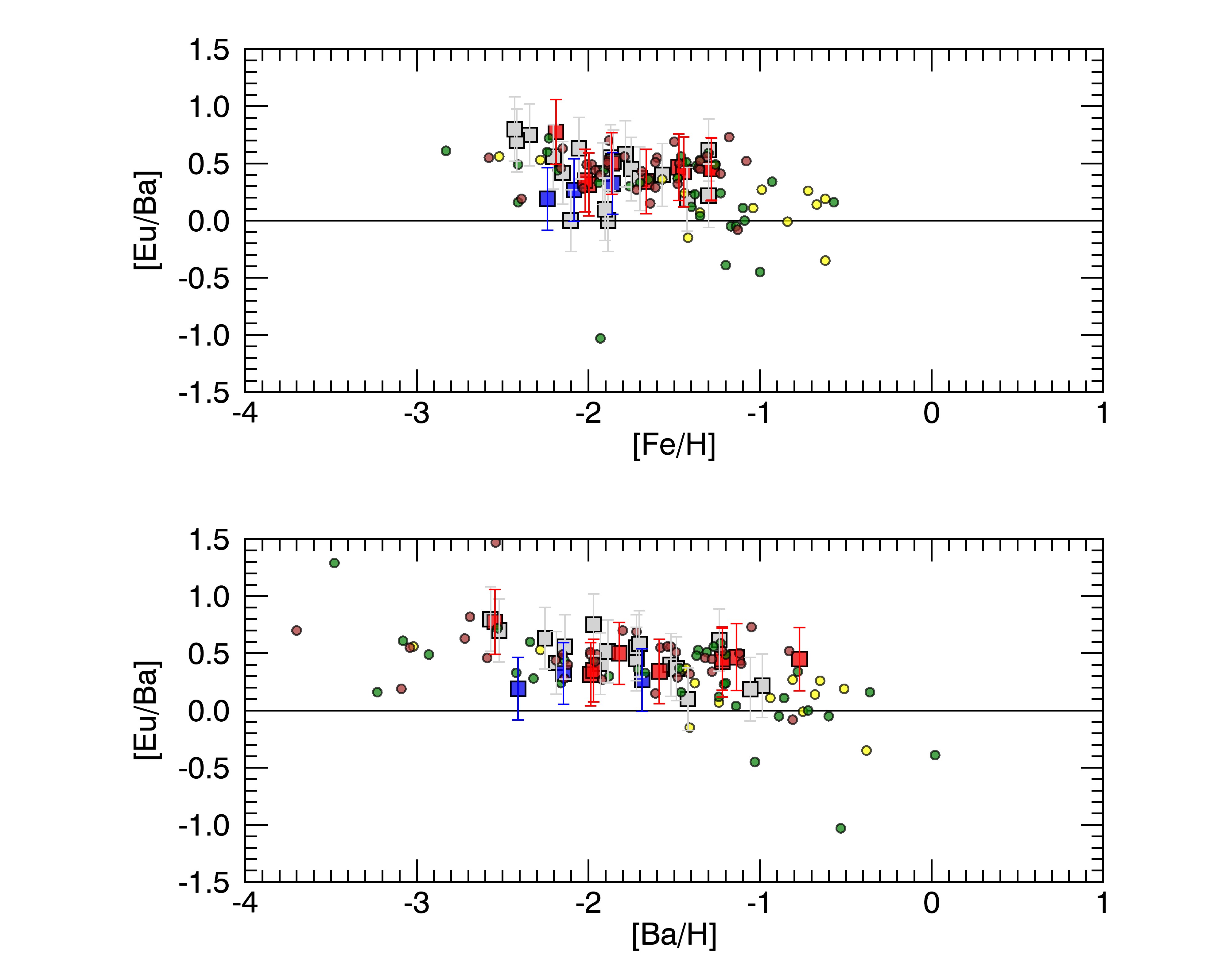}
 \end{subfigure} \\
 \caption{Abundance  ratios   of [Sr/Ba] and  [Eu/Ba]   as a function of [Fe/H]  (panels one and three) and as a function of [Ba/H] (panels two and four) for the stars in our sample, shown as square symbols.   We added the stars from  \citet{ishigaki2013},  represented as circles.}
\label{fig:lAbratio_plot1}
 \end{figure}

In Fig. \ref{fig:lAbratio_plot1},  we plotted the abundances ratios [Sr/Ba] and [Eu/Ba] as a function of [Fe/H] and [Ba/H] for our sample of stars.  We added the results from \citet{ishigaki2013}.
 The symbols are the same as in Fig. \ref{fig:lAbu_plot1}.

 The [Eu/Ba]  is representative of the relative production  of the r-process   relative  to the s-process  in these metal-poor stars. The results show the well-known constant ratio [Eu/Ba] 
with a level of $\simeq + 0.6$ dex, which is slightly  lower than the pure r-component value of   $\simeq + 0.7$ dex   computed by \citet{arlandini1999}.  
 It is interesting to note that the  median value found for the GSE  sample is larger  ( <[Eu/Ba]> = 0.45 dex ) than the one found for the Sequoia stars ( <[Eu/Ba]> = 0.27 dex), indicating a different level of enrichment of the matter that formed these two systems. 
 This difference will be studied in light of the models of galactic chemical evolution in  section \ref{sec:GCE}.

The [Sr/Ba] versus [Fe/H] and  [Sr/Ba] versus [Ba/H] are particularly interesting. From the top, the  second  panel  of Fig. \ref{fig:lAbratio_plot1} shows the variation in [Sr/Ba] as a function of [Ba/H].  The results for the GSE stars show a tight relation between [Sr/Ba] and [Ba/H], with a regression coefficient of  -0.94. 
This result, which needs to be confirmed with a larger sample, shows how the chemical enrichment of the GSE has evolved over  a range of 2 dex   in [Ba/H].  The Sequoia stars follow the same trend; that is, a decreasing [Sr/Ba] as [Ba/H] increases, with a slight difference. 
Indeed,  at a given [Ba/H], 
the [Sr/Ba] ratio found in the Sequoia  stars seems higher than the value  found in GSE stars.  This difference in Galactic chemical evolution of these two components will be addressed in  Sec. \ref{sec:GCE}.

\section{Gaia Sausage Enceladus and Sequoia substructures \label{sec:streams}}

The aim of this section is to compare  our results for the stars belonging to GSE and Sequoia with existing literature results based on high-resolution spectroscopy. 
Until now, there have only been a handful  of  papers in which detailed neutron-capture abundances have been computed. 
Spectroscopic surveys have been used to  characterise the chemical properties of halo substructures. In general, these studies 
concentrate on a small number of elements and mostly light metals \citep[e.g.][]{hasselquist2021,buder2022}.
A more extended analysis of a larger number of elements (13 elements) has been made by \citet{horta2022}  using APOGEE data. Unfortunately,  this study contains only cerium as a neutron-capture element. 
A few studies based on follow-up programs of smaller samples  using  high-resolution spectrographs have include some neutron-capture elements in their analysis. 
 \citet{limberg2021}  has analysed  a sample of stars belonging to the Helmi stream  and derived Eu and Ba abundances for a couple of  stars in their sample.   
 \citet{aguado2021} derived the abundances of Sr, Ba, and Eu in a sample of GSE and Sequoia stars.
 \citet{matsuno2022}  studied a sample of stars belonging to Sequoia and determined the abundances of several elements,  among them Sr and Y.

\begin{figure}
 \begin{subfigure}{.5\textwidth}
 \centering
\includegraphics[width=1.0\linewidth]{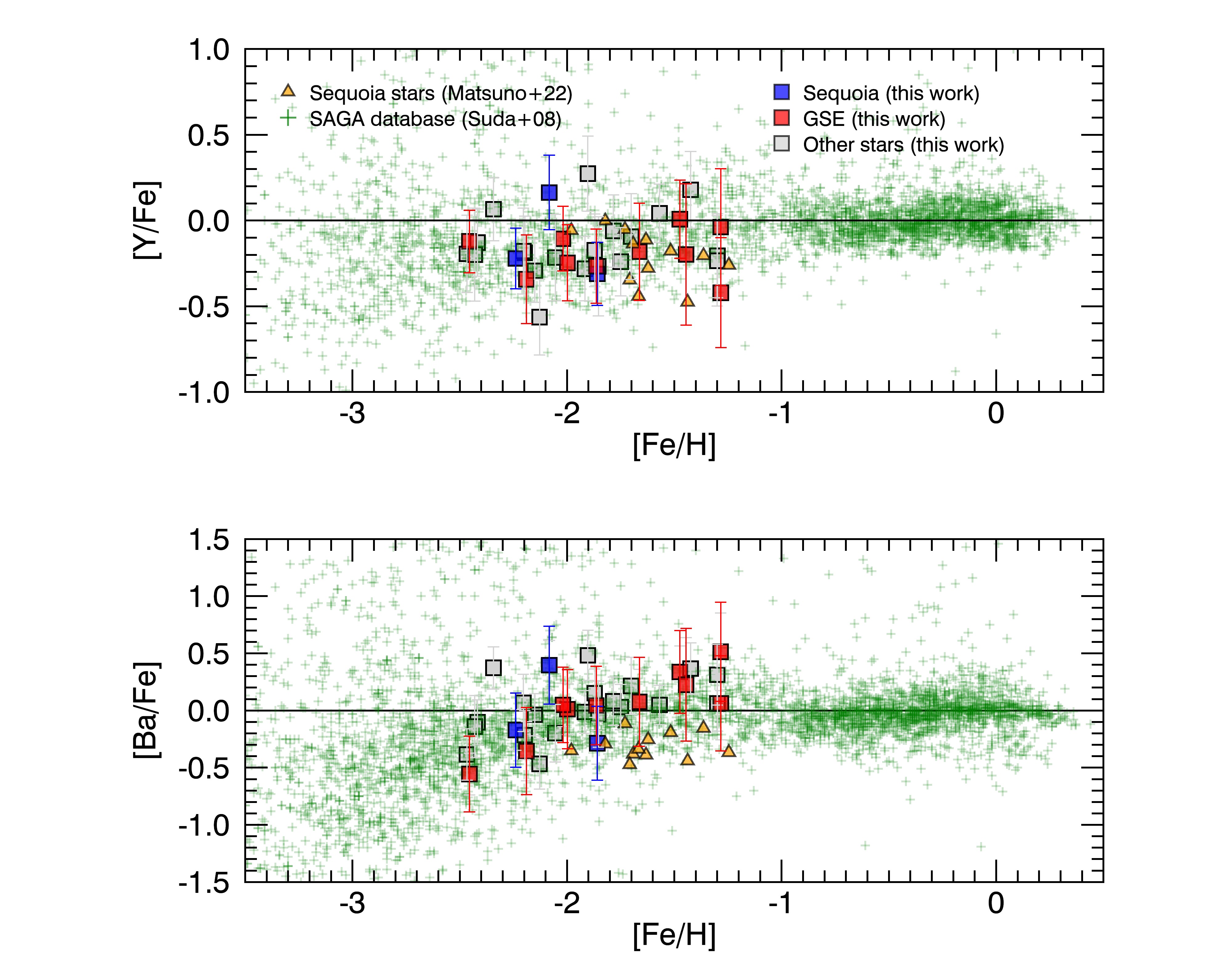}
 \end{subfigure} \\%
\begin{subfigure}{.5\textwidth}
 \centering
\includegraphics[width=1.0\linewidth]{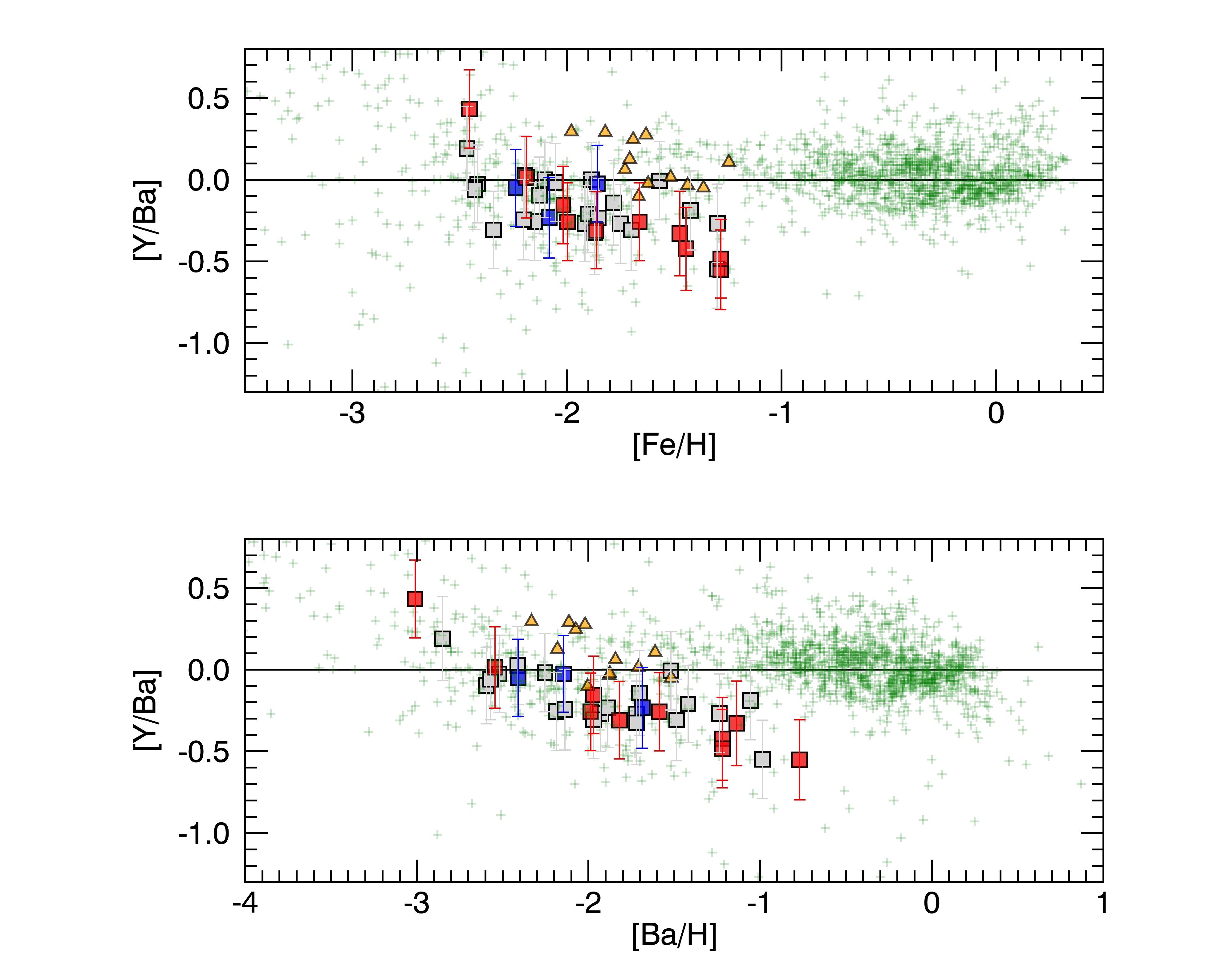}
 \end{subfigure} \\
 \caption{Abundance  ratios   of [Y/Fe] and  [Ba/Fe] as a function of [Fe/H] for the stars in our sample shown as square symbols.  \citet{matsuno2022} results for Sequoia stars have been added and shown as orange triangles. Green symbols represent   stars from the SAGA database \citep{suda2008}.}
\label{fig:lAbund_streams}
 \end{figure}

Fig. \ref{fig:lAbund_streams}  represents [Y/Fe] and [Y/Ba] as a function of [Fe/H] and [Ba/H] for our sample of stars. The  stars are shown as square symbols with the same colour scheme as in Fig. \ref{fig:lAbratio_plot1}. 
We have added the results for the Sequoia stars analysed by \citet{matsuno2022}. We find results compatible with a constant [Y/Fe], although with a significant dispersion,  over a range of metallicities from [Fe/H] $\simeq -2.4$ dex to [Fe/H] 
$\simeq -1.2$ dex if  both samples are combined.  For the [Ba/Fe] versus [Fe/H],  the trend could also be interpreted as a constant value with a large dispersion, with a mean value of around --0.2 dex. 
The [Y/Ba] ratio gives very different results. The values found by \citet{matsuno2022}  appear to be significantly higher than our results.  

\begin{figure}
 \begin{subfigure}{.5\textwidth}
 \centering
\includegraphics[width=1.0\linewidth]{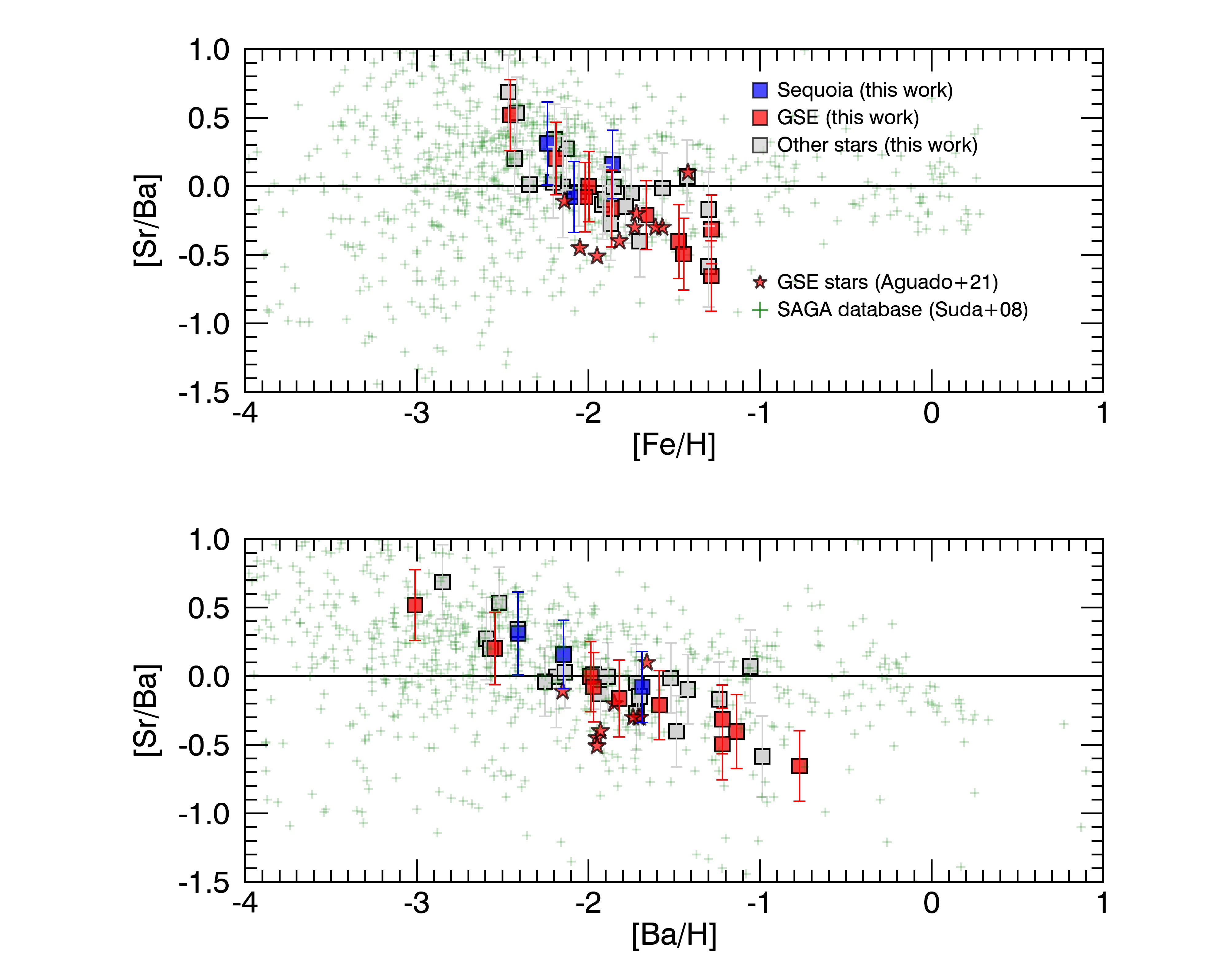}
 \end{subfigure} \\%
\begin{subfigure}{.5\textwidth}
 \centering
\includegraphics[width=1.0\linewidth]{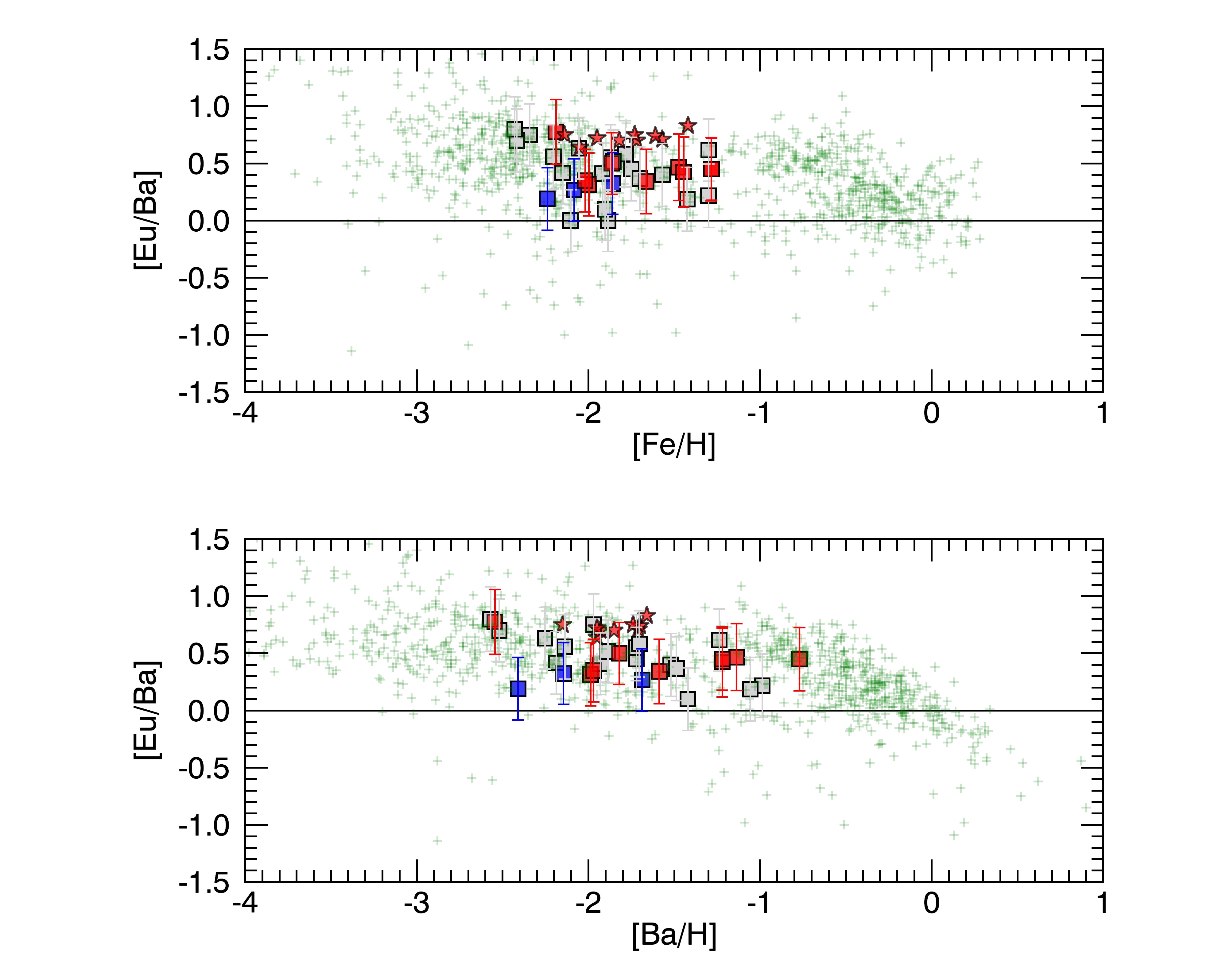}
 \end{subfigure} \\
 \caption{Abundance  ratios   of [Y/Fe] and  [Ba/Fe] as a function of [Fe/H]  for the stars in our sample are shown as square symbols. \citet{aguado2021} results for GSE stars have been  added and shown as red stars.  Green symbols represent  halo stars from  the SAGA database \citep{suda2008}.}
\label{fig:lAbund_streams2}
 \end{figure}

In Fig. \ref{fig:lAbund_streams2},   we plot the [Sr/Ba] and [Eu/Ba] ratios  as a function of [Fe/H] and [Ba/H] for our sample of stars.   We have added the results for the GSE stars analysed by \citet{aguado2021}.   We do not confirm the high [Eu/Ba] that they found in their sample.
It is interesting to note that the [Eu/Ba] ratio found  by \citet{francois2007}  
was not as high as that found in the stars of \citet{aguado2021}. 
 Concerning the [Sr/Ba] ratio versus [Fe/H],  we found ratios in reasonable agreement  with their values. However, our results seem to indicate a tight relation between the [Sr/Ba] and [Fe/H] ratio, whereas they found a ratio compatible with a constant value of $\simeq -0.30$ dex  and with a dispersion of  $\simeq  0.20$ dex.

This first limited sample of the MINCE project  does not allow us to draw firm conclusions about the characteristic abundances  of the well-identified galactic sub-components of our Galaxy, but clearly shows
the potential of the study of neutron-capture elements in intermediate-metallicity stars.
Analysis of a larger sample is needed to understand these differences better.

\section{Reference chemical evolution models  \label{sec:GCE}}

 One of the more striking features of the results obtained for the abundances of neutron-capture elements is the strong correlation of the data for GSE in the [Sr/Ba] versus [Ba/H] plane. 
With this first sample of MINCE stars, our aim is  to investigate this feature by means of a stochastic chemical evolution model, based on the original code developed in \citet{Cesc08}. In this work, we try to reproduce  the evolution of GSE and we use the same parameters as in  the chemical evolution model used in MINCE I, but within a stochastic framework.   GSE chemical evolution parameters were tuned to match the metallicity distribution function of GSE stars described in \citet{Cescutti20}. Overall, its evolution is similar to that of a dwarf galaxy with a mass of about 3\% of the Milky Way; however, given its galactic winds and an inefficient star formation ending more than 5 Gyr ago, its stellar content is only around 1\% of the Galactic one. These results are similar to the ones obtained for the chemical evolution of GSE in \citet{Vincenzo15}. In the stochastic model, we have to consider a typical volume in which the gas is always well mixed. For this model, the fixed volume has a radius of 150 pc and a total infalling mass of 2.6 10$^5$ \msun.   
The nucleosynthesis considered is the same as in \citet{Cescutti14}, so we take into account these prescriptions for the neutron-capture elements:
the r-process enrichment from magneto-rotational-driven supernovae (SNe)
\citep{Winteler12, Nishimura15}; the s-process production from asymptotic giant branch (AGB) stars described in \citet{Cristallo09, Cristallo11}; and the s-process production from rotating massive stars taken from \citet{Frisch12, Frisch15}. It is worth underlining that the use of the yields of \citet{Limongi18}, although with different velocities, would not alter the results \citep{Prantzos18, Rizzuti19, Rizzuti21}. Similarly, considering neutron star mergers under specific characteristics that allow them an almost prompt enrichment of the Galaxy produces results similar to magneto-rotational-driven SNe \citep{Cescutti15, Cavallo21, Cavallo23}. 
Iron is produced both by massive stars, with a fixed  production assumed to be 0.07\,\msun\
 \citep[see][]{Limongi18} and SNe Ia, with the classic delay time adopted by \citet{Matteucci1986} and the \citet{Iwamoto1999} yields. 

\subsection{Results of [Sr/Fe], [Ba/Fe] and [Eu/Fe] versus [Fe/H] for the Gaia Sausage Enceladus galaxy}

\begin{figure}
 \centering
\includegraphics[width=1.0\linewidth]{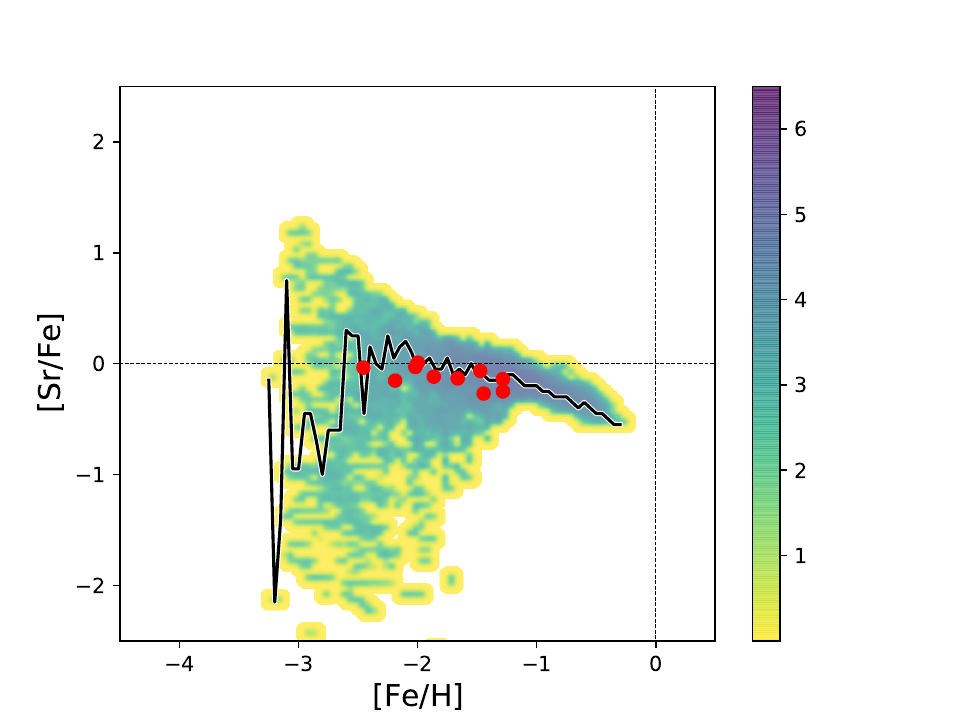}
\caption{Abundance ratios   of [Sr/Fe] vs [Fe/H].
The red dots are the abundances of the GSE stars in our sample; the colour-coded area shows the results of the stochastic model. The colour coding is described in the bar as the number density of long-living stars in log-scale. The black line shows the most probable value of the model in [Sr/Fe] at each bin in [Fe/H].}
\label{modelSrFe}
 \end{figure}

\begin{figure}
 \centering
\includegraphics[width=1.0\linewidth]{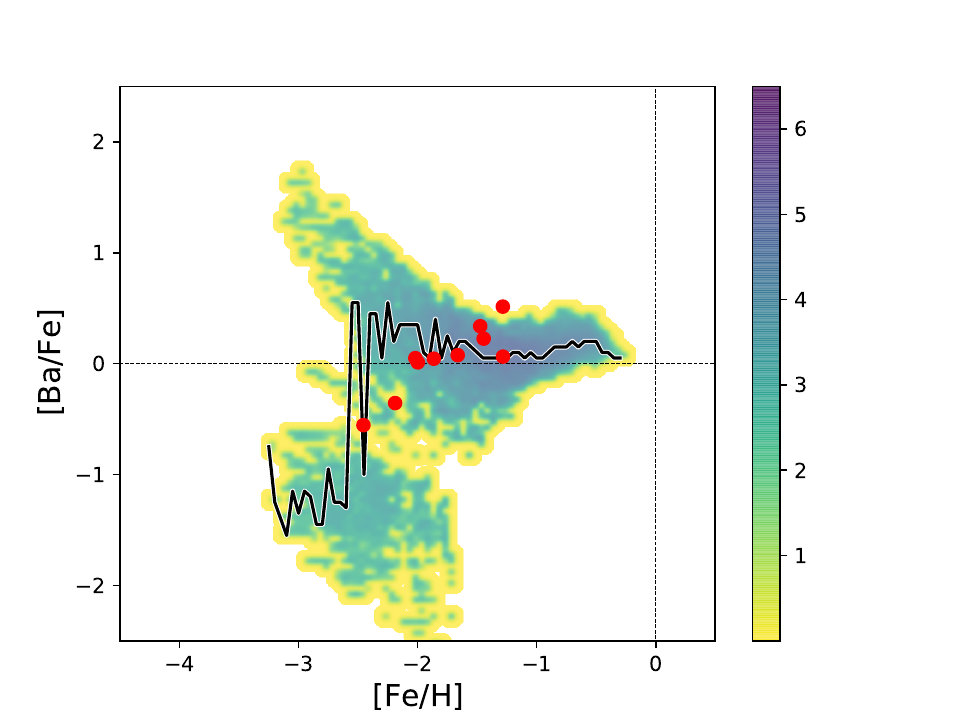}
\caption{Same as Fig. \ref{modelSrFe}, but for the abundance  ratios   of [Ba/Fe] vs [Fe/H].}
\label{modelBaFe}
 \end{figure}

\begin{figure}
 \centering
\includegraphics[width=1.0\linewidth]{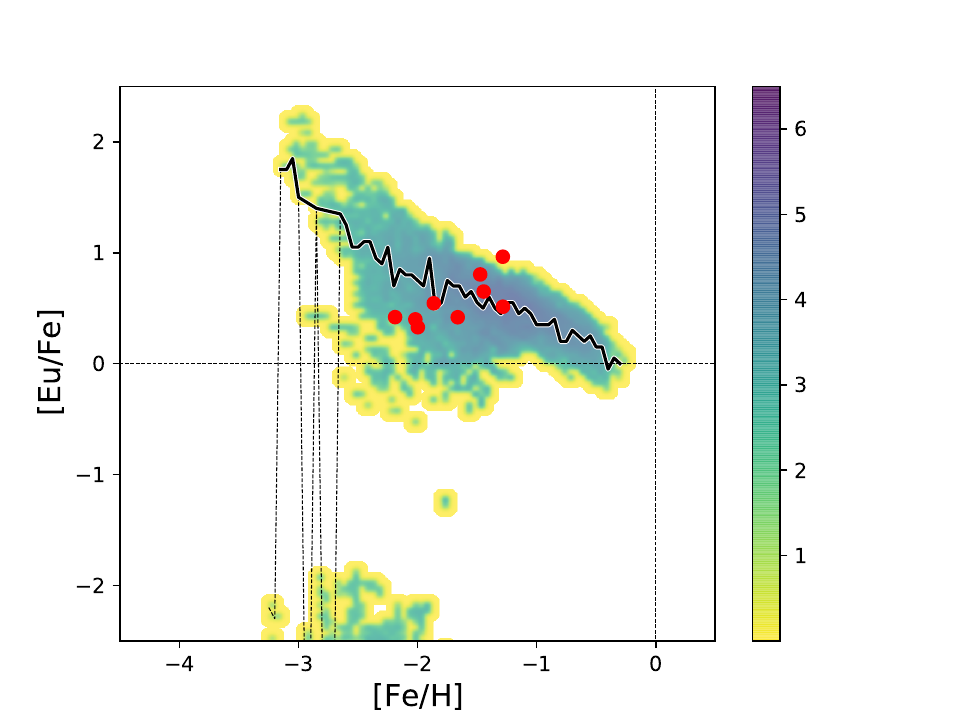}
\caption{Same as Fig. \ref{modelBaFe}, but for the abundance  ratios   of [Eu/Fe] vs [Fe/H]. The black line shows the most probable value of the model in [Eu/Fe] at each bin in [Fe/H]. We exclude the abundances with [Eu/Fe]$<-$2, since these extremely low abundances in europium are hardly measurable. The line obtained including these extreme values is displayed with a tiny dashed line.}
\label{modelEuFe}
 \end{figure}

In Figures \ref{modelSrFe},  \ref{modelBaFe}, and \ref{modelEuFe} we present the comparison between the model results and the abundances measured for GSE. Small adjustments were applied to reproduce at best the data for barium and strontium; the model results for barium were modified by +0.2 dex and strontium by $-$0.1 dex. These adjustments are probably due to our assumptions about nucleosynthesis;  for example, they can be explained by a more efficient production of barium in rotating massive stars or a more complex production by the r-process events. In any case, these elements are mostly produced by the s-process in AGB stars at solar metallicity, so these tiny adjustments would not significantly vary the predictions for the solar abundances. Another possibility is that the 1D LTE chemical abundances are affected by offsets not considered due to NLTE and/or 3D corrections.  

The abundances of [Sr/Fe] versus [Fe/H] of the GSE stars are all well explained by the model results in Fig. \ref{modelSrFe}; they are also located well within the boundaries of the colour distribution, in the region of the [Sr/Fe] versus [Fe/H] plane where we expect the larger fractions of GSE.
We also note a severe decrease in the dispersion at [Fe/H] $\geq -1.2$. More data will be useful in this area; that is, slightly above the metallicity considered by MINCE stars. We also see that the dispersion of our sample seems quite reduced for the stars in our sample.
The stochastic volumes considered may be too small for this system; given the amount of data collected, we prefer to keep this parameter fixed. We also present in each plot a line connecting the most likely value obtained by the model at each metallicity (either in [Fe/H] or [Ba/H]). The purpose of this line is to provide the most likely observed abundances, and it is clear that overall the model is closely following the abundances measured in the MINCE sample. 
For barium in Fig.\ref{modelBaFe} and europium in Fig. \ref{modelEuFe}, the comparisons between measured abundances and model results are similar. The small cloud of the model results, showing a very low europium abundance ([Eu/Fe]$< -$2), is due to the tiny amount of europium produced by the rotating massive stars and corresponds to the cloud presented by barium in the range $-$2.5$\geq$ [Ba/Fe] $\geq-$1.0.
Most of the data points are well inside the model predictions; we do not find any stars sitting in the r-process rich star locus -- the tail at high [Eu/Fe] (or  [Ba/Fe]) at [Fe/H]$\sim -2.5$. This is expected given the rarity of these objects, the small sample for GSE, and considering also that, by construction, the MINCE sample does not include extremely metal-poor stars. We have one star in the MINCE sample that is actually in this zone (BD\,$+$21\,4759), but it is not shown here since it does not belong to GSE. 
Nonetheless, we also note that we have an outlier (BD\,$-$04\,18); this star is just at the border of the models and is (practically) a europium-rich star, with [Eu II/Fe I]$=$1.16 and [Eu II/Fe II]$=$0.96. We cannot easily explain this, but we note that the [Ba/Eu] ratio of this star is close to the pure r-process ratio and that it is actually the star with the highest abundance of both barium and europium, given also that it is one of the most metal-rich within the sample. 
Could this be the signature of an extremely high enrichment by a delayed source of the r-process (a neutron star merger)? Clearly, a single system is not enough to support this claim; still, it is a rather peculiar object in terms of barium and europium enrichment,  and we will search for more of these objects in future MINCE works.

\subsection{Results for [Sr/Ba] and [Eu/Ba] versus [Ba/H]}

\begin{figure}
 \centering
\includegraphics[width=1.0\linewidth]{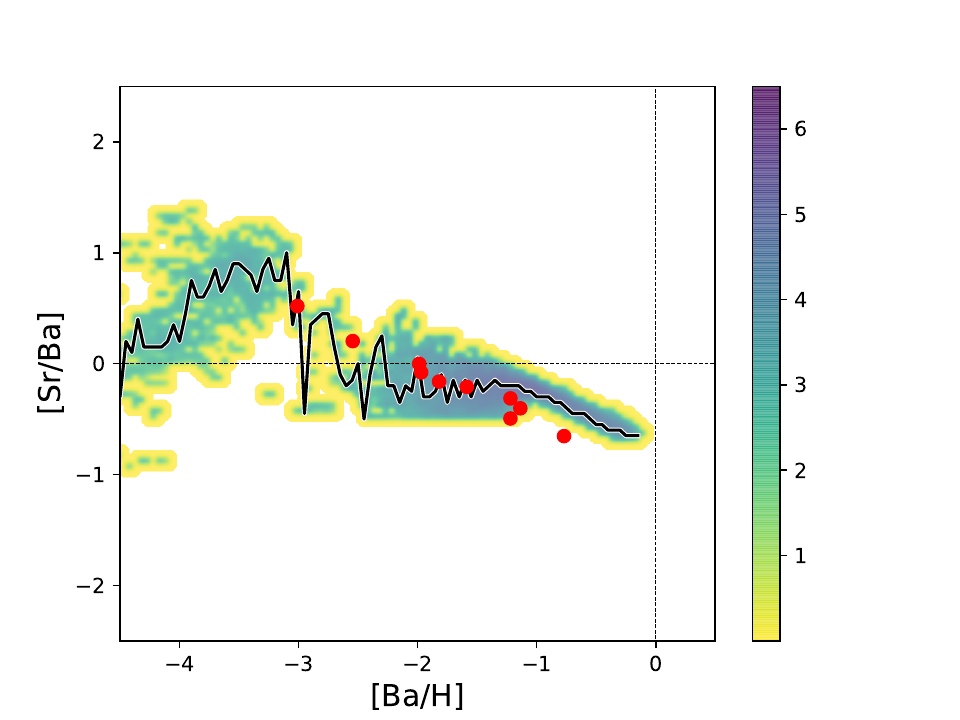}
\caption{Abundance  ratios of [Sr/Ba] vs [Ba/H].
The red dots are the abundances of the GSE stars in our sample; the colour-coded area shows the results of the stochastic model. The colour coding is described in the colour bar as the number density of long-lived stars on a log-scale.}
\label{modelSrBa}
 \end{figure}
 \begin{figure}
 \centering
\includegraphics[width=1.0\linewidth]{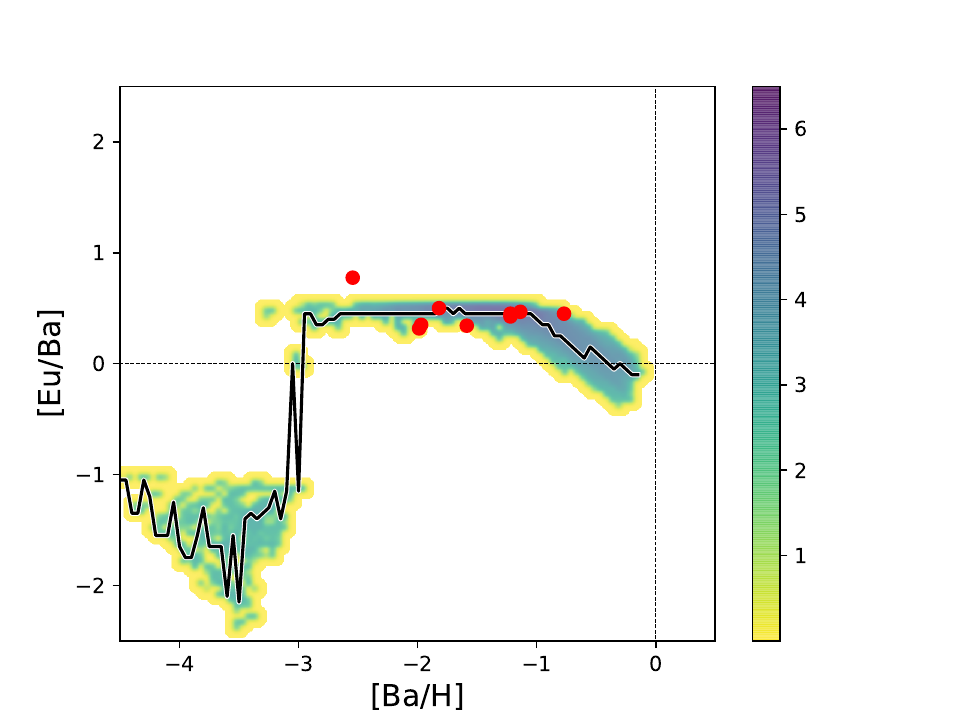}
\caption{Same as Fig.\ref{modelSrBa}, but for the abundance  ratios   of [Eu/Ba] vs [Ba/H].}
\label{modelEuBa}
 \end{figure}
 
In Fig. \ref{modelSrBa}, we present the model and the stellar abundance results for GSE in the [Sr/Ba] vs [Ba/H] plane.
As was mentioned previously, the results of the stellar abundances are striking, as they show a distinct correlation between strontium and barium in this plane; similar to the [Ba/Fe] versus [Fe/H] case, the model reproduces fairly well most of the data, except the most barium-rich star of the sample  BD$+$04 18. It should be added that the dispersion of the model does not predict the observed correlation; on the other hand, the most probable stellar abundances predicted by the model are close to the observational trend. 
In this plot, the area with [Ba/H]$<-$3  is enriched only by the rotating massive stars, with a low barium enrichment and a dispersion between [Sr/Ba] depending on the mass of the stars. The dispersion at  $-$2.5$<$[Ba/H]$<-$1.5  is created mostly by the pollution of the r-process events, but there is also a mild contribution coming from the rotating massive stars that produces the smaller variation in the  [Sr/Ba] ratio. If we remove the s-process contribution from rotating massive stars and consider only r-process enrichment, our model will produce a constant [Sr/Ba] ratio. 
Therefore, to reproduce this observed correlation in [Sr/Ba], rotating massive stars represent a key ingredient. A possibility, not considered at the moment,  is that the r-process production is more complex and there is an interplay between strontium and barium. Finally, the region at [Ba/H]$>-$1 is determined by the enrichment from AGB stars, which at low metallicities tend to produce a low [Sr/Ba] ratio.

Fig. \ref{modelEuBa} shows that in the plane [Eu/Ba] versus [Ba/H] all of the stars are located on a narrow plateau associated with the assumed r-process production. We recall that the model assumes a variation in the barium of 0.2 dex, as was mentioned above;  otherwise, the pure r-process would produce a ratio of [Eu/Ba]=0.7dex.
In this plane, the only star that does not sit within the model prediction is TYC 4221-640-1. We also underline again that the enrichment by rotating massive stars produces the small spread in the model results in the range $-$2.5$<$[Ba/H]$<-$1.5. The spread is smaller than the one in [Sr/Ba] because the theoretical predictions of barium enrichment of this source are quite small compared to those of strontium; according to the model, the region with [Eu/Ba]$<-$1 is free from r-process pollution. In the abundance region [Ba/H] $\leq -3 $, we will eventually find the (almost) Eu-free stars mentioned in \citet{Cescutti15, Cavallo21}.  Finally, the region at [Ba/H]$>-1$ is created by the enrichment of AGB stars, producing a decreasing trend.  

Overall, the model can explain the behavior found in the sample of GSE stars, but certainly further measurements are needed to strengthen these results or to disprove them.

\section{Conclusions \label{sec:Conclusions}}

This article is a follow-up on the work of MINCE I, which described the method adopted in the MINCE project to select  the sample, determine the stellar atmosphere, and measure at intermediate-low metallicities the chemical abundances of several $\alpha$ elements and iron peak elements, for a first sample of stars. 
\begin{itemize}

\item{} We could determine the abundances for  up to ten neutron-capture elements (Sr, Y, Zr, Ba, La, Ce, Pr, Nd, Sm, and Eu) in 33 stars. The  general trends  of  [n-capture element/Fe]  versus [Fe/H] abundance ratios are in agreement with the results found in the literature.

\item{} When our sample  is divided into galactic sub-components depending on their kinematics identified in MINCE I, we find that the variation  in the  [Sr/Ba] versus [Ba/H] ratio for the stars belonging to the GSE accretion event shows a tight anti-correlation.  
\item{}The results for the Sequoia stars, although based on a 
very limited sample,  shows a [Sr/Ba] systematically higher than the [Sr/Ba] found for the GSE stars at a given [Ba/H], hinting at a different chemical history.   The variations in 
the [n-capture/Fe] abundance ratios for GSE, Sequoia, and the rest of the sample as a function of [Fe/H] do not show any systematic differences.  
 \item{}

 Stochastic chemical evolution models were computed in order to help us understand the evolution of the GSE chemical composition of Sr and Ba with these characteristics: 
the same chemical evolution parameters of \citet{Cescutti20}, so compatible to those of a satellite galaxy with an original mass of 3\,\%  of the Milky Way, with an inefficient star formation ending more than 5 Gyr ago, and a stellar content that is around 1\,\% of the Galactic stellar mass; 
the same nucleosynthesis for Sr and Ba considered in \citet{Cescutti14}, so an r-process enrichment from magneto-rotational-driven SNe 
\citep{Winteler12} and an s-process production from AGB stars described in \citet{Cristallo11}; and an s-process production from rotating massive stars taken from \citet{Frisch15}.
The preliminary conclusions are that this stochastic chemical evolution model's predictions for Sr and Ba are compatible with this first sample of MINCE stars.
\end{itemize}


\begin{acknowledgements}
We gratefully acknowledge support from the French National Research Agency (ANR) funded project ``Pristine'' (ANR-18-CE31-0017).
This work has made use of data from the European Space Agency (ESA) mission
{\it Gaia} \footnote{https://www.cosmos.esa.int/gaia} to , processed by the {\it Gaia}
Data Processing and Analysis Consortium (DPAC,
\footnote{https://www.cosmos.esa.int/web/gaia/dpac/consortium}). Funding for the DPAC
has been provided by national institutions, in particular the institutions
participating in the {\it Gaia} Multilateral Agreement.
This work was also partially supported by the European Union (ChETEC-INFRA, project no. 101008324)
This research has made use of the SIMBAD database, operated at CDS, Strasbourg, France. ES received funding from the European Union’s Horizon 2020 research and innovation program under SPACE-H2020 grant agreement number 101004214 (EXPLORE project). 
GC and LM acknowledge the
grant PRIN project n.2022X4TM3H "Cosmic POT" from Ministero
dell’Universitá e la Ricerca (MUR).
Funding for the Stellar Astrophysics Centre is provided by The Danish National Research Foundation (Grant agreement no.: DNRF106). We also acknowledge support from the CRC ELEMENTS.
\end{acknowledgements}

\bibliographystyle{aa}
\bibliography{49539_Arxiv}

\begin{thebibliography}{89}
\expandafter\ifx\csname natexlab\endcsname\relax\def\natexlab#1{#1}\fi

\bibitem[{{Aguado} {et~al.}(2021){Aguado}, {Belokurov}, {Myeong}, {Evans},
  {Kobayashi}, {Sbordone}, {Chanam{\'e}}, {Navarrete}, \&
  {Koposov}}]{aguado2021}
{Aguado}, D.~S., {Belokurov}, V., {Myeong}, G.~C., {et~al.} 2021, \apjl, 908,
  L8

\bibitem[{{Alvarez} \& {Plez}(1998)}]{alvarez1998}
{Alvarez}, R. \& {Plez}, B. 1998, \aap, 330, 1109

\bibitem[{{Amarsi} {et~al.}(2020){Amarsi}, {Lind}, {Osorio}, {Nordlander},
  {Bergemann}, {Reggiani}, {Wang}, {Buder}, {Asplund}, {Barklem}, {Wehrhahn},
  {Sk{\'u}lad{\'o}ttir}, {Kobayashi}, {Karakas}, {Gao}, {Bland-Hawthorn}, {de
  Silva}, {Kos}, {Lewis}, {Martell}, {Sharma}, {Simpson}, {Zucker},
  {{\v{C}}otar}, {Horner}, \& {GALAH Collaboration}}]{amarsi2020}
{Amarsi}, A.~M., {Lind}, K., {Osorio}, Y., {et~al.} 2020, \aap, 642, A62

\bibitem[{{Anders} {et~al.}(2019){Anders}, {Khalatyan}, {Chiappini}, {Queiroz},
  {Santiago}, {Jordi}, {Girardi}, {Brown}, {Matijevi{\v{c}}}, {Monari},
  {Cantat-Gaudin}, {Weiler}, {Khan}, {Miglio}, {Carrillo}, {Romero-G{\'o}mez},
  {Minchev}, {de Jong}, {Antoja}, {Ramos}, {Steinmetz}, \& {Enke}}]{Anders19}
{Anders}, F., {Khalatyan}, A., {Chiappini}, C., {et~al.} 2019, \aap, 628, A94

\bibitem[{{Arlandini} {et~al.}(1999){Arlandini}, {K{\"a}ppeler}, {Wisshak},
  {Gallino}, {Lugaro}, {Busso}, \& {Straniero}}]{arlandini1999}
{Arlandini}, C., {K{\"a}ppeler}, F., {Wisshak}, K., {et~al.} 1999, \apj, 525,
  886

\bibitem[{{Asplund} {et~al.}(2009){Asplund}, {Grevesse}, {Sauval}, \&
  {Scott}}]{asplund2009}
{Asplund}, M., {Grevesse}, N., {Sauval}, A.~J., \& {Scott}, P. 2009, \araa, 47,
  481

\bibitem[{{Barb{\'a}} {et~al.}(2019){Barb{\'a}}, {Minniti}, {Geisler},
  {Alonso-Garc{\'\i}a}, {Hempel}, {Monachesi}, {Arias}, \&
  {G{\'o}mez}}]{barba19}
{Barb{\'a}}, R.~H., {Minniti}, D., {Geisler}, D., {et~al.} 2019, \apjl, 870,
  L24

\bibitem[{{Belokurov} {et~al.}(2018){Belokurov}, {Erkal}, {Evans}, {Koposov},
  \& {Deason}}]{Belokurov18}
{Belokurov}, V., {Erkal}, D., {Evans}, N.~W., {Koposov}, S.~E., \& {Deason},
  A.~J. 2018, \mnras, 478, 611

\bibitem[{{Bensby} {et~al.}(2014){Bensby}, {Feltzing}, \& {Oey}}]{bensby14}
{Bensby}, T., {Feltzing}, S., \& {Oey}, M.~S. 2014, \aap, 562, A71

\bibitem[{{Bonifacio} {et~al.}(2021){Bonifacio}, {Monaco}, {Salvadori},
  {Caffau}, {Spite}, {Sbordone}, {Spite}, {Ludwig}, {Di Matteo}, {Haywood},
  {Fran{\c{c}}ois}, {Koch-Hansen}, {Christlieb}, \& {Zaggia}}]{toposVI}
{Bonifacio}, P., {Monaco}, L., {Salvadori}, S., {et~al.} 2021, \aap, 651, A79

\bibitem[{{Bouchy} \& {Sophie Team}(2006)}]{2006tafp.conf..319B}
{Bouchy}, F. \& {Sophie Team}. 2006, in Tenth Anniversary of 51 Peg-b: Status
  of and prospects for hot Jupiter studies, ed. L.~{Arnold}, F.~{Bouchy}, \&
  C.~{Moutou}, 319--325

\bibitem[{{Buder} {et~al.}(2022){Buder}, {Lind}, {Ness}, {Feuillet}, {Horta},
  {Monty}, {Buck}, {Nordlander}, {Bland-Hawthorn}, {Casey}, {de Silva},
  {D'Orazi}, {Freeman}, {Hayden}, {Kos}, {Martell}, {Lewis}, {Lin},
  {Schlesinger}, {Sharma}, {Simpson}, {Stello}, {Zucker}, {Zwitter},
  {Ciuc{\u{a}}}, {Horner}, {Kobayashi}, {Ting}, {Wyse}, \& {Wyse}}]{buder2022}
{Buder}, S., {Lind}, K., {Ness}, M.~K., {et~al.} 2022, \mnras, 510, 2407

\bibitem[{{Calura} \& {Menci}(2009)}]{Calura09}
{Calura}, F. \& {Menci}, N. 2009, \mnras, 400, 1347

\bibitem[{{Cavallo} {et~al.}(2021){Cavallo}, {Cescutti}, \&
  {Matteucci}}]{Cavallo21}
{Cavallo}, L., {Cescutti}, G., \& {Matteucci}, F. 2021, \mnras, 503, 1

\bibitem[{{Cavallo} {et~al.}(2023){Cavallo}, {Cescutti}, \&
  {Matteucci}}]{Cavallo23}
{Cavallo}, L., {Cescutti}, G., \& {Matteucci}, F. 2023, \aap, 674, A130

\bibitem[{{Cayrel} {et~al.}(2004){Cayrel}, {Depagne}, {Spite}, {Hill}, {Spite},
  {Fran{\c{c}}ois}, {Plez}, {Beers}, {Primas}, {Andersen}, {Barbuy},
  {Bonifacio}, {Molaro}, \& {Nordstr{\"o}m}}]{cayrel2004}
{Cayrel}, R., {Depagne}, E., {Spite}, M., {et~al.} 2004, \aap, 416, 1117

\bibitem[{{Cescutti}(2008)}]{Cesc08}
{Cescutti}, G. 2008, \aap, 481, 691

\bibitem[{{Cescutti} {et~al.}(2022){Cescutti}, {Bonifacio}, {Caffau}, {Monaco},
  {Franchini}, {Lombardo}, {Matas Pinto}, {Lucertini}, {Fran{\c{c}}ois},
  {Spitoni}, {Lallement}, {Sbordone}, {Mucciarelli}, {Spite}, {Hansen}, {Di
  Marcantonio}, {Ku{\v{c}}inskas}, {Dobrovolskas}, {Korn}, {Valentini},
  {Magrini}, {Cristallo}, \& {Matteucci}}]{cescutti2022}
{Cescutti}, G., {Bonifacio}, P., {Caffau}, E., {et~al.} 2022, \aap, 668, A168

\bibitem[{{Cescutti} \& {Chiappini}(2014)}]{Cescutti14}
{Cescutti}, G. \& {Chiappini}, C. 2014, \aap, 565, A51

\bibitem[{{Cescutti} {et~al.}(2008){Cescutti}, {Matteucci}, {Lanfranchi}, \&
  {McWilliam}}]{Cescutti08}
{Cescutti}, G., {Matteucci}, F., {Lanfranchi}, G.~A., \& {McWilliam}, A. 2008,
  \aap, 491, 401

\bibitem[{{Cescutti} {et~al.}(2020){Cescutti}, {Molaro}, \& {Fu}}]{Cescutti20}
{Cescutti}, G., {Molaro}, P., \& {Fu}, X. 2020, \memsai, 91, 153

\bibitem[{{Cescutti} {et~al.}(2015){Cescutti}, {Romano}, {Matteucci},
  {Chiappini}, \& {Hirschi}}]{Cescutti15}
{Cescutti}, G., {Romano}, D., {Matteucci}, F., {Chiappini}, C., \& {Hirschi},
  R. 2015, \aap, 577, A139

\bibitem[{{Cosentino} {et~al.}(2012){Cosentino}, {Lovis}, {Pepe}, {Collier
  Cameron}, {Latham}, {Molinari}, {Udry}, {Bezawada}, {Black}, {Born},
  {Buchschacher}, {Charbonneau}, {Figueira}, {Fleury}, {Galli}, {Gallie},
  {Gao}, {Ghedina}, {Gonzalez}, {Gonzalez}, {Guerra}, {Henry}, {Horne},
  {Hughes}, {Kelly}, {Lodi}, {Lunney}, {Maire}, {Mayor}, {Micela}, {Ordway},
  {Peacock}, {Phillips}, {Piotto}, {Pollacco}, {Queloz}, {Rice}, {Riverol},
  {Riverol}, {San Juan}, {Sasselov}, {Segransan}, {Sozzetti}, {Sosnowska},
  {Stobie}, {Szentgyorgyi}, {Vick}, \& {Weber}}]{HARPSN}
{Cosentino}, R., {Lovis}, C., {Pepe}, F., {et~al.} 2012, in Society of
  Photo-Optical Instrumentation Engineers (SPIE) Conference Series, Vol. 8446,
  Ground-based and Airborne Instrumentation for Astronomy IV, ed. I.~S.
  {McLean}, S.~K. {Ramsay}, \& H.~{Takami}, 84461V

\bibitem[{{Cristallo} {et~al.}(2011){Cristallo}, {Piersanti}, {Straniero},
  {Gallino}, {Dom{\'{\i}}nguez}, {Abia}, {Di Rico}, {Quintini}, \&
  {Bisterzo}}]{Cristallo11}
{Cristallo}, S., {Piersanti}, L., {Straniero}, O., {et~al.} 2011, \apjs, 197,
  17

\bibitem[{{Cristallo} {et~al.}(2009){Cristallo}, {Straniero}, {Gallino},
  {Piersanti}, {Dom{\'{\i}}nguez}, \& {Lederer}}]{Cristallo09}
{Cristallo}, S., {Straniero}, O., {Gallino}, R., {et~al.} 2009, \apj, 696, 797

\bibitem[{{Den Hartog} {et~al.}(2003){Den Hartog}, {Lawler}, {Sneden}, \&
  {Cowan}}]{denhartog2003}
{Den Hartog}, E.~A., {Lawler}, J.~E., {Sneden}, C., \& {Cowan}, J.~J. 2003,
  \apjs, 148, 543

\bibitem[{{Donati} {et~al.}(2006){Donati}, {Catala}, {Landstreet}, \&
  {Petit}}]{Espadons}
{Donati}, J.~F., {Catala}, C., {Landstreet}, J.~D., \& {Petit}, P. 2006, in
  Astronomical Society of the Pacific Conference Series, Vol. 358, Solar
  Polarization 4, ed. R.~{Casini} \& B.~W. {Lites}, 362

\bibitem[{{Edvardsson} {et~al.}(1993){Edvardsson}, {Andersen}, {Gustafsson},
  {Lambert}, {Nissen}, \& {Tomkin}}]{edvardsson1993}
{Edvardsson}, B., {Andersen}, J., {Gustafsson}, B., {et~al.} 1993, \aap, 275,
  101

\bibitem[{{Feuillet} {et~al.}(2021){Feuillet}, {Sahlholdt}, {Feltzing}, \&
  {Casagrande}}]{feuillet21}
{Feuillet}, D.~K., {Sahlholdt}, C.~L., {Feltzing}, S., \& {Casagrande}, L.
  2021, \mnras, 508, 1489

\bibitem[{{Fran{\c{c}}ois} {et~al.}(2007){Fran{\c{c}}ois}, {Depagne}, {Hill},
  {Spite}, {Spite}, {Plez}, {Beers}, {Andersen}, {James}, {Barbuy}, {Cayrel},
  {Bonifacio}, {Molaro}, {Nordstr{\"o}m}, \& {Primas}}]{francois2007}
{Fran{\c{c}}ois}, P., {Depagne}, E., {Hill}, V., {et~al.} 2007, \aap, 476, 935

\bibitem[{{Frischknecht} {et~al.}(2016){Frischknecht}, {Hirschi}, {Pignatari},
  {Maeder}, {Meynet}, {Chiappini}, {Thielemann}, {Rauscher}, {Georgy}, \&
  {Ekstr{\"o}m}}]{Frisch15}
{Frischknecht}, U., {Hirschi}, R., {Pignatari}, M., {et~al.} 2016, \mnras, 456,
  1803

\bibitem[{{Frischknecht} {et~al.}(2012){Frischknecht}, {Hirschi}, \&
  {Thielemann}}]{Frisch12}
{Frischknecht}, U., {Hirschi}, R., \& {Thielemann}, F.-K. 2012, \aap, 538, L2

\bibitem[{{Gaia Collaboration} {et~al.}(2021){Gaia Collaboration}, {Brown},
  {Vallenari}, {Prusti}, {de Bruijne}, {Babusiaux}, {Biermann}, {Creevey},
  {Evans}, {Eyer}, {Hutton}, {Jansen}, {Jordi}, {Klioner}, {Lammers},
  {Lindegren}, {Luri}, {Mignard}, {Panem}, {Pourbaix}, {Randich}, {Sartoretti},
  {Soubiran}, {Walton}, {Arenou}, {Bailer-Jones}, {Bastian}, {Cropper},
  {Drimmel}, {Katz}, {Lattanzi}, {van Leeuwen}, {Bakker}, {Cacciari},
  {Casta{\~n}eda}, {De Angeli}, {Ducourant}, {Fabricius}, {Fouesneau},
  {Fr{\'e}mat}, {Guerra}, {Guerrier}, {Guiraud}, {Jean-Antoine Piccolo},
  {Masana}, {Messineo}, {Mowlavi}, {Nicolas}, {Nienartowicz}, {Pailler},
  {Panuzzo}, {Riclet}, {Roux}, {Seabroke}, {Sordo}, {Tanga}, {Th{\'e}venin},
  {Gracia-Abril}, {Portell}, {Teyssier}, {Altmann}, {Andrae}, {Bellas-Velidis},
  {Benson}, {Berthier}, {Blomme}, {Brugaletta}, {Burgess}, {Busso}, {Carry},
  {Cellino}, {Cheek}, {Clementini}, {Damerdji}, {Davidson}, {Delchambre},
  {Dell'Oro}, {Fern{\'a}ndez-Hern{\'a}ndez}, {Galluccio}, {Garc{\'\i}a-Lario},
  {Garcia-Reinaldos}, {Gonz{\'a}lez-N{\'u}{\~n}ez}, {Gosset}, {Haigron},
  {Halbwachs}, {Hambly}, {Harrison}, {Hatzidimitriou}, {Heiter},
  {Hern{\'a}ndez}, {Hestroffer}, {Hodgkin}, {Holl}, {Jan{\ss}en}, {Jevardat de
  Fombelle}, {Jordan}, {Krone-Martins}, {Lanzafame}, {L{\"o}ffler}, {Lorca},
  {Manteiga}, {Marchal}, {Marrese}, {Moitinho}, {Mora}, {Muinonen}, {Osborne},
  {Pancino}, {Pauwels}, {Petit}, {Recio-Blanco}, {Richards}, {Riello},
  {Rimoldini}, {Robin}, {Roegiers}, {Rybizki}, {Sarro}, {Siopis}, {Smith},
  {Sozzetti}, {Ulla}, {Utrilla}, {van Leeuwen}, {van Reeven}, {Abbas}, {Abreu
  Aramburu}, {Accart}, {Aerts}, {Aguado}, {Ajaj}, {Altavilla}, {{\'A}lvarez},
  {{\'A}lvarez Cid-Fuentes}, {Alves}, {Anderson}, {Anglada Varela}, {Antoja},
  {Audard}, {Baines}, {Baker}, {Balaguer-N{\'u}{\~n}ez}, {Balbinot}, {Balog},
  {Barache}, {Barbato}, {Barros}, {Barstow}, {Bartolom{\'e}}, {Bassilana},
  {Bauchet}, {Baudesson-Stella}, {Becciani}, {Bellazzini}, {Bernet}, {Bertone},
  {Bianchi}, {Blanco-Cuaresma}, {Boch}, {Bombrun}, {Bossini}, {Bouquillon},
  {Bragaglia}, {Bramante}, {Breedt}, {Bressan}, {Brouillet}, {Bucciarelli},
  {Burlacu}, {Busonero}, {Butkevich}, {Buzzi}, {Caffau}, {Cancelliere},
  {C{\'a}novas}, {Cantat-Gaudin}, {Carballo}, {Carlucci}, {Carnerero},
  {Carrasco}, {Casamiquela}, {Castellani}, {Castro-Ginard}, {Castro Sampol},
  {Chaoul}, {Charlot}, {Chemin}, {Chiavassa}, {Cioni}, {Comoretto}, {Cooper},
  {Cornez}, {Cowell}, {Crifo}, {Crosta}, {Crowley}, {Dafonte}, {Dapergolas},
  {David}, {David}, {de Laverny}, {De Luise}, {De March}, {De Ridder}, {de
  Souza}, {de Teodoro}, {de Torres}, {del Peloso}, {del Pozo}, {Delbo},
  {Delgado}, {Delgado}, {Delisle}, {Di Matteo}, {Diakite}, {Diener},
  {Distefano}, {Dolding}, {Eappachen}, {Edvardsson}, {Enke}, {Esquej}, {Fabre},
  {Fabrizio}, {Faigler}, {Fedorets}, {Fernique}, {Fienga}, {Figueras},
  {Fouron}, {Fragkoudi}, {Fraile}, {Franke}, {Gai}, {Garabato},
  {Garcia-Gutierrez}, {Garc{\'\i}a-Torres}, {Garofalo}, {Gavras}, {Gerlach},
  {Geyer}, {Giacobbe}, {Gilmore}, {Girona}, {Giuffrida}, {Gomel}, {Gomez},
  {Gonzalez-Santamaria}, {Gonz{\'a}lez-Vidal}, {Granvik},
  {Guti{\'e}rrez-S{\'a}nchez}, {Guy}, {Hauser}, {Haywood}, {Helmi}, {Hidalgo},
  {Hilger}, {H{\l}adczuk}, {Hobbs}, {Holland}, {Huckle}, {Jasniewicz},
  {Jonker}, {Juaristi Campillo}, {Julbe}, {Karbevska}, {Kervella}, {Khanna},
  {Kochoska}, {Kontizas}, {Kordopatis}, {Korn}, {Kostrzewa-Rutkowska},
  {Kruszy{\'n}ska}, {Lambert}, {Lanza}, {Lasne}, {Le Campion}, {Le Fustec},
  {Lebreton}, {Lebzelter}, {Leccia}, {Leclerc}, {Lecoeur-Taibi}, {Liao},
  {Licata}, {Lindstr{\o}m}, {Lister}, {Livanou}, {Lobel}, {Madrero Pardo},
  {Managau}, {Mann}, {Marchant}, {Marconi}, {Marcos Santos}, {Marinoni},
  {Marocco}, {Marshall}, {Martin Polo}, {Mart{\'\i}n-Fleitas}, {Masip},
  {Massari}, {Mastrobuono-Battisti}, {Mazeh}, {McMillan}, {Messina},
  {Michalik}, {Millar}, {Mints}, {Molina}, {Molinaro}, {Moln{\'a}r},
  {Montegriffo}, {Mor}, {Morbidelli}, {Morel}, {Morris}, {Mulone}, {Munoz},
  {Muraveva}, {Murphy}, {Musella}, {Noval}, {Ord{\'e}novic}, {Orr{\`u}},
  {Osinde}, {Pagani}, {Pagano}, {Palaversa}, {Palicio}, {Panahi}, {Pawlak},
  {Pe{\~n}alosa Esteller}, {Penttil{\"a}}, {Piersimoni}, {Pineau}, {Plachy},
  {Plum}, {Poggio}, {Poretti}, {Poujoulet}, {Pr{\v{s}}a}, {Pulone}, {Racero},
  {Ragaini}, {Rainer}, {Raiteri}, {Rambaux}, {Ramos}, {Ramos-Lerate}, {Re
  Fiorentin}, {Regibo}, {Reyl{\'e}}, {Ripepi}, {Riva}, {Rixon}, {Robichon},
  {Robin}, {Roelens}, {Rohrbasser}, {Romero-G{\'o}mez}, {Rowell}, {Royer},
  {Rybicki}, {Sadowski}, {Sagrist{\`a} Sell{\'e}s}, {Sahlmann}, {Salgado},
  {Salguero}, {Samaras}, {Sanchez Gimenez}, {Sanna}, {Santove{\~n}a},
  {Sarasso}, {Schultheis}, {Sciacca}, {Segol}, {Segovia}, {S{\'e}gransan},
  {Semeux}, {Shahaf}, {Siddiqui}, {Siebert}, {Siltala}, {Slezak}, {Smart},
  {Solano}, {Solitro}, {Souami}, {Souchay}, {Spagna}, {Spoto}, {Steele},
  {Steidelm{\"u}ller}, {Stephenson}, {S{\"u}veges}, {Szabados}, {Szegedi-Elek},
  {Taris}, {Tauran}, {Taylor}, {Teixeira}, {Thuillot}, {Tonello}, {Torra},
  {Torra}, {Turon}, {Unger}, {Vaillant}, {van Dillen}, {Vanel}, {Vecchiato},
  {Viala}, {Vicente}, {Voutsinas}, {Weiler}, {Wevers}, {Wyrzykowski}, {Yoldas},
  {Yvard}, {Zhao}, {Zorec}, {Zucker}, {Zurbach}, \&
  {Zwitter}}]{Gaia2021A&A...649A...1G}
{Gaia Collaboration}, {Brown}, A.~G.~A., {Vallenari}, A., {et~al.} 2021, \aap,
  649, A1

\bibitem[{{Gaia Collaboration} {et~al.}(2016){Gaia Collaboration}, {Prusti},
  {de Bruijne}, {Brown}, {Vallenari}, {Babusiaux}, {Bailer-Jones}, {Bastian},
  {Biermann}, {Evans}, {Eyer}, {Jansen}, {Jordi}, {Klioner}, {Lammers},
  {Lindegren}, {Luri}, {Mignard}, {Milligan}, {Panem}, {Poinsignon},
  {Pourbaix}, {Randich}, {Sarri}, {Sartoretti}, {Siddiqui}, {Soubiran},
  {Valette}, {van Leeuwen}, {Walton}, {Aerts}, {Arenou}, {Cropper}, {Drimmel},
  {H{\o}g}, {Katz}, {Lattanzi}, {O'Mullane}, {Grebel}, {Holland}, {Huc},
  {Passot}, {Bramante}, {Cacciari}, {Casta{\~n}eda}, {Chaoul}, {Cheek}, {De
  Angeli}, {Fabricius}, {Guerra}, {Hern{\'a}ndez}, {Jean-Antoine-Piccolo},
  {Masana}, {Messineo}, {Mowlavi}, {Nienartowicz}, {Ord{\'o}{\~n}ez-Blanco},
  {Panuzzo}, {Portell}, {Richards}, {Riello}, {Seabroke}, {Tanga},
  {Th{\'e}venin}, {Torra}, {Els}, {Gracia-Abril}, {Comoretto},
  {Garcia-Reinaldos}, {Lock}, {Mercier}, {Altmann}, {Andrae}, {Astraatmadja},
  {Bellas-Velidis}, {Benson}, {Berthier}, {Blomme}, {Busso}, {Carry},
  {Cellino}, {Clementini}, {Cowell}, {Creevey}, {Cuypers}, {Davidson}, {De
  Ridder}, {de Torres}, {Delchambre}, {Dell'Oro}, {Ducourant}, {Fr{\'e}mat},
  {Garc{\'\i}a-Torres}, {Gosset}, {Halbwachs}, {Hambly}, {Harrison}, {Hauser},
  {Hestroffer}, {Hodgkin}, {Huckle}, {Hutton}, {Jasniewicz}, {Jordan},
  {Kontizas}, {Korn}, {Lanzafame}, {Manteiga}, {Moitinho}, {Muinonen},
  {Osinde}, {Pancino}, {Pauwels}, {Petit}, {Recio-Blanco}, {Robin}, {Sarro},
  {Siopis}, {Smith}, {Smith}, {Sozzetti}, {Thuillot}, {van Reeven}, {Viala},
  {Abbas}, {Abreu Aramburu}, {Accart}, {Aguado}, {Allan}, {Allasia},
  {Altavilla}, {{\'A}lvarez}, {Alves}, {Anderson}, {Andrei}, {Anglada Varela},
  {Antiche}, {Antoja}, {Ant{\'o}n}, {Arcay}, {Atzei}, {Ayache}, {Bach},
  {Baker}, {Balaguer-N{\'u}{\~n}ez}, {Barache}, {Barata}, {Barbier}, {Barblan},
  {Baroni}, {Barrado y Navascu{\'e}s}, {Barros}, {Barstow}, {Becciani},
  {Bellazzini}, {Bellei}, {Bello Garc{\'\i}a}, {Belokurov}, {Bendjoya},
  {Berihuete}, {Bianchi}, {Bienaym{\'e}}, {Billebaud}, {Blagorodnova},
  {Blanco-Cuaresma}, {Boch}, {Bombrun}, {Borrachero}, {Bouquillon}, {Bourda},
  {Bouy}, {Bragaglia}, {Breddels}, {Brouillet}, {Br{\"u}semeister},
  {Bucciarelli}, {Budnik}, {Burgess}, {Burgon}, {Burlacu}, {Busonero}, {Buzzi},
  {Caffau}, {Cambras}, {Campbell}, {Cancelliere}, {Cantat-Gaudin}, {Carlucci},
  {Carrasco}, {Castellani}, {Charlot}, {Charnas}, {Charvet}, {Chassat},
  {Chiavassa}, {Clotet}, {Cocozza}, {Collins}, {Collins}, {Costigan}, {Crifo},
  {Cross}, {Crosta}, {Crowley}, {Dafonte}, {Damerdji}, {Dapergolas}, {David},
  {David}, {De Cat}, {de Felice}, {de Laverny}, {De Luise}, {De March}, {de
  Martino}, {de Souza}, {Debosscher}, {del Pozo}, {Delbo}, {Delgado},
  {Delgado}, {di Marco}, {Di Matteo}, {Diakite}, {Distefano}, {Dolding}, {Dos
  Anjos}, {Drazinos}, {Dur{\'a}n}, {Dzigan}, {Ecale}, {Edvardsson}, {Enke},
  {Erdmann}, {Escolar}, {Espina}, {Evans}, {Eynard Bontemps}, {Fabre},
  {Fabrizio}, {Faigler}, {Falc{\~a}o}, {Farr{\`a}s Casas}, {Faye}, {Federici},
  {Fedorets}, {Fern{\'a}ndez-Hern{\'a}ndez}, {Fernique}, {Fienga}, {Figueras},
  {Filippi}, {Findeisen}, {Fonti}, {Fouesneau}, {Fraile}, {Fraser}, {Fuchs},
  {Furnell}, {Gai}, {Galleti}, {Galluccio}, {Garabato}, {Garc{\'\i}a-Sedano},
  {Gar{\'e}}, {Garofalo}, {Garralda}, {Gavras}, {Gerssen}, {Geyer}, {Gilmore},
  {Girona}, {Giuffrida}, {Gomes}, {Gonz{\'a}lez-Marcos},
  {Gonz{\'a}lez-N{\'u}{\~n}ez}, {Gonz{\'a}lez-Vidal}, {Granvik}, {Guerrier},
  {Guillout}, {Guiraud}, {G{\'u}rpide}, {Guti{\'e}rrez-S{\'a}nchez}, {Guy},
  {Haigron}, {Hatzidimitriou}, {Haywood}, {Heiter}, {Helmi}, {Hobbs},
  {Hofmann}, {Holl}, {Holland}, {Hunt}, {Hypki}, {Icardi}, {Irwin}, {Jevardat
  de Fombelle}, {Jofr{\'e}}, {Jonker}, {Jorissen}, {Julbe}, {Karampelas},
  {Kochoska}, {Kohley}, {Kolenberg}, {Kontizas}, {Koposov}, {Kordopatis},
  {Koubsky}, {Kowalczyk}, {Krone-Martins}, {Kudryashova}, {Kull}, {Bachchan},
  {Lacoste-Seris}, {Lanza}, {Lavigne}, {Le Poncin-Lafitte}, {Lebreton},
  {Lebzelter}, {Leccia}, {Leclerc}, {Lecoeur-Taibi}, {Lemaitre}, {Lenhardt},
  {Leroux}, {Liao}, {Licata}, {Lindstr{\o}m}, {Lister}, {Livanou}, {Lobel},
  {L{\"o}ffler}, {L{\'o}pez}, {Lopez-Lozano}, {Lorenz}, {Loureiro},
  {MacDonald}, {Magalh{\~a}es Fernandes}, {Managau}, {Mann}, {Mantelet},
  {Marchal}, {Marchant}, {Marconi}, {Marie}, {Marinoni}, {Marrese},
  {Marschalk{\'o}}, {Marshall}, {Mart{\'\i}n-Fleitas}, {Martino}, {Mary},
  {Matijevi{\v{c}}}, {Mazeh}, {McMillan}, {Messina}, {Mestre}, {Michalik},
  {Millar}, {Miranda}, {Molina}, {Molinaro}, {Molinaro}, {Moln{\'a}r},
  {Moniez}, {Montegriffo}, {Monteiro}, {Mor}, {Mora}, {Morbidelli}, {Morel},
  {Morgenthaler}, {Morley}, {Morris}, {Mulone}, {Muraveva}, {Musella},
  {Narbonne}, {Nelemans}, {Nicastro}, {Noval}, {Ord{\'e}novic},
  {Ordieres-Mer{\'e}}, {Osborne}, {Pagani}, {Pagano}, {Pailler}, {Palacin},
  {Palaversa}, {Parsons}, {Paulsen}, {Pecoraro}, {Pedrosa}, {Pentik{\"a}inen},
  {Pereira}, {Pichon}, {Piersimoni}, {Pineau}, {Plachy}, {Plum}, {Poujoulet},
  {Pr{\v{s}}a}, {Pulone}, {Ragaini}, {Rago}, {Rambaux}, {Ramos-Lerate},
  {Ranalli}, {Rauw}, {Read}, {Regibo}, {Renk}, {Reyl{\'e}}, {Ribeiro},
  {Rimoldini}, {Ripepi}, {Riva}, {Rixon}, {Roelens}, {Romero-G{\'o}mez},
  {Rowell}, {Royer}, {Rudolph}, {Ruiz-Dern}, {Sadowski}, {Sagrist{\`a}
  Sell{\'e}s}, {Sahlmann}, {Salgado}, {Salguero}, {Sarasso}, {Savietto},
  {Schnorhk}, {Schultheis}, {Sciacca}, {Segol}, {Segovia}, {Segransan},
  {Serpell}, {Shih}, {Smareglia}, {Smart}, {Smith}, {Solano}, {Solitro},
  {Sordo}, {Soria Nieto}, {Souchay}, {Spagna}, {Spoto}, {Stampa}, {Steele},
  {Steidelm{\"u}ller}, {Stephenson}, {Stoev}, {Suess}, {S{\"u}veges}, {Surdej},
  {Szabados}, {Szegedi-Elek}, {Tapiador}, {Taris}, {Tauran}, {Taylor},
  {Teixeira}, {Terrett}, {Tingley}, {Trager}, {Turon}, {Ulla}, {Utrilla},
  {Valentini}, {van Elteren}, {Van Hemelryck}, {van Leeuwen}, {Varadi},
  {Vecchiato}, {Veljanoski}, {Via}, {Vicente}, {Vogt}, {Voss}, {Votruba},
  {Voutsinas}, {Walmsley}, {Weiler}, {Weingrill}, {Werner}, {Wevers},
  {Whitehead}, {Wyrzykowski}, {Yoldas}, {{\v{Z}}erjal}, {Zucker}, {Zurbach},
  {Zwitter}, {Alecu}, {Allen}, {Allende Prieto}, {Amorim},
  {Anglada-Escud{\'e}}, {Arsenijevic}, {Azaz}, {Balm}, {Beck}, {Bernstein},
  {Bigot}, {Bijaoui}, {Blasco}, {Bonfigli}, {Bono}, {Boudreault}, {Bressan},
  {Brown}, {Brunet}, {Bunclark}, {Buonanno}, {Butkevich}, {Carret}, {Carrion},
  {Chemin}, {Ch{\'e}reau}, {Corcione}, {Darmigny}, {de Boer}, {de Teodoro}, {de
  Zeeuw}, {Delle Luche}, {Domingues}, {Dubath}, {Fodor}, {Fr{\'e}zouls},
  {Fries}, {Fustes}, {Fyfe}, {Gallardo}, {Gallegos}, {Gardiol}, {Gebran},
  {Gomboc}, {G{\'o}mez}, {Grux}, {Gueguen}, {Heyrovsky}, {Hoar}, {Iannicola},
  {Isasi Parache}, {Janotto}, {Joliet}, {Jonckheere}, {Keil}, {Kim},
  {Klagyivik}, {Klar}, {Knude}, {Kochukhov}, {Kolka}, {Kos}, {Kutka}, {Lainey},
  {LeBouquin}, {Liu}, {Loreggia}, {Makarov}, {Marseille}, {Martayan},
  {Martinez-Rubi}, {Massart}, {Meynadier}, {Mignot}, {Munari}, {Nguyen},
  {Nordlander}, {Ocvirk}, {O'Flaherty}, {Olias Sanz}, {Ortiz}, {Osorio},
  {Oszkiewicz}, {Ouzounis}, {Palmer}, {Park}, {Pasquato}, {Peltzer}, {Peralta},
  {P{\'e}turaud}, {Pieniluoma}, {Pigozzi}, {Poels}, {Prat}, {Prod'homme},
  {Raison}, {Rebordao}, {Risquez}, {Rocca-Volmerange}, {Rosen}, {Ruiz-Fuertes},
  {Russo}, {Sembay}, {Serraller Vizcaino}, {Short}, {Siebert}, {Silva},
  {Sinachopoulos}, {Slezak}, {Soffel}, {Sosnowska}, {Strai{\v{z}}ys}, {ter
  Linden}, {Terrell}, {Theil}, {Tiede}, {Troisi}, {Tsalmantza}, {Tur},
  {Vaccari}, {Vachier}, {Valles}, {Van Hamme}, {Veltz}, {Virtanen}, {Wallut},
  {Wichmann}, {Wilkinson}, {Ziaeepour}, \&
  {Zschocke}}]{Gaia2016A&A...595A...1G}
{Gaia Collaboration}, {Prusti}, T., {de Bruijne}, J.~H.~J., {et~al.} 2016,
  \aap, 595, A1

\bibitem[{{Gallagher} {et~al.}(2020){Gallagher}, {Bergemann}, {Collet}, {Plez},
  {Leenaarts}, {Carlsson}, {Yakovleva}, \& {Belyaev}}]{gallagher2020}
{Gallagher}, A.~J., {Bergemann}, M., {Collet}, R., {et~al.} 2020, \aap, 634,
  A55

\bibitem[{{Garc{\'\i}a-Bethencourt} {et~al.}(2023){Garc{\'\i}a-Bethencourt},
  {Brook}, {Grand}, \& {Kawata}}]{Garcia23}
{Garc{\'\i}a-Bethencourt}, G., {Brook}, C.~B., {Grand}, R. J.~J., \& {Kawata},
  D. 2023, \mnras, 526, 1190

\bibitem[{{Grevesse} \& {Sauval}(2000)}]{grevesse2000}
{Grevesse}, N. \& {Sauval}, A.~J. 2000, in Origin of Elements in the Solar
  System, Implications of Post-1957 Observations, ed. O.~{Manuel}, 261

\bibitem[{{Gustafsson} {et~al.}(1975){Gustafsson}, {Bell}, {Eriksson}, \&
  {Nordlund}}]{gustafsson1975}
{Gustafsson}, B., {Bell}, R.~A., {Eriksson}, K., \& {Nordlund}, A. 1975, \aap,
  42, 407

\bibitem[{{Gustafsson} {et~al.}(2008){Gustafsson}, {Edvardsson}, {Eriksson},
  {J{\o}rgensen}, {Nordlund}, \& {Plez}}]{gustafsson2008}
{Gustafsson}, B., {Edvardsson}, B., {Eriksson}, K., {et~al.} 2008, \aap, 486,
  951

\bibitem[{{Gustafsson} {et~al.}(2003){Gustafsson}, {Edvardsson}, {Eriksson},
  {Mizuno-Wiedner}, {J{\o}rgensen}, \& {Plez}}]{gustafsson2003}
{Gustafsson}, B., {Edvardsson}, B., {Eriksson}, K., {et~al.} 2003, in
  Astronomical Society of the Pacific Conference Series, Vol. 288, Stellar
  Atmosphere Modeling, ed. I.~{Hubeny}, D.~{Mihalas}, \& K.~{Werner}, 331

\bibitem[{{Hannaford} {et~al.}(1982){Hannaford}, {Lowe}, {Grevesse}, {Biemont},
  \& {Whaling}}]{hannaford1982}
{Hannaford}, P., {Lowe}, R.~M., {Grevesse}, N., {Biemont}, E., \& {Whaling}, W.
  1982, \apj, 261, 736

\bibitem[{{Hansen} {et~al.}(2013){Hansen}, {Bergemann}, {Cescutti},
  {Fran{\c{c}}ois}, {Arcones}, {Karakas}, {Lind}, \& {Chiappini}}]{hansen2013}
{Hansen}, C.~J., {Bergemann}, M., {Cescutti}, G., {et~al.} 2013, \aap, 551, A57

\bibitem[{{Hasselquist} {et~al.}(2021){Hasselquist}, {Hayes}, {Lian},
  {Weinberg}, {Zasowski}, {Horta}, {Beaton}, {Feuillet}, {Garro}, {Gallart},
  {Smith}, {Holtzman}, {Minniti}, {Lacerna}, {Shetrone}, {J{\"o}nsson},
  {Cioni}, {Fillingham}, {Cunha}, {O'Connell}, {Fern{\'a}ndez-Trincado},
  {Mu{\~n}oz}, {Schiavon}, {Almeida}, {Anguiano}, {Beers}, {Bizyaev},
  {Brownstein}, {Cohen}, {Frinchaboy}, {Garc{\'\i}a-Hern{\'a}ndez}, {Geisler},
  {Lane}, {Majewski}, {Nidever}, {Nitschelm}, {Povick}, {Price-Whelan},
  {Roman-Lopes}, {Rosado}, {Sobeck}, {Stringfellow}, {Valenzuela}, {Villanova},
  \& {Vincenzo}}]{hasselquist2021}
{Hasselquist}, S., {Hayes}, C.~R., {Lian}, J., {et~al.} 2021, \apj, 923, 172

\bibitem[{{Haywood} {et~al.}(2018){Haywood}, {Di Matteo}, {Lehnert}, {Snaith},
  {Khoperskov}, \& {G{\'o}mez}}]{Haywood18}
{Haywood}, M., {Di Matteo}, P., {Lehnert}, M.~D., {et~al.} 2018, \apj, 863, 113

\bibitem[{{Helmi}(2020)}]{Helmi20}
{Helmi}, A. 2020, \araa, 58, 205

\bibitem[{{Helmi} {et~al.}(2018){Helmi}, {Babusiaux}, {Koppelman}, {Massari},
  {Veljanoski}, \& {Brown}}]{Helmi18}
{Helmi}, A., {Babusiaux}, C., {Koppelman}, H.~H., {et~al.} 2018, \nat, 563, 85

\bibitem[{{Horta} {et~al.}(2022){Horta}, {Schiavon}, {Mackereth}, {Weinberg},
  {Hasselquist}, {Feuillet}, {O'Connell}, {Anguiano}, {Allende-Prieto},
  {Beaton}, {Bizyaev}, {Cunha}, {Geisler}, {Garc{\'\i}a-Hern{\'a}ndez},
  {Holtzman}, {J{\"o}nsson}, {Lane}, {Majewski}, {M{\'e}sz{\'a}ros}, {Minniti},
  {Nitschelm}, {Shetrone}, {Smith}, \& {Zasowski}}]{horta2022}
{Horta}, D., {Schiavon}, R.~P., {Mackereth}, J.~T., {et~al.} 2022, \mnras
  [\eprint[arXiv]{2204.04233}]

\bibitem[{{Ishigaki} {et~al.}(2013){Ishigaki}, {Aoki}, \&
  {Chiba}}]{ishigaki2013}
{Ishigaki}, M.~N., {Aoki}, W., \& {Chiba}, M. 2013, \apj, 771, 67

\bibitem[{{Ivans} {et~al.}(2006){Ivans}, {Simmerer}, {Sneden}, {Lawler},
  {Cowan}, {Gallino}, \& {Bisterzo}}]{ivans2006}
{Ivans}, I.~I., {Simmerer}, J., {Sneden}, C., {et~al.} 2006, \apj, 645, 613

\bibitem[{{Ivarsson} {et~al.}(2001){Ivarsson}, {Litz{\'e}n}, \&
  {Wahlgren}}]{ivarsson2001}
{Ivarsson}, S., {Litz{\'e}n}, U., \& {Wahlgren}, G.~M. 2001, \physscr, 64, 455

\bibitem[{{Iwamoto} {et~al.}(1999){Iwamoto}, {Brachwitz}, {Nomoto},
  {Kishimoto}, {Umeda}, {Hix}, \& {Thielemann}}]{Iwamoto1999}
{Iwamoto}, K., {Brachwitz}, F., {Nomoto}, K., {et~al.} 1999, \apjs, 125, 439

\bibitem[{{Jean-Baptiste} {et~al.}(2017){Jean-Baptiste}, {Di Matteo},
  {Haywood}, {G{\'o}mez}, {Montuori}, {Combes}, \& {Semelin}}]{Jean-Baptiste17}
{Jean-Baptiste}, I., {Di Matteo}, P., {Haywood}, M., {et~al.} 2017, \aap, 604,
  A106

\bibitem[{{Korotin} {et~al.}(2015){Korotin}, {Andrievsky}, {Hansen}, {Caffau},
  {Bonifacio}, {Spite}, {Spite}, \& {Fran{\c{c}}ois}}]{korotin2015}
{Korotin}, S.~A., {Andrievsky}, S.~M., {Hansen}, C.~J., {et~al.} 2015, \aap,
  581, A70

\bibitem[{{Kruijssen} {et~al.}(2020){Kruijssen}, {Pfeffer}, {Chevance},
  {Bonaca}, {Trujillo-Gomez}, {Bastian}, {Reina-Campos}, {Crain}, \&
  {Hughes}}]{Kruijssen2020}
{Kruijssen}, J.~M.~D., {Pfeffer}, J.~L., {Chevance}, M., {et~al.} 2020, \mnras,
  498, 2472

\bibitem[{{Lawler} {et~al.}(2001){Lawler}, {Bonvallet}, \&
  {Sneden}}]{lawler2001}
{Lawler}, J.~E., {Bonvallet}, G., \& {Sneden}, C. 2001, \apj, 556, 452

\bibitem[{{Lawler} {et~al.}(2006){Lawler}, {Den Hartog}, {Sneden}, \&
  {Cowan}}]{lawler2006}
{Lawler}, J.~E., {Den Hartog}, E.~A., {Sneden}, C., \& {Cowan}, J.~J. 2006,
  \apjs, 162, 227

\bibitem[{{Li} {et~al.}(2007){Li}, {Chatelain}, {Holt}, {Rehse}, {Rosner}, \&
  {Scholl}}]{li2007}
{Li}, R., {Chatelain}, R., {Holt}, R.~A., {et~al.} 2007, \physscr, 76, 577

\bibitem[{{Limberg} {et~al.}(2021){Limberg}, {Santucci}, {Rossi}, {Queiroz},
  {Chiappini}, {Souza}, {Perottoni}, {P{\'e}rez-Villegas}, \&
  {Barbosa}}]{limberg2021}
{Limberg}, G., {Santucci}, R.~M., {Rossi}, S., {et~al.} 2021, \apjl, 913, L28

\bibitem[{{Limongi} \& {Chieffi}(2018)}]{Limongi18}
{Limongi}, M. \& {Chieffi}, A. 2018, \apjs, 237, 13

\bibitem[{{Ljung} {et~al.}(2006){Ljung}, {Nilsson}, {Asplund}, \&
  {Johansson}}]{ljung2006}
{Ljung}, G., {Nilsson}, H., {Asplund}, M., \& {Johansson}, S. 2006, \aap, 456,
  1181

\bibitem[{{Lodders} {et~al.}(2009){Lodders}, {Palme}, \& {Gail}}]{Lodders09}
{Lodders}, K., {Palme}, H., \& {Gail}, H.~P. 2009, Landolt B\&ouml;rnstein, 4B,
  712

\bibitem[{{Malhan} {et~al.}(2021){Malhan}, {Yuan}, {Ibata}, {Arentsen},
  {Bellazzini}, \& {Martin}}]{malhan2021}
{Malhan}, K., {Yuan}, Z., {Ibata}, R.~A., {et~al.} 2021, \apj, 920, 51

\bibitem[{{Mashonkina} {et~al.}(2017){Mashonkina}, {Jablonka}, {Sitnova},
  {Pakhomov}, \& {North}}]{mashonkina17}
{Mashonkina}, L., {Jablonka}, P., {Sitnova}, T., {Pakhomov}, Y., \& {North}, P.
  2017, \aap, 608, A89

\bibitem[{{Mashonkina} {et~al.}(2008){Mashonkina}, {Zhao}, {Gehren}, {Aoki},
  {Bergemann}, {Noguchi}, {Shi}, {Takada-Hidai}, \& {Zhang}}]{mashonkina2008}
{Mashonkina}, L., {Zhao}, G., {Gehren}, T., {et~al.} 2008, \aap, 478, 529

\bibitem[{{Matsuno} {et~al.}(2022){Matsuno}, {Koppelman}, {Helmi}, {Aoki},
  {Ishigaki}, {Suda}, {Yuan}, \& {Hattori}}]{matsuno2022}
{Matsuno}, T., {Koppelman}, H.~H., {Helmi}, A., {et~al.} 2022, \aap, 661, A103

\bibitem[{{Matteucci} \& {Greggio}(1986)}]{Matteucci1986}
{Matteucci}, F. \& {Greggio}, L. 1986, \aap, 154, 279

\bibitem[{{McWilliam}(1998)}]{mcwilliam98}
{McWilliam}, A. 1998, \aj, 115, 1640

\bibitem[{{Murphy} {et~al.}(2022){Murphy}, {Yates}, \& {Mohamed}}]{Murphy22}
{Murphy}, G.~G., {Yates}, R.~M., \& {Mohamed}, S.~S. 2022, \mnras, 510, 1945

\bibitem[{{Myeong} {et~al.}(2019){Myeong}, {Vasiliev}, {Iorio}, {Evans}, \&
  {Belokurov}}]{myeong19}
{Myeong}, G.~C., {Vasiliev}, E., {Iorio}, G., {Evans}, N.~W., \& {Belokurov},
  V. 2019, \mnras, 488, 1235

\bibitem[{{Naidu} {et~al.}(2020){Naidu}, {Conroy}, {Bonaca}, {Johnson}, {Ting},
  {Caldwell}, {Zaritsky}, \& {Cargile}}]{naidu2020}
{Naidu}, R.~P., {Conroy}, C., {Bonaca}, A., {et~al.} 2020, \apj, 901, 48

\bibitem[{{Nishimura} {et~al.}(2015){Nishimura}, {Takiwaki}, \&
  {Thielemann}}]{Nishimura15}
{Nishimura}, N., {Takiwaki}, T., \& {Thielemann}, F.-K. 2015, \apj, 810, 109

\bibitem[{{Palmeri} {et~al.}(2000){Palmeri}, {Quinet}, {Wyart}, \&
  {Bi{\'e}mont}}]{palmeri2000}
{Palmeri}, P., {Quinet}, P., {Wyart}, J.~F., \& {Bi{\'e}mont}, E. 2000,
  \physscr, 61, 323

\bibitem[{{Placco} {et~al.}(2021){Placco}, {Sneden}, {Roederer}, {Lawler}, {Den
  Hartog}, {Hejazi}, {Maas}, \& {Bernath}}]{placco2021}
{Placco}, V.~M., {Sneden}, C., {Roederer}, I.~U., {et~al.} 2021, Research Notes
  of the American Astronomical Society, 5, 92

\bibitem[{{Plez}(2012)}]{plez2012}
{Plez}, B. 2012, {Turbospectrum: Code for spectral synthesis}, Astrophysics
  Source Code Library, record ascl:1205.004

\bibitem[{{Plez} {et~al.}(1992){Plez}, {Brett}, \& {Nordlund}}]{plez1992}
{Plez}, B., {Brett}, J.~M., \& {Nordlund}, {\r{A}}. 1992, in Instabilities in
  Evolved Super- and Hypergiants, ed. C.~{de Jager} \& H.~{Nieuwenhuijzen}, 119

\bibitem[{{Prantzos} {et~al.}(2018){Prantzos}, {Abia}, {Limongi}, {Chieffi}, \&
  {Cristallo}}]{Prantzos18}
{Prantzos}, N., {Abia}, C., {Limongi}, M., {Chieffi}, A., \& {Cristallo}, S.
  2018, \mnras, 476, 3432

\bibitem[{{Rizzuti} {et~al.}(2019){Rizzuti}, {Cescutti}, {Matteucci},
  {Chieffi}, {Hirschi}, \& {Limongi}}]{Rizzuti19}
{Rizzuti}, F., {Cescutti}, G., {Matteucci}, F., {et~al.} 2019, \mnras, 489,
  5244

\bibitem[{{Rizzuti} {et~al.}(2021){Rizzuti}, {Cescutti}, {Matteucci},
  {Chieffi}, {Hirschi}, {Limongi}, \& {Saro}}]{Rizzuti21}
{Rizzuti}, F., {Cescutti}, G., {Matteucci}, F., {et~al.} 2021, \mnras, 502,
  2495

\bibitem[{{Roederer} {et~al.}(2014){Roederer}, {Preston}, {Thompson},
  {Shectman}, {Sneden}, {Burley}, \& {Kelson}}]{roederer2014}
{Roederer}, I.~U., {Preston}, G.~W., {Thompson}, I.~B., {et~al.} 2014, \aj,
  147, 136

\bibitem[{{Ryabchikova} {et~al.}(2015){Ryabchikova}, {Piskunov}, {Kurucz},
  {Stempels}, {Heiter}, {Pakhomov}, \& {Barklem}}]{Ryabchikova2015}
{Ryabchikova}, T., {Piskunov}, N., {Kurucz}, R.~L., {et~al.} 2015, \physscr,
  90, 054005

\bibitem[{{Sbordone} {et~al.}(2014){Sbordone}, {Caffau}, {Bonifacio}, \&
  {Duffau}}]{Sbordone14}
{Sbordone}, L., {Caffau}, E., {Bonifacio}, P., \& {Duffau}, S. 2014, \aap, 564,
  A109

\bibitem[{{Schlafly} \& {Finkbeiner}(2011)}]{schlafy}
{Schlafly}, E.~F. \& {Finkbeiner}, D.~P. 2011, \apj, 737, 103

\bibitem[{{Simonetti} {et~al.}(2019){Simonetti}, {Matteucci}, {Greggio}, \&
  {Cescutti}}]{Simonetti19}
{Simonetti}, P., {Matteucci}, F., {Greggio}, L., \& {Cescutti}, G. 2019,
  \mnras, 486, 2896

\bibitem[{{Sneden} {et~al.}(2009){Sneden}, {Lawler}, {Cowan}, {Ivans}, \& {Den
  Hartog}}]{sneden2009}
{Sneden}, C., {Lawler}, J.~E., {Cowan}, J.~J., {Ivans}, I.~I., \& {Den Hartog},
  E.~A. 2009, \apjs, 182, 80

\bibitem[{{Suda} {et~al.}(2008){Suda}, {Katsuta}, {Yamada}, {Suwa}, {Ishizuka},
  {Komiya}, {Sorai}, {Aikawa}, \& {Fujimoto}}]{suda2008}
{Suda}, T., {Katsuta}, Y., {Yamada}, S., {et~al.} 2008, \pasj, 60, 1159

\bibitem[{{Telting} {et~al.}(2014){Telting}, {Avila}, {Buchhave}, {Frandsen},
  {Gandolfi}, {Lindberg}, {Stempels}, {Prins}, \& {NOT staff}}]{FIES}
{Telting}, J.~H., {Avila}, G., {Buchhave}, L., {et~al.} 2014, Astronomische
  Nachrichten, 335, 41

\bibitem[{{Villanova} {et~al.}(2019){Villanova}, {Monaco}, {Geisler},
  {O'Connell}, {Minniti}, {Assmann}, \& {Barb{\'a}}}]{villanova19}
{Villanova}, S., {Monaco}, L., {Geisler}, D., {et~al.} 2019, \apj, 882, 174

\bibitem[{{Vincenzo} {et~al.}(2019){Vincenzo}, {Spitoni}, {Calura},
  {Matteucci}, {Silva Aguirre}, {Miglio}, \& {Cescutti}}]{Vincenzo15}
{Vincenzo}, F., {Spitoni}, E., {Calura}, F., {et~al.} 2019, \mnras, 487, L47

\bibitem[{{Winteler} {et~al.}(2012){Winteler}, {K{\"a}ppeli}, {Perego},
  {Arcones}, {Vasset}, {Nishimura}, {Liebend{\"o}rfer}, \&
  {Thielemann}}]{Winteler12}
{Winteler}, C., {K{\"a}ppeli}, R., {Perego}, A., {et~al.} 2012, \apjl, 750, L22

\end{thebibliography}

\begin{appendix} 

\section{Line list and abundances} 
\begin{table}[!ht]
\label{tab:linelist1}
\caption{ Line list }
\begin{tabular}{lrlcr}
\hline
\hline
Elt & Ion & Wavelength & $\chi_{exc}$ & \loggf \\
\hline
Sr  & \ion{I}  &      4607.327 & 0.000 & +0.230 \\
Y   & \ion{II}  &   4854.863 & 0.992 & -0.380  \\
Y   & \ion{II}  &   4883.684  & 1.084  & 0.070  \\
Y   & \ion{II}  &   4900.120  &1.033 & -0.090  \\ 
Y   & \ion{II}  &   5087.416 & 1.084 & -0.170  \\
Y   & \ion{II}  & 5119.112 & 0.992 & -1.360  \\
Y   & \ion{II}  & 5123.211 & 0.992 & -0.830  \\
Y   & \ion{II}  &  5200.406 & 0.992  &-0.570  \\
Y   & \ion{II}  &  5402.774 & 1.839  &-0.510  \\
Y   & \ion{II}  &   5662.925 & 1.944  & 0.160  \\
Zr   & \ion{II}  & 4317.299&  0.713&  -1.450 \\  
Zr   & \ion{II}  &4613.946 & 0.972  &-1.540  \\  
Zr   & \ion{II}  & 5112.270 & 1.665  &-0.850 \\  
Zr   & \ion{II}  &  5350.350 & 1.773 & -1.160 \\  
La   & \ion{II}  &   4662.4774 & 0.000 & -2.951  \\  
La   & \ion{II}  & 4662.4814 &0.000  &-2.511      \\
La   & \ion{II}  & 4662.4852& 0.000  &-2.240    \\
La   & \ion{II}  &   4662.4903& 0.000  &-2.252   \\
La   & \ion{II}  &  4662.4914& 0.000  &-2.136    \\
La   & \ion{II}  & 4662.4924& 0.000  &-2.256    \\
 La   & \ion{II}  &4662.5024 &0.000  &-2.511    \\
 La   & \ion{II}  & 4662.5044 &0.000  &-2.056   \\
 La   & \ion{II}  &4662.5063 &0.000  &-1.763  \\
La   & \ion{II}  & 4920.976 & 0.126  &-0.580  \\
La   & \ion{II}  & 4921.776 & 0.244  &-0.450  \\
La   & \ion{II}  &5114.5115& 0.235  &-1.624    \\
La   & \ion{II}  &  5114.5284 &0.235  &-1.820  \\
La   & \ion{II}  & 5114.5553 &0.235  &-1.820   \\
La   & \ion{II}  & 5114.5723& 0.235  &-3.006   \\
La   & \ion{II}  & 5114.5855 &0.235  &-1.824   \\
 La   & \ion{II}  &5114.6073 &0.235  &-1.824   \\
La   & \ion{II}  & 5114.6205 &0.235  &-2.079   \\
La   & \ion{II}  &  5122.9798 & 0.321  &-1.536     \\
La   & \ion{II}  &  5122.9799 & 0.321  &-2.106    \\
La   & \ion{II}  &  5122.9859 & 0.321  &-2.106    \\
La   & \ion{II}  &  5122.9860 & 0.321  &-1.934  \\
La   & \ion{II}  &  5122.9864 & 0.321  &-1.948    \\
La   & \ion{II}  &  5122.9911 & 0.321  &-1.948   \\
La   & \ion{II}  &  5122.9915 & 0.321  &-2.630    \\
La   & \ion{II}  &  5122.9919 & 0.321  &-1.959   \\
La   & \ion{II}  &  5122.9954 & 0.321  &-1.959    \\
La   & \ion{II}  &  5122.9958 & 0.321  &-4.059    \\
La   & \ion{II}  &  5122.9962 & 0.321  &-2.132   \\
La   & \ion{II}  &  5122.9986 & 0.321  &-2.132    \\
La   & \ion{II}  &  5122.9990 & 0.321  &-2.308    \\
La   & \ion{II}  &  5290.818  & 0.000  &-1.650   \\
Ce   & \ion{II}  &   4539.853 & 1.645& -2.050  \\
Ce   & \ion{II}  &  4562.282 & 1.327 & -2.120   \\
Ce   & \ion{II}  & 4628.161 & 0.516  & 0.200  \\
Ce   & \ion{II}  & 5187.460 & 0.495  &-2.300  \\
 \hline 
 \end{tabular}
 \end{table}

\begin{table}
\begin{tabular}{lrlcr}
\hline
\hline
 Pr & \ion{II}  &  5219.0096 & 0.795 &  -1.990  \\
 Pr & \ion{II}  &  5219.0245 & 0.795 &  -1.943  \\
 Pr & \ion{II}  &  5219.0382 & 0.795 &  -1.994  \\  
 Pr & \ion{II}  &  5219.0422 & 0.795 &  -0.720  \\  
 Pr & \ion{II}  &  5219.0507 & 0.795 &  -2.194  \\  
 Pr & \ion{II}  &  5219.0531 & 0.795 &  -0.794  \\
 Pr & \ion{II}  &  5219.0629 & 0.795 &  -0.866   \\
 Pr & \ion{II}  &  5219.0715 & 0.795 &  -0.934   \\
 Pr & \ion{II}  &  5219.0789 & 0.795 &  -0.992   \\
 Pr & \ion{II}  &  5219.0852 & 0.795 &  -1.035  \\
 Pr & \ion{II}  &  5219.1018 & 0.795 &  -2.188  \\
 Pr & \ion{II}  &  5219.1064 & 0.795 & -1.990  \\
 Pr & \ion{II}  &  5219.1099 & 0.795 & -1.943   \\
 Pr & \ion{II}  &  5219.1122 & 0.795 & -1.994  \\
 Pr & \ion{II}  &  5219.1134 & 0.795 & -2.194   \\
Pr & \ion{II}  &  5220.0178 & 0.795  & -3.464  \\
Pr & \ion{II}  &  5220.0339 & 0.795  & -3.410 \\
Pr & \ion{II}  &  5220.0475 & 0.795  & -1.892  \\
Pr & \ion{II}  &  5220.0486 & 0.795  & -3.602   \\
Pr & \ion{II}  &  5220.0599 & 0.795  & -1.693   \\
Pr & \ion{II}  &  5220.0711 & 0.795  & -1.645  \\
Pr & \ion{II}  &  5220.0809 & 0.795  & -1.696  \\
Pr & \ion{II} &  5220.0894 & 0.795  & -1.895   \\
Pr & \ion{II}  &  5220.0996 & 0.795  & -0.368  \\
Pr & \ion{II}  &  5220.1071 & 0.795  & -0.424   \\
Pr & \ion{II} &  5220.1132 & 0.795  & -0.481   \\
Pr & \ion{II} &  5220.1181 & 0.795  & -0.540   \\
Pr & \ion{II} &  5220.1217 & 0.795  & -0.598   \\
Pr & \ion{II}  &  5220.1239 & 0.795  & -0.656   \\
Pr & \ion{II}&  5259.6145 &0.633  &-3.727   \\
Pr & \ion{II}&  5259.6329 &0.633  &-3.418   \\
Pr & \ion{II}&  5259.6498 &0.633 & -3.356   \\
Pr & \ion{II}&  5259.6653 &0.633 & -3.539   \\
Pr & \ion{II}&  5259.6667 &0.633 & -1.961   \\
Pr & \ion{II}&  5259.6789 &0.633  &-1.763   \\
Pr & \ion{II}&  5259.6897 &0.633  &-1.716   \\
Pr & \ion{II}&  5259.6991 &0.633  &-1.767   \\
Pr & \ion{II}&  5259.7070 &0.633  &-1.965   \\
Pr & \ion{II}&  5259.7251 &0.633  &-0.538   \\
Pr & \ion{II}&  5259.7312 &0.633  &-0.603   \\
Pr & \ion{II}&  5259.7358 &0.633  &-0.669   \\
Pr & \ion{II}&  5259.7390 &0.633  &-0.737   \\
Pr & \ion{II}&  5259.7408 &0.633  &-0.806   \\
Pr & \ion{II}&  5259.7411 &0.633  &-0.874   \\
Pr & \ion{II}&  5322.6702 &0.482  &-3.392   \\
Pr & \ion{II}&  5322.6704 &0.482  &-3.320   \\
Pr & \ion{II}&  5322.6714 &0.482  &-3.710   \\
Pr & \ion{II}&  5322.6718 &0.482 & -3.488   \\
Pr & \ion{II}&  5322.7044 &0.482  &-2.073   \\
Pr & \ion{II}&  5322.7102 &0.482  &-1.878   \\
Pr & \ion{II}&  5322.7173 &0.482  &-1.826   \\
Pr & \ion{II}&  5322.7257 &0.482  &-1.871   \\
Pr & \ion{II}&  5322.7297 &0.482  &-1.164   \\
Pr & \ion{II}&  5322.7354 &0.482  &-2.066   \\
Pr & \ion{II}&  5322.7427 &0.482  &-1.082   \\
Pr & \ion{II}&  5322.7571 &0.482  &-0.998   \\
Pr & \ion{II}&  5322.7727 &0.482  &-0.915   \\
Pr & \ion{II}&  5322.7897 &0.482  &-0.836   \\
Pr & \ion{II}&  5322.8079 &0.482  &-0.760   \\
Pr & \ion{II}&  6165.891   &0.923  &-0.299   \\
\hline 
\hline 
 \end{tabular}
 \end{table}
 
 \begin{table}
\begin{tabular}{lrlcr}
\hline 
\hline 
Nd & \ion{II}& 4501.810  &0.205&  -0.690 \\
Nd & \ion{II}& 4859.026  &0.321 & -0.440  \\
Nd & \ion{II}& 4959.115  &0.064  &-0.800 \\
Nd & \ion{II}& 5076.580 & 0.742 & -0.386  \\
Nd & \ion{II}& 5255.502 & 0.205  &-0.670 \\
Nd & \ion{II}& 5293.160 & 0.823  & 0.100  \\
Nd & \ion{II}& 5319.810 & 0.550  &-0.140 \\
Sm & \ion{II}&   4566.200  &0.333 & -0.590  \\
Sm & \ion{II}&   4615.440 & 0.544  &-0.690 \\
Sm & \ion{II}&   4669.640 & 0.277  &-0.530 \\
Sm & \ion{II}&   4674.590 & 0.185  &-0.560 \\
 Sm & \ion{II}&  4676.900 & 0.040  &-0.870  \\
 Sm & \ion{II}&  4791.579 & 0.104  &-1.440    \\
 Sm & \ion{II}&  4913.259 & 0.659  &-0.930 \\
 Eu& \ion{II} &  4435.4571 &0.207 & -0.696   \\    
 Eu& \ion{II} &  4435.4650 &0.207  &-1.708 \\
 Eu& \ion{II} &  4435.4725 &0.207  &-3.034 \\
 Eu& \ion{II} &  4435.5254 &0.207  &-0.816 \\
 Eu& \ion{II} &  4435.5329 &0.207  &-1.525 \\
 Eu& \ion{II} &  4435.5395 &0.207  &-2.689  \\
 Eu& \ion{II} &  4435.5823 &0.207  &-0.947  \\
 Eu& \ion{II} &  4435.5889 &0.207  &-1.491  \\
 Eu& \ion{II} &  4435.5944 &0.207  &-2.577  \\
 Eu& \ion{II} &  4435.6273 &0.207  &-1.093  \\
 Eu& \ion{II} &  4435.6328 &0.207  &-1.550  \\
 Eu& \ion{II} &  4435.6369 &0.207  &-2.689   \\
 Eu& \ion{II} &  4435.6602 &0.207  &-1.256   \\
 Eu& \ion{II} &  4435.6643 &0.207  &-1.733      \\      
 Eu& \ion{II} &  4435.6808 &0.207  &-1.432  \\
 Eu& \ion{II} &  4522.4767 &0.207 & -2.159   \\
 Eu& \ion{II} &  4522.4887 &0.207 & -1.266   \\
 Eu& \ion{II} &  4522.5292 &0.207 & -1.984   \\
 Eu& \ion{II} &  4522.5394 &0.207 & -1.474   \\
 Eu& \ion{II} &  4522.5515 &0.207 & -2.159   \\
 Eu& \ion{II} &  4522.5724 &0.207 & -1.962   \\
 Eu& \ion{II} &  4522.5806 &0.207 & -1.711   \\
 Eu& \ion{II} &  4522.5908 &0.207 & -1.984   \\
 Eu& \ion{II} &  4522.6064 &0.207 & -2.038   \\
 Eu& \ion{II} &  4522.6123 &0.207 & -1.980   \\
 Eu& \ion{II} &  4522.6205 &0.207 & -1.962   \\
 Eu& \ion{II} &  4522.6312 &0.207 & -2.247   \\
 Eu& \ion{II} &  4522.6349 &0.207 & -2.256   \\
 Eu& \ion{II} &  4522.6408 &0.207 & -2.038   \\
 Eu& \ion{II} &  4522.6483 &0.207 & -2.344   \\
 Eu& \ion{II} &  4522.6520 &0.207 & -2.247   \\
 Eu& \ion{II} &  6645.0727 &1.379 & -1.823   \\
 Eu& \ion{II} &  6645.0744 &1.379 & -0.517   \\
 Eu& \ion{II} &  6645.0749 &1.379 & -3.452   \\
 Eu& \ion{II} &  6645.0876 &1.379  &-0.593   \\
 Eu& \ion{II} &  6645.0898 &1.379 & -1.628   \\
 Eu& \ion{II} &  6645.0945 &1.379 & -3.151   \\
 Eu& \ion{II} &  6645.0974 &1.379 & -0.672   \\
 Eu& \ion{II} &  6645.1021 &1.379 & -1.583   \\
 Eu& \ion{II} &  6645.1047 &1.379 & -0.755   \\
 Eu& \ion{II} &  6645.1081 &1.379 & -3.079   \\
 Eu& \ion{II} &  6645.1101 &1.379 & -0.839   \\
 Eu& \ion{II} &  6645.1107 &1.379 & -1.635   \\
 Eu& \ion{II} &  6645.1144 &1.379 & -0.921   \\
 Eu& \ion{II} &  6645.1164 &1.379 & -1.830   \\
 Eu& \ion{II} &  6645.1170 &1.379 & -3.236   \\
 \hline 
 \end{tabular}
 \end{table}
 
 \begin{table}
\begin{tabular}{lrrrrr}
\hline 
\hline 
Ba & \ion{II }  &134 &   4554.034 &  0.000  &  +0.170    \\                                  
Ba & \ion{II } & 134 &   4934.100 &  0.000  &  -1.157    \\                                  
Ba & \ion{II }  &134 &   5853.690 &  0.604  &  -1.010    \\                                  
Ba & \ion{II }  &134 &   6141.730 &  0.704  &  -0.077    \\                                  
Ba & \ion{II }  &134 &   6496.910 &  0.604  &  -0.380    \\   
Ba & \ion{II }  &135 &   4554.003 &  0.000  & -0.636     \\                                  
Ba & \ion{II }  &135 &   4554.004 &  0.000  & -1.033     \\                                  
Ba & \ion{II }  &135 &   4554.004 &  0.000  & -0.636     \\                                  
Ba & \ion{II }  &135 &   4554.050 &  0.000  & -0.189     \\                                  
Ba & \ion{II }  &135 &   4554.053 &  0.000  & -0.636     \\                                  
Ba & \ion{II }  &135 &   4554.054 &  0.000  & -1.337     \\                                  
Ba & \ion{II }  &135 &   4934.059 &  0.000  & -1.662     \\                                  
Ba & \ion{II }  &135 &   4934.070 &  0.000  & -2.362     \\                                  
Ba & \ion{II }  &135 &   4934.118 &  0.000  & -1.662     \\                                  
Ba & \ion{II }  &135 &   4934.129 &  0.000  & -1.662     \\                                  
Ba & \ion{II }  &135 &   5853.687 &  0.604  & -2.066     \\                                  
Ba & \ion{II }  &135 &   5853.688 &  0.604  & -2.009     \\                                  
Ba & \ion{II }  &135 &   5853.689 &  0.604  & -2.215     \\                                  
Ba & \ion{II }  &135 &   5853.690 &  0.604  & -2.620     \\                                  
Ba & \ion{II }  &135 &   5853.690 &  0.604  & -1.914     \\                                  
Ba & \ion{II }  &135 &   5853.690 &  0.604  & -1.466     \\                                  
Ba & \ion{II }  &135 &   5853.691 &  0.604  & -2.215     \\                                  
Ba & \ion{II }  &135 &   5853.693 &  0.604  & -2.009     \\                                  
Ba & \ion{II }  &135 &   5853.694 &  0.604  & -2.066     \\                                  
Ba & \ion{II }  &135 &   6141.725 &  0.704  & -2.456     \\                                  
Ba & \ion{II }  &135 &   6141.727 &  0.704  & -1.311     \\                                  
Ba & \ion{II }  &135 &   6141.728 &  0.704  & -2.284     \\                                  
Ba & \ion{II }  &135 &   6141.729 &  0.704  & -1.214     \\                                  
Ba & \ion{II }  &135 &   6141.729 &  0.704  & -0.503     \\                                  
Ba & \ion{II }  &135 &   6141.731 &  0.704  & -1.327     \\ 
Ba & \ion{II }  &135 &   6141.731 &  0.704  & -0.709     \\                                  
Ba & \ion{II }  &135 &   6141.732 &  0.704  & -1.281     \\                                  
Ba & \ion{II }  &135 &   6141.732 &  0.704  & -0.959     \\                                  
Ba & \ion{II }  &135 &   6496.899 &  0.604  & -1.886     \\                                  
Ba & \ion{II }  &135 &   6496.902 &  0.604  & -1.186     \\                                  
Ba & \ion{II }  &135 &   6496.906 &  0.604  & -0.739     \\                                  
Ba & \ion{II }  &135 &   6496.916 &  0.604  & -1.583     \\                                  
Ba & \ion{II }  &135 &   6496.917 &  0.604  & -1.186     \\                                  
Ba & \ion{II }  &135 &   6496.920 &  0.604  & -1.186     \\    
\hline 
\hline 
 \end{tabular}
 \end{table}
 
 \begin{table*}
\begin{tabular}{lrrrrr}
\hline 
\hline 
Ba & \ion{II }  &136 &   4554.034 &  0.000  &  +0.170    \\                                  
Ba & \ion{II } &136 &   4934.100 &  0.000  &  -1.157    \\                                  
Ba & \ion{II }  &136 &   5853.690 &  0.604  &  -1.010    \\                                  
Ba & \ion{II }  &136 &   6141.730 &  0.704  &  -0.077    \\                                  
Ba & \ion{II }  &136 &   6496.910 &  0.604  &  -0.380    \\     
 Ba & \ion{II } &137  &   4554.001 &  0.000  & -0.636     \\                                  
Ba & \ion{II } &137  &   4554.002 &  0.000  & -1.033     \\                                  
Ba & \ion{II } &137  &   4554.002 &  0.000  & -0.636   \\                                   
Ba & \ion{II } &137  &   4554.053 &  0.000  & -0.189   \\                                   
Ba & \ion{II } &137  &   4554.056 &  0.000  & -0.636   \\                                   
Ba & \ion{II } &137  &   4554.057 &  0.000  & -1.337   \\                                   
Ba & \ion{II } &137  &   4934.054 &  0.000  & -1.662   \\                                    
Ba & \ion{II } &137  &   4934.066 &  0.000  & -2.362   \\                                    
Ba & \ion{II } &137  &   4934.121 &  0.000  & -1.662   \\                                    
Ba & \ion{II } &137  &   4934.132 &  0.000  & -1.662   \\                                    
Ba & \ion{II } &137  &   5853.686 &  0.604  & -2.066   \\                                  
Ba & \ion{II } &137  &   5853.687 &  0.604  & -2.009   \\                                   
Ba & \ion{II } &137  &   5853.689 &  0.604  & -2.215   \\                                   
Ba & \ion{II } &137  &   5853.690 &  0.604  & -2.620   \\                                   
Ba & \ion{II } &137  &   5853.690 &  0.604  & -1.914   \\                                   
Ba & \ion{II } &137  &   5853.690 &  0.604  & -1.466   \\                                   
Ba & \ion{II } &137  &   5853.692 &  0.604  & -2.215   \\                                   
Ba & \ion{II } &137  &   5853.693 &  0.604  & -2.009   \\                                   
Ba & \ion{II } &137  &   5853.694 &  0.604  & -2.066   \\                                                   
Ba & \ion{II } &137  &   6141.725 &  0.704  & -2.456   \\                                   
Ba & \ion{II } &137  &   6141.727 &  0.704  & -1.311   \\                                   
Ba & \ion{II } &137  &   6141.728 &  0.704  & -2.284   \\                                   
Ba & \ion{II } &137  &   6141.729 &  0.704  & -1.214   \\                                   
Ba & \ion{II } &137  &   6141.729 &  0.704  & -0.503   \\                                   
Ba & \ion{II } &137  &   6141.731 &  0.704  & -1.327   \\                                   
Ba & \ion{II } &137  &   6141.731 &  0.704  & -0.709   \\                                   
Ba & \ion{II } &137  &   6141.732 &  0.704  & -0.959   \\                                   
Ba & \ion{II } &137  &   6141.733 &  0.704  & -1.281   \\                                   
Ba & \ion{II } &137  &   6496.898 &  0.604  & -1.886   \\                                      
Ba & \ion{II } &137  &   6496.901 &  0.604  & -1.186   \\                                      
Ba & \ion{II } &137  &   6496.906 &  0.604  & -0.739   \\                                      
Ba & \ion{II } &137  &   6496.916 &  0.604  & -1.583   \\                                      
Ba & \ion{II } &137  &   6496.918 &  0.604  & -1.186   \\ 
Ba & \ion{II } &137  &   6496.922 &  0.604  & -1.186   \\  
Ba & \ion{II } &138  &  4554.036  &  0.000  & +0.170   \\                                     
Ba & \ion{II } &138  &  4934.100  &  0.000  & -1.157   \\                                     
Ba & \ion{II } &138  &  5853.690  &  0.604  & -1.010   \\                                     
Ba & \ion{II } &138  &  6141.730  &  0.704  & -0.077   \\                                     
Ba & \ion{II } &138  &  6496.910  &  0.604  & -0.380   \\  
\hline                                    
\end{tabular}
\end{table*}

\begin{table*}

\caption{ Abundances }
\label{tab:abund}
\begin{tabular}{lcccccccccccc}
\hline
Star              &   [Fe1/H]  &   $\sigma$(Fe1) & [Fe2/H]  &   $\sigma$(Fe2)  &  [Sr/H]  &   $\sigma$(Sr) &   [Y/H] &    $\sigma$(Y)   &  [Zr/H]   &    $\sigma$(Zr) &  [Ba/H]  &   $\sigma$(Ba)   \\
\hline 
Sequoia              &            &          &          &           &          &         &         &           &           &          &          &          \\
 \hline                                                                                                                                                           
HD115575          &    -1.99   &   0.09   &   -1.86  &    0.12   &   -2.40  &   0.00  &   -2.16 &    0.05   &   -1.53   &    0.04  &   -2.16  &    0.08   \\                     
TYC 4267-2023-1   &    -1.74   &   0.14   &   -2.08  &    0.16   &   -1.65  &   0.05  &   -1.91 &    0.08   &   -1.36   &    0.09  &   -1.70  &    0.09   \\
BD+31 2143        &    -2.36   &   0.10   &   -2.24  &    0.12   &   -2.30  &   0.16  &   -2.45 &    0.03   &   -1.98   &    0.05  &   -2.42  &    0.00   \\
\hline
GSE               &            &          &          &           &          &         &         &           &           &          &          &           \\
\hline
BD+20 3298        &    -1.95   &   0.10   &   -1.86  &    0.17   &   -2.23  &   0.13  &   -2.12 &    0.05   &   -1.55   &    0.03  &   -1.83  &    0.02   \\
TYC 1008-1200-1   &    -2.23   &   0.10   &   -2.00  &    0.17   &   -2.45  &   0.05  &   -2.24 &    0.06   &   -1.77   &    0.10  &   -2.00  &    0.05   \\
HD 238439         &    -2.09   &   0.11   &   -2.02  &    0.13   &   -2.40  &   0.00  &   -2.12 &    0.04   &   -1.58   &    0.04  &   -1.98  &    0.09   \\
HD 142614         &    -1.45   &   0.11   &   -1.47  &    0.18   &   -1.80  &   0.00  &   -1.46 &    0.06   &   -0.86   &    0.02  &   -1.15  &    0.10   \\
BD +04 18         &    -1.48   &   0.12   &   -1.28  &    0.31   &   -1.90  &   0.00  &   -1.31 &    0.06   &   -0.80   &    0.04  &   -0.78  &    0.04   \\
BD+39 3309        &    -2.58   &   0.13   &   -2.46  &    0.13   &   -2.30  &   0.00  &   -2.57 &    0.02   &    ----   &    ----  &   -3.02  &    0.07   \\
TYC 2824-1963-1   &    -1.60   &   0.14   &   -1.45  &    0.38   &   -2.15  &   0.00  &   -1.63 &    0.08   &   -1.08   &    0.02  &   -1.23  &    0.00   \\
TYC 4001-1161-1   &    -1.62   &   0.11   &   -1.28  &    0.29   &   -2.15  &   0.00  &   -1.69 &    0.06   &   -1.05   &    0.06  &   -1.23  &    0.03   \\
TYC 4221-640-1    &    -2.27   &   0.11   &   -2.19  &    0.22   &   -2.70  &   0.00  &   -2.52 &    0.04   &   -1.94   &    0.05  &   -2.56  &    0.06   \\  \
TYC 4-369-1       &    -1.84   &   0.11   &   -1.66  &    0.25   &   -2.25  &   0.00  &   -1.84 &    0.06   &   -1.32   &    0.05  &   -1.60  &    0.00   \\
\hline
other             &            &          &          &         &          &         &         &           &           &          &          &           \\
\hline

BD -00 4538        &    -1.90   &   0.09   &   -1.75  &    0.14   &   -2.00  &   0.12  &   -1.98 &    0.05   &   -1.43   &    0.04  &   -1.73  &    0.05   \\
BD +03 4904       &    -2.57   &   0.12   &   -2.47  &    0.17   &   -2.53  &   0.03  &   -2.65 &    0.06   &   -1.97   &    0.00  &   -2.86  &    0.06   \\
BD +07 4625       &    -1.93   &   0.10   &   -1.92  &    0.13   &   -2.28  &   0.06  &   -2.19 &    0.04   &   -1.64   &    0.06  &   -1.94  &    0.00   \\
BD+11 2896        &    -1.41   &   0.12   &   -1.30  &    0.20   &   -1.70  &   0.10  &   -1.49 &    0.06   &   -0.90   &    0.04  &   -1.25  &    0.00   \\
BD +21 4759       &    -2.50   &   0.14   &   -2.34  &    0.12   &   -2.40  &   0.00  &   -2.27 &    0.04   &   -1.82   &    0.04  &   -1.98  &    0.03   \\
BD +25 4520       &    -2.28   &   0.10   &   -2.13  &    0.17   &   -2.52  &   0.16  &   -2.68 &    0.05   &   -1.94   &    0.06  &   -2.61  &    ----   \\
BD +32 2483       &    -2.25   &   0.11   &   -2.20  &    0.16   &   -2.35  &   0.05  &   -2.38 &    0.05   &   -1.82   &    0.04  &   -2.42  &    0.04   \\
BD +35 4847       &    -1.91   &   0.10   &   -1.90  &    0.17   &   -1.80  &   0.00  &   -1.62 &    0.04   &   -1.25   &    0.06  &   -1.43  &    0.06   \\
BD +48 2167       &    -2.28   &   0.11   &   -2.15  &    0.12   &   -1.95  &   0.10  &   -2.43 &    0.04   &   -1.90   &    0.02  &   -2.20  &    0.04   \\
BD -07 3523       &    -1.95   &   0.10   &   -1.87  &    0.16   &   -2.35  &   0.00  &   -2.04 &    0.04   &   -1.52   &    0.05  &   -1.73  &    0.10   \\
BD +06 2880       &    -1.45   &   0.11   &   -1.30  &    0.23   &   -1.85  &   0.15  &   -1.53 &    0.04   &   -0.98   &    0.02  &   -1.00  &    0.02   \\
HD 139423         &    -1.70   &   0.11   &   -1.78  &    0.15   &   -2.05  &   0.00  &   -1.84 &    0.07   &   -1.23   &    0.04  &   -1.71  &    0.07   \\
HD 208316         &    -1.61   &   0.10   &   -1.57  &    0.10   &   -1.85  &   0.00  &   -1.52 &    0.04   &   -1.00   &    0.03  &   -1.53  &    0.04   \\
HD 354750         &    -2.35   &   0.11   &   -2.42  &    0.14   &   -2.25  &   0.05  &   -2.54 &    0.03   &    ----   &    ----  &   -2.53  &    0.02   \\
TYC 2588-1386-1   &    -1.73   &   0.12   &   -1.70  &    0.20   &   -2.20  &   0.00  &   -1.79 &    0.06   &   -1.28   &    0.02  &   -1.50  &    0.07   \\
TYC 3085-119-1$^a$    &    -1.51   &   0.10   &   -1.42  &    0.17   &   -1.31  &   0.07  &   -1.24 &    0.06   &   -0.78   &    0.07  &   -1.07  &    0.05   \\
TYC 33-446-1      &    -2.22   &   0.12   &   -2.20  &    0.20   &   -2.38  &   0.03  &   -2.37 &    0.06   &   -1.70   &    0.05  &   -2.15  &    0.06   \\
TYC 3944-698-1    &    -2.18   &   0.13   &   -2.06  &    0.21   &   -2.70  &   0.00  &   -2.26 &    0.07   &   -1.79   &    0.02  &   -2.26  &    0.04   \\
TYC 4331-136-1    &    -2.53   &   0.11   &   -2.43  &    0.23   &   -2.75  &   0.00  &   -2.62 &    0.07   &   -1.92   &    0.05  &   -2.58  &    0.02   \\
TYC 4584-784-1    &    -2.03   &   0.11   &   -1.85  &    0.25   &   -2.35  &   0.00  &   -2.11 &    0.08   &   -1.49   &    0.06  &   -1.90  &    0.00   \\
\hline
\\
  \multicolumn{4}{l}{$^a$ Thick-disc star}\\
\end{tabular}
\end{table*}

\begin{table*}
\label{tab:abund}
\begin{tabular}{lcccccccccccc}

\hline
Star    & [La/H]  &  $\sigma$(La)  & [Ce/H]  &  $\sigma$(Ce)  &  [Pr/H]   &   $\sigma$(Pr)&  [Nd/H] &    $\sigma$(Nd)&   [Sm/H] &   $\sigma$(Sm)&  [Eu/H] &   $\sigma$(Eu)    \\
 \hline  

Sequoia              &            &          &          &           &          &         &         &           &           &          &          &          \\
 \hline                                                                                                                                                                                                                                                                                                   
HD115575        &  -2.06  &  0.06   &  -2.19  &   0.06  &   -2.17   &  0.04  &   -2.02 &    0.03 &    -1.84 &   0.05 &   -1.83  &  0.00   \\
TYC 4267-2023-1 &  -1.64  &  0.07   &  -1.81  &   0.07  &   -1.74   &  0.09  &   -1.61 &    0.05 &    -1.47 &   0.04 &   -1.43  &  0.00   \\
BD+31 2143      &  -2.42  &  0.05   &  -2.50  &   0.04  &   -2.19   &  0.08  &   -2.37 &    0.05 &    -2.10 &   0.04 &   -2.23  &  0.00   \\
\hline
GSE             &         &         &         &         &           &        &         &         &          &        &          &         \\
\hline
BD+20 3298      &  -1.77  &  0.05   &  -1.96  &   0.07  &   -1.53   &  0.10  &   -1.73 &    0.03 &    -1.55 &   0.09 &   -1.33  &  0.00   \\
TYC 1008-1200-1 &  -2.01  &  0.04   &  -2.20  &   0.04  &   -1.95   &  0.03  &   -1.93 &    0.04 &    -1.74 &   0.02 &   -1.68  &  0.05   \\
HD 238439       &  -2.03  &  0.05   &  -2.15  &   0.11  &   -2.02   &  0.04  &   -2.01 &    0.04 &    -1.86 &   0.03 &   -1.63  &  0.00   \\
HD 142614       &  -1.11  &  0.05   &  -1.34  &   0.07  &   -1.12   &  0.04  &   -1.07 &    0.05 &    -0.79 &   0.06 &   -0.68  &  0.05   \\
BD +04 18       &  -0.90  &  0.05   &  -1.19  &   0.13  &   -0.62   &  0.09  &   -0.82 &    0.09 &    -0.57 &   0.06 &   -0.33  &  0.00   \\
BD+39 3309      &   ----  &  ----   &   ----  &   ----  &    ----   &  ----  &    ---- &    ---- &     ---- &   ---- &    ----  &  ----   \\
TYC 2824-1963-1 &  -1.28  &  0.04   &  -1.57  &   0.02  &   -1.22   &  0.04  &   -1.22 &    0.04 &    -1.00 &   0.10 &   -0.81  &  0.13   \\
TYC 4001-1161-1 &  -1.39  &  0.08   &  -1.53  &   0.06  &   -1.36   &  0.06  &   -1.28 &    0.05 &    -1.04 &   0.12 &   -0.78  &  0.00   \\
TYC 4221-640-1  &  -2.47  &  0.03   &  -2.63  &   0.08  &   -2.32   &  0.00  &   -2.29 &    0.07 &    -1.99 &   0.08 &   -1.78  &  0.00   \\
TYC 4-369-1     &  -1.54  &  0.04   &  -1.67  &   0.05  &   -1.56   &  0.04  &   -1.47 &    0.05 &    -1.34 &   0.04 &   -1.26  &  0.07   \\
\hline
other           &         &         &         &         &           &        &         &         &          &        &          &         \\
\hline
          
BD -00 4538      &  -1.65  &  0.03   &  -1.73  &   0.05  &   -1.50   &  0.06  &   -1.62 &    0.04 &    -1.37 &   0.03 &   -1.28  &  0.04   \\
BD +03 4904     &  -2.61  &  0.02   &  -2.73  &   0.02  &    ----   &  ----  &   -2.52 &    0.04 &    -2.01 &   0.05 &    ----  &  ----   \\
BD +07 4625     &  -1.75  &  0.07   &  -1.89  &   0.04  &    ----   &  0.19  &   -1.71 &    0.05 &    -1.52 &   0.04 &   -1.53  &  0.00   \\
BD+11 2896      &  -1.13  &  0.07   &  -1.38  &   0.07  &   -0.92   &  0.12  &   -1.05 &    0.02 &    -0.80 &   0.11 &   -0.63  &  0.00   \\
BD +21 4759     &  -1.85  &  0.03   &  -2.07  &   0.06  &    ----   &  0.06  &   -1.73 &    0.05 &    -1.47 &   0.05 &   -1.23  &  0.00   \\
BD +25 4520     &  -2.50  &  0.04   &  -2.72  &   0.10  &    ----   &  ----  &   -2.43 &    0.04 &    -2.16 &   0.05 &    ----  &  ----   \\
BD +32 2483     &  -2.28  &  0.04   &  -2.45  &   0.04  &   -2.12   &  0.05  &   -2.22 &    0.04 &    -2.04 &   0.06 &    ----  &  ----   \\
BD +35 4847     &  -1.43  &  0.08   &  -1.69  &   0.05  &   -1.38   &  0.08  &   -1.56 &    0.05 &    -1.45 &   0.06 &   -1.33  &  0.00   \\
BD +48 2167     &  -2.22  &  0.04   &  -2.38  &   0.06  &   -1.97   &  0.10  &   -2.21 &    0.06 &    -1.93 &   0.08 &   -1.78  &  0.00   \\
BD -07 3523     &  -1.71  &  0.02   &  -1.88  &   0.06  &   -1.45   &  0.10  &   -1.61 &    0.04 &    -1.42 &   0.03 &   -1.18  &  0.00   \\
BD +06 2880     &  -1.15  &  0.04   &  -1.48  &   0.10  &   -1.03   &  0.10  &   -1.10 &    0.06 &    -0.95 &   0.05 &   -0.78  &  0.04   \\
HD 139423       &  -1.49  &  0.04   &  -1.72  &   0.02  &   -1.41   &  0.04  &   -1.39 &    0.04 &    -1.18 &   0.02 &   -1.13  &  0.05   \\
HD 208316       &  -1.49  &  0.04   &  -1.68  &   0.09  &   -1.51   &  0.04  &   -1.47 &    0.04 &    -1.30 &   0.05 &   -1.13  &  0.00   \\
HD 354750       &  -2.26  &  0.05   &  -2.53  &   0.08  &    ----   &  ----  &   -2.16 &    0.05 &    -1.94 &   0.06 &   -1.83  &  0.00   \\
TYC 2588-1386-1 &  -1.47  &  0.06   &  -1.69  &   0.13  &   -1.52   &  0.04  &   -1.41 &    0.07 &    -1.24 &   0.03 &   -1.13  &  0.00   \\
TYC 3085-119-1$^a$  &  -1.04  &  0.03   &  -1.19  &   0.07  &   -1.15   &  0.02  &   -1.11 &    0.04 &    -0.93 &   0.06 &   -0.88  &  0.05   \\
TYC 33-446-1    &  -2.04  &  0.05   &  -2.15  &   0.04  &   -1.97   &  0.08  &   -1.93 &    0.03 &    -1.86 &   0.07 &   -0.58  &  1.05   \\
TYC 3944-698-1  &  -2.20  &  0.02   &  -2.33  &   0.05  &   -2.02   &  0.00  &   -2.16 &    0.03 &    -1.94 &   0.08 &   -1.63  &  0.00   \\
TYC 4331-136-1  &  -2.34  &  0.03   &  -2.57  &   0.06  &   -2.32   &  0.04  &   -2.29 &    0.06 &    -2.02 &   0.04 &   -1.78  &  0.05   \\
TYC 4584-784-1  &  -1.76  &  0.06   &  -1.94  &   0.08  &   -1.70   &  0.05  &   -1.72 &    0.04 &    -1.55 &   0.05 &   -1.38  &  0.05   \\
\hline
\\
  \multicolumn{4}{l}{$^a$ Thick-disc star}\\
\end{tabular}
\end{table*}

\end{appendix}






%




%



\end{document}